\documentclass[aps,pre,reprint,superscriptaddress,nofootinbib]{revtex4-2}

\usepackage{physics}
\usepackage{graphicx}% Include figure files
\usepackage{dcolumn}% Align table columns on decimal point
\usepackage{bm}% bold math
\usepackage{bbold}
\usepackage{mathtools}
\usepackage{hyperref}

\DeclareUnicodeCharacter{2061}{}

\newcommand{\nyuphysics}{Center for Soft Matter Research, Department of Physics, New York University, New York 10003, USA}
\newcommand{\nyuchemistry}{Department of Chemistry, New York University, New York 10003, USA}
\newcommand{\nyusimons}{Simons Center for Computational Physical Chemistry, Department of Chemistry, New York University, New York 10003, USA}
\newcommand{\nyucourant}{Courant Institute of Mathematical Sciences, New York University, New York 10003, USA}
\newcommand{\nyucns}{Center for Neural Science, New York University, New York 10003, USA}

\begin{document}

\title{The Basins of Attraction of Soft Sphere Packings Are Not Fractal}

\author{Praharsh Suryadevara}
\thanks{P.S. and M.C. contributed equally to this work.}
\affiliation{\nyuphysics}

\author{Mathias Casiulis}
\affiliation{\nyuphysics}
\affiliation{\nyusimons}

\author{Stefano Martiniani}
\email{sm7683@nyu.edu}
\affiliation{\nyuphysics}
\affiliation{\nyusimons}
\affiliation{\nyuchemistry}
\affiliation{\nyucns}
\affiliation{\nyucourant}


\begin{abstract}
  The energy landscape picture is a central tool to study many-body systems.
  In particular, the energy landscapes of glass-forming liquids, jammed packings, constraint satisfaction problems, or neural networks contain a plethora of minima corresponding to competing states.
  Due to their complexity, these landscapes resist analytical treatment and must be studied numerically.
  We focus on jammed soft spheres, a paradigmatic model of glasses and granulars, to expose the limitations of standard numerical methods in resolving the true structure of energy landscapes.
  We show that the ODE solver with the best time-for-error trade-off, outperforming commonly used steepest-descent solvers by several orders of magnitude, is the C-language Variable-coefficients ODE (CVODE) solver.
 Using this numerical approach, we provide unequivocal evidence that optimizers widely used in computational studies destroy all semblance of the true landscape geometry, even in moderately low dimensions.
  Employing a range of geometric indicators, both low- and high-dimensional, we show that earlier claims on the fractality of basins of attraction of minima originated from the use of inadequate mapping strategies.
  In reality, the basins of attraction of soft sphere packings are smooth structures with well-defined length scales, a result that likely extends to a much broader family of problems.
\end{abstract}

\date{\today}

%%%%%%%%%%%% Make Title %%%%%%%%
\maketitle

%%%%%%%%%%%%%%%%%%%%%%%%% Main text %%%%%%%%%%%%%%%%%%%%%%%%%%

\begin{figure}[h!]
    \centering
    \includegraphics[width = 0.96\columnwidth]{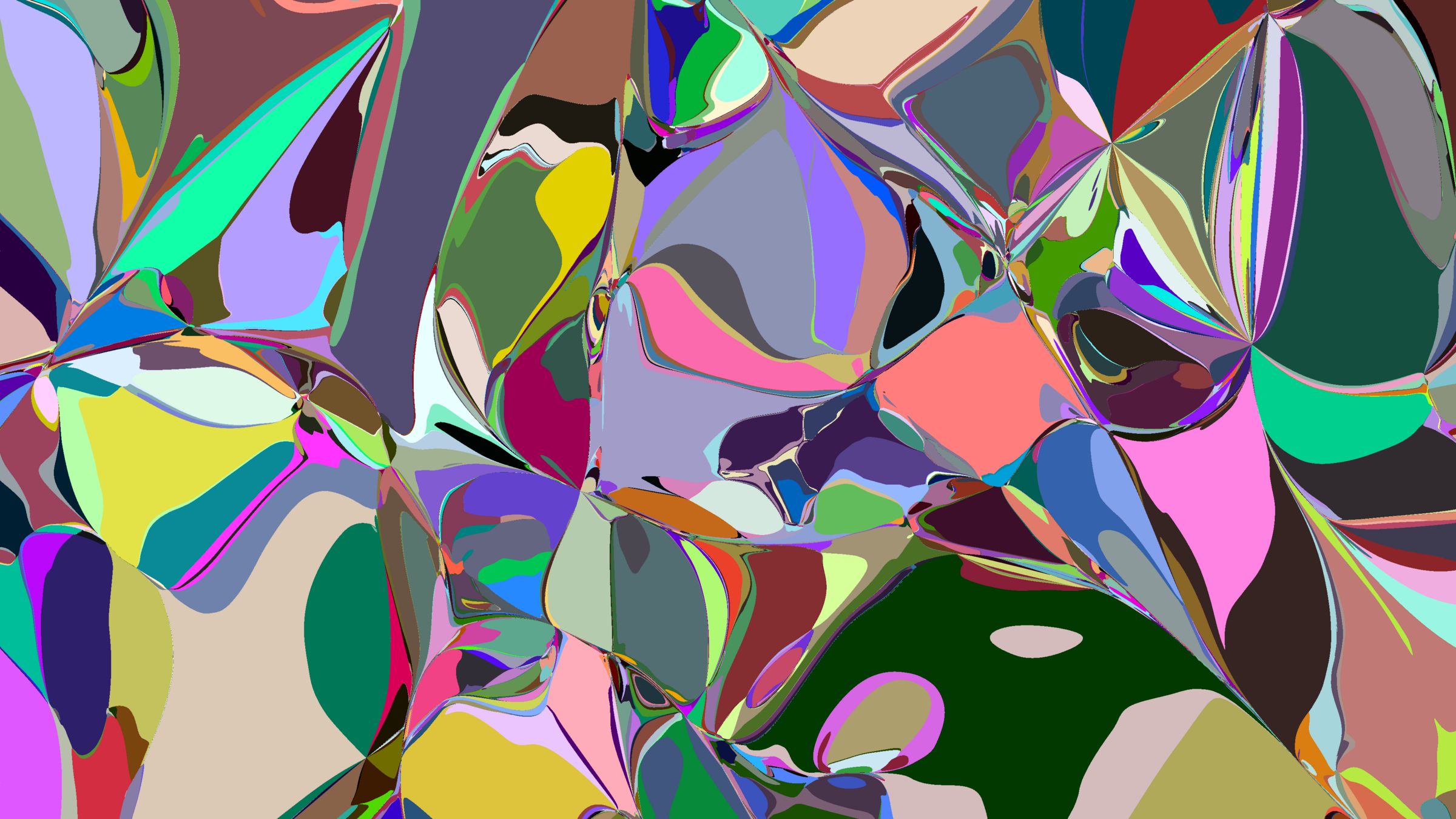}
    \includegraphics[width = 0.96\columnwidth]{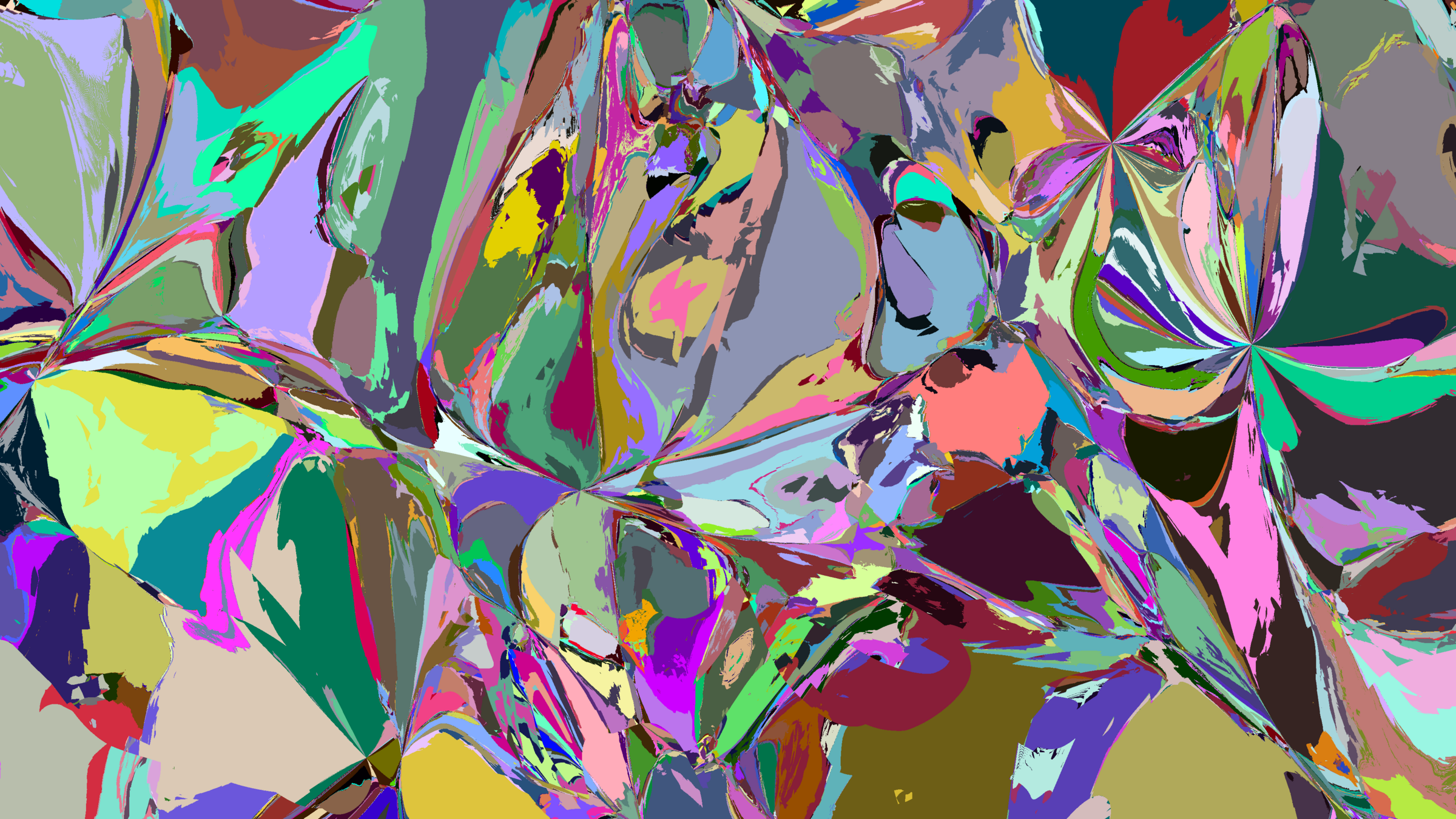}
    \caption{\textbf{Slicing the energy landscape.}
    $1350 \times 2400$ pixels in a random $2d$ plane in the configuration space of $N = 16$ disks.
    At each pixel, we use CVODE (top) and  FIRE (bottom), to identify which basin of attraction it belongs to.
    Each basin is uniquely encoded by one color across both panels.
    }
    \label{fig:LandscapeSlices}
\end{figure}

The energy landscape framework, which maps the configuration of a system of $N$ particles interacting through conservative forces in $d$ dimensions to a single point on a $dN$-dimensional potential energy surface, is a central tool in the study of many-body systems~\cite{Onuchic1997,Stillinger2015,Aldeghi2022}.
In glassy systems, mean-field theories predict that the (free) energy landscape is characteristically rough, giving rise to rich physical behavior~\cite{Parisi2010,Charbonneau2014,Biroli2016,Folena2020}.
Many parallels have been drawn between glassy landscapes and gradient descent problems that display many minima, notably in constraint satisfaction~\cite{Krzakala2007,Franz2017,Sclocchi2022} and machine learning~\cite{Choromanska2015, BaityJesi2019, Franz2019,Mannelli2019,Fabbricatore2022,Baldassi2023,Kamali2023,Winter2025}.
In particular, in neural networks, the structure and geometry of the loss landscape determine both learning dynamics and generalization error~\cite{Wu2017,Li2018,Baldassi2020,Pittorino2021,Feng2021,Baldassi2023,Pouplin2023a,Lourie2025}.

Many studies have examined the number and arrangement of low-lying minima in these landscapes~\cite{Stillinger1982, Stillinger1984, Heuer1997,Fyodorov2004,Xu2011,Fyodorov2012,Charbonneau2014,Ros2019,Ros2023,Hagh2024}, as well as the probabilities of finding states near each minimum~\cite{Martiniani2016, Martiniani2017a, Martiniani2019, Martiniani2017, Asenjo2014, Sciortino1999, Sciortino2005, Xu2011}.
These properties are naturally probed by studying the \textit{basins of attraction}~\cite{Stillinger2015}, the sets of configurations that flow to a given minimum under steepest-descent dynamics, \textit{i.e.} the dynamics followed after an instantaneous quench to zero temperature.
Basin volumes are proportional to the probability of relaxing to a given minimum from uniformly distributed initial conditions; they are thus estimators of configurational entropies~\cite{Xu2011, Asenjo2014, Martiniani2016}.
The geometry and connectivity of the basins, in turn, govern relaxation pathways~\cite{Thirumalaiswamy2022,Pacco2025,Charbonneau2025}.

Fractality of the energy landscape has been discussed in several contexts, including, the fractal organization of basins in the Gardner phase,~\cite{Charbonneau2014,Dennis2020, Altieri2021}, and the fractal structure of relaxation paths~\cite{Lidar1999,Hwang2016,Li2021, Liu2021a, Thirumalaiswamy2022}.
Inspired by chaotic dynamical systems~\cite{Strogatz,McDonald1985,Grebogi1987,Nusse1996,Sprott2015,Levi2018,Daza2022,Bollt2023}, it has further been suggested that the basins of attraction themselves may be fractal in neural networks~\cite{SohlDickstein2024,Ly2025}, constraint satisfaction problems~\cite{Ercsey-Ravasz2011,Varga2016} and energy landscapes at large~\cite{Wales1992, Chakravarty2005, Massen2007,Asenjo2013,Bautista2023,Interiano-Alberto2024,Hagh2024}.

Counter-intuitively, such claims cannot easily be dismissed on the basis of the functional form of the interactions between particles. Indeed, the smoothness of the velocity field in a continuous-time dynamical system is not enough to rule out fractal basins of attraction. Simple examples illustrating this point exist in the dynamical-systems literature~\cite{Rossler1979,Sprott1997,Sprott2015}. Specifically, fractality has been reported in certain regimes of constraint satisfaction problems that map onto gradient descent dynamics with smooth velocities and only fixed points as stable attractors~\cite{Ercsey-Ravasz2011}, showing that even gradient descent on a smooth potential can, in principle, exhibit fractal basins of attraction.

These observations and the many claims of fractality of the landscape of jammed soft particles thus behove a systematic study of the geometry of their basins of attraction.

\textit{Model --} In this work, we focus on a paradigmatic model of repulsive soft disks~\cite{Wales1992, Heuer1997,Saksaengwijit2003,Bogdan2006, Xu2011,Ashwin2012,Asenjo2013,Asenjo2014,Hwang2016,Martiniani2016,Martiniani2016a, Martiniani2017a, Wales2018, Dennis2020, Boltz2021,Thirumalaiswamy2022,Hagh2024}, namely two-dimensional polydisperse collections of $N$ particles in periodic squares with sidelength $L$ interacting via a Hertzian repulsive potential (see App.~\ref{app:model}).
Introducing the packing fraction $\phi$, the system undergoes a jamming transition at $\phi_J \approx 0.84$~\cite{OHern2003}, see App.~\ref{app:jamming-point}.
For $\phi < \phi_J$, the energy is minimized in flat regions with $E = 0$, or liquid states, where all contacts between particles can be removed at no energy cost.
For $\phi \geq \phi_J$, most minima of the energy comprise a backbone of particles that are collectively stuck in place, such that displacing any backbone particle increases $E$, and only a few \textit{rattlers} remain free to move without affecting the energy.
Unless otherwise specified, we henceforth work at $\phi = 0.9$.

Characterizing basins of attraction is notoriously difficult due to the high dimensionality of configuration space, $D=2(N-1)$.
In high dimensions, even simple geometric objects like hypercubes are costly to measure, as their volume concentrates in a thin layer, deep into the exponentially many narrow corners~\cite{ArtsteinAvidan2015,Casiulis2023}.
Basins of attraction are typically far more complex, exhibiting non-trivial, non-convex geometries~\cite{Martiniani2016a, Martiniani2017} (Fig.~\ref{fig:LandscapeSlices} and video in SM~\cite{[{See Supplemental Material at }][{ for a detailed discussion of the choice of ODE solver and of its parameters, a complete description of numerical methods, a video of slices of basins obtained by moving along a third orthogonal direction, and additional data on the effect of both $N$ and $\phi$ on the landscape.}]supp}), but these geometric intricacies are only one aspect of the numerical challenge posed by measuring basins of attraction.

Determining whether a given point lies inside a basin requires integrating the steepest-descent ordinary differential equation (ODE) in $D$ dimensions.
In soft spheres close to jamming, the landscape exhibits nearly flat directions around minima, leading to ill-conditioned Hessians and ``stiff'' ODEs, meaning that they are prone to instability when using explicit methods like the forward-Euler scheme, and that an appropriate implicit time stepping method is required.
As integrating such ODEs accurately is computationally intensive~\cite{Bautista2026}, most prior work performed energy quenches using heuristic optimizers.
These have most often been momentum-based methods, that approximate steepest descent by adding an inertial term, like the Fast Inertial Relaxation Engine (FIRE),~\cite{Bitzek2006} or quasi-Newton schemes, that estimate the local curvature of the landscape in addition to the gradient to accelerate convergence, like the Limited-memory Broyden–Fletcher–Goldfarb–Shanno algorithm (L-BFGS)~\cite{Liu1989} (see SM Sec.~1.A. for precise dynamics~\cite{supp}).

In this work, we show that optimizers destroy geometric features of the energy landscape of soft disks and bias the sampling of minima, even at moderate $N$ (Fig.~\ref{fig:LandscapeSlices}).
In particular, basins of attraction appear fractal when using popular yet inadequate optimization algorithms, but this mirage is dispelled when using appropriate methods.

\textit{Unscrambling the energy landscape --}
We benchmark common ODE solvers~\cite{Rackauckas2017,Rackauckas2019,supp} for Hertzian disks, and find that the best time-for-error is achieved by the C-language Variable-coefficients ODE (CVODE) solver~\cite{Hindmarsh2005,Gardner2022} (which also performs well in potentials other than Hertzian~\cite{supp}).
For each set $(N, \phi)$, we tighten the tolerance of CVODE until the minima associated with a collection of random points in configuration space stop changing~\cite{supp}.
Armed with this fast, accurate map of configuration space, we measure how much error is introduced by using optimizers to identify basins.

First, we uniformly draw a random $2d$ plane in configuration space~\cite{supp}.
On that plane, we define a grid, and use each pixel as an initial condition for steepest descent.
Pixels are then colored according to the minimum that they asymptotically reach, \textit{i.e.} which basin they belong to.
In Fig.~\ref{fig:LandscapeSlices}, we show the result using CVODE (top) and FIRE~\cite{Bitzek2006} (bottom), the most widely used optimizer in jammed systems~\cite{Goodrich2014,Martiniani2017a,Artiaco2020,Dennis2020,Charbonneau2021,Rissone2021,Hagh2024} (we use a strictly downhill variant of FIRE with limited stepsize~\cite{Asenjo2014,supp}).
Minima are matched across the two panels based on the metric distance between them (after rattler removal~\cite{supp}).
Even in moderate dimension, $D = 30$, FIRE scrambles the basins, creates discontinuities in their shapes, and alters their sizes.
The apparent roughness of basins in $2d$ cuts was argued to be a feature of basins, and an indicator of the fractal nature of their geometry~\cite{Wales1992,Asenjo2013,SohlDickstein2024}.
These features are in fact artifacts of inaccurate noiseless relaxations, that map points to the wrong basins of attraction~\footnote{This is reminiscent of fractals generated by the iterative Newton-Raphson method on equations that yield roots of unity~\cite{Epureanu1998}.}.

\begin{figure}[h!]
    \centering
    \includegraphics[height=0.46\columnwidth]{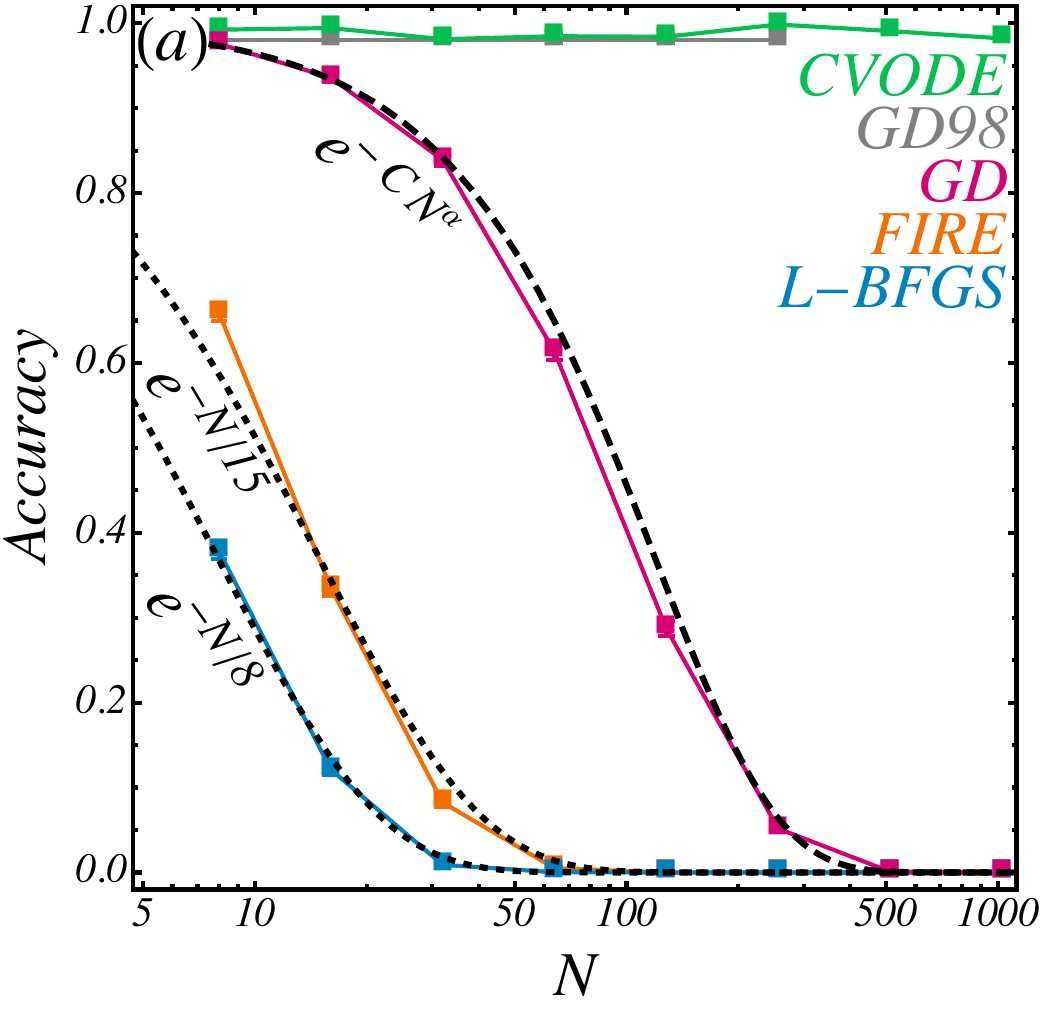}
    \includegraphics[height=0.46\columnwidth]{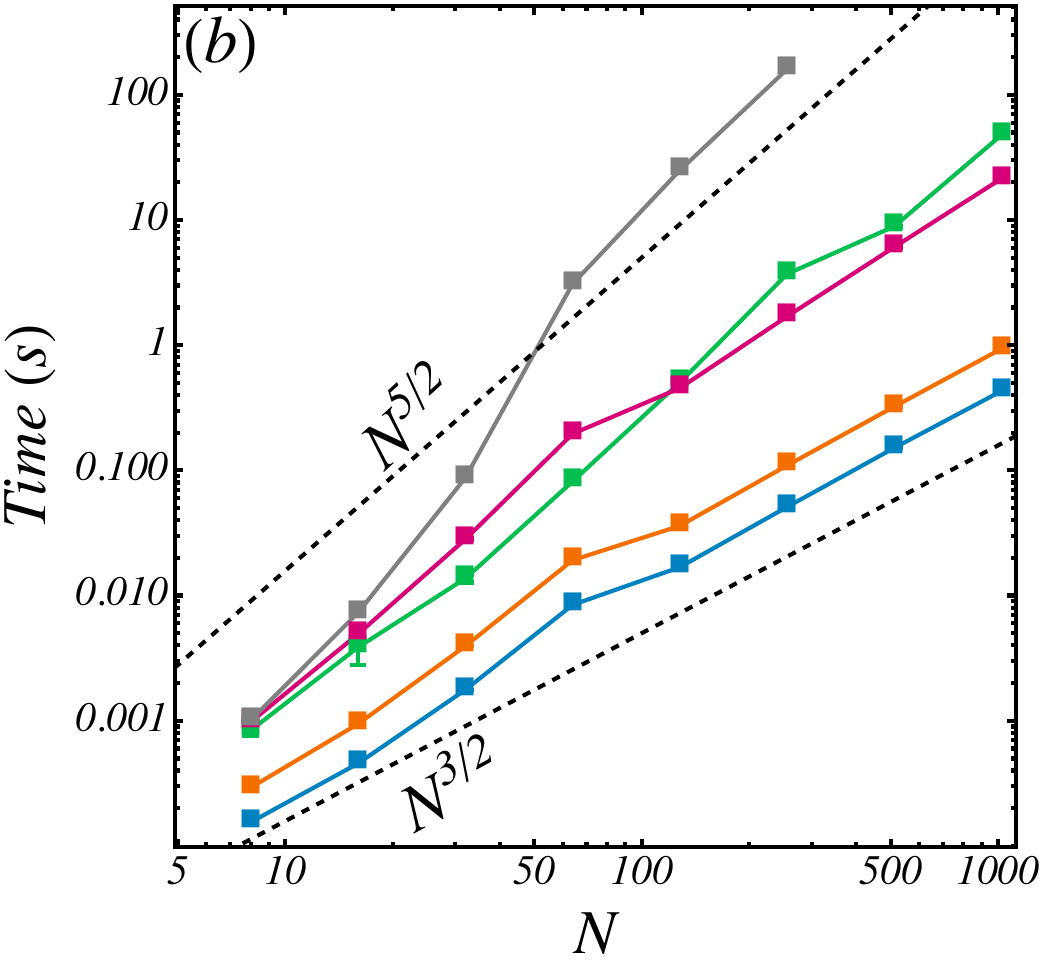} \\
    \includegraphics[width = 0.24\columnwidth]{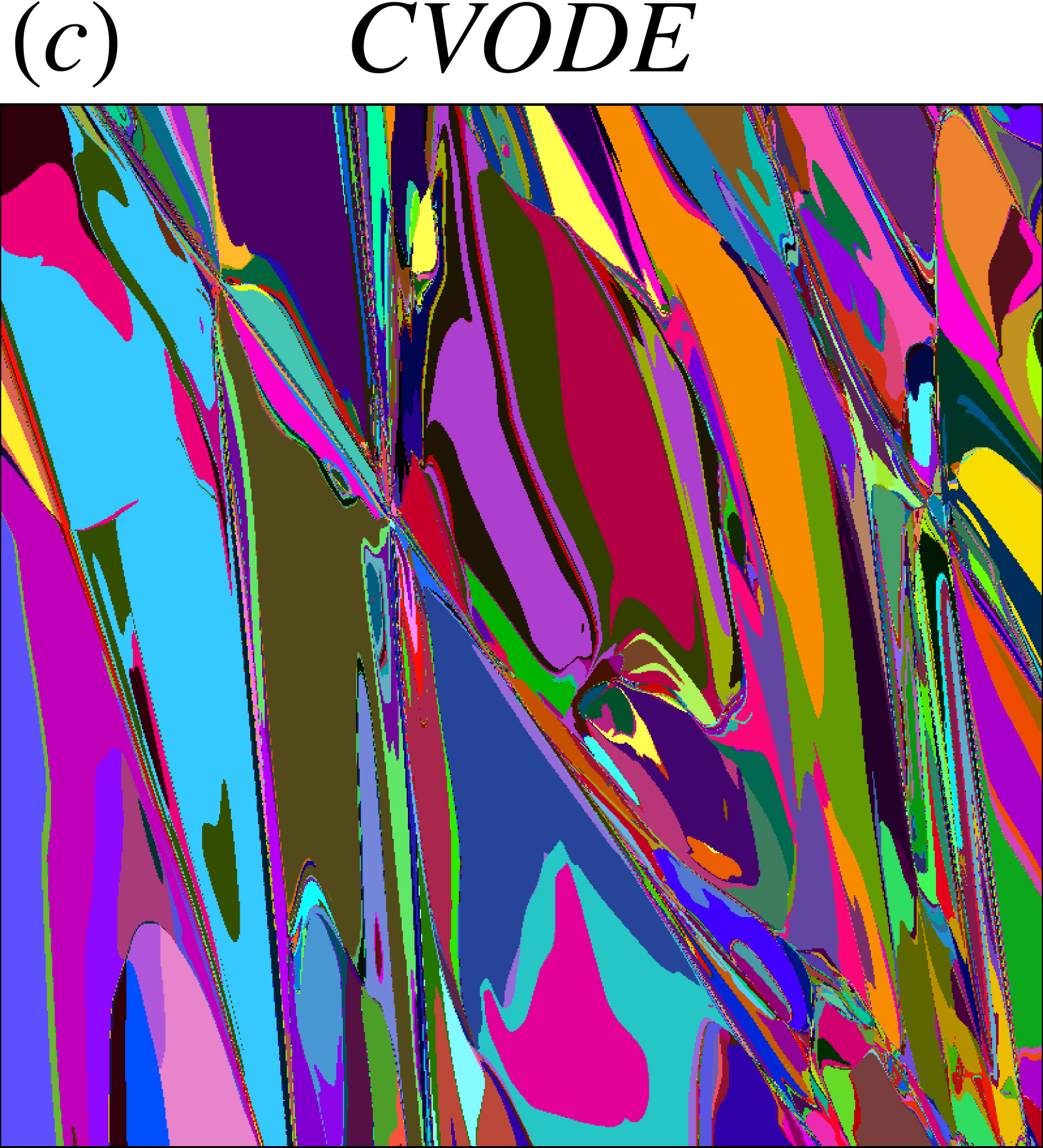}
    \includegraphics[width = 0.24\columnwidth]{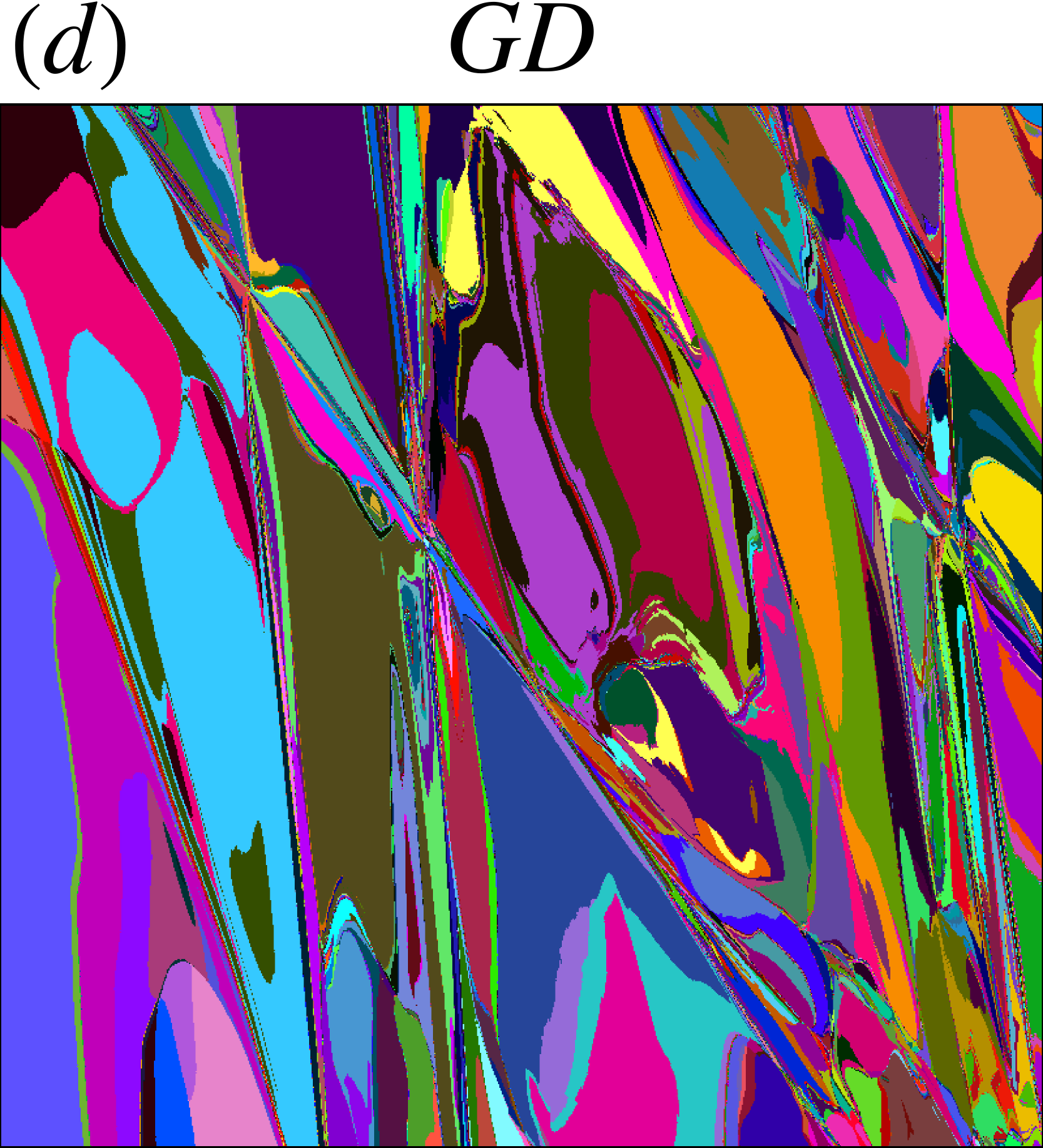}
    \includegraphics[width = 0.24\columnwidth]{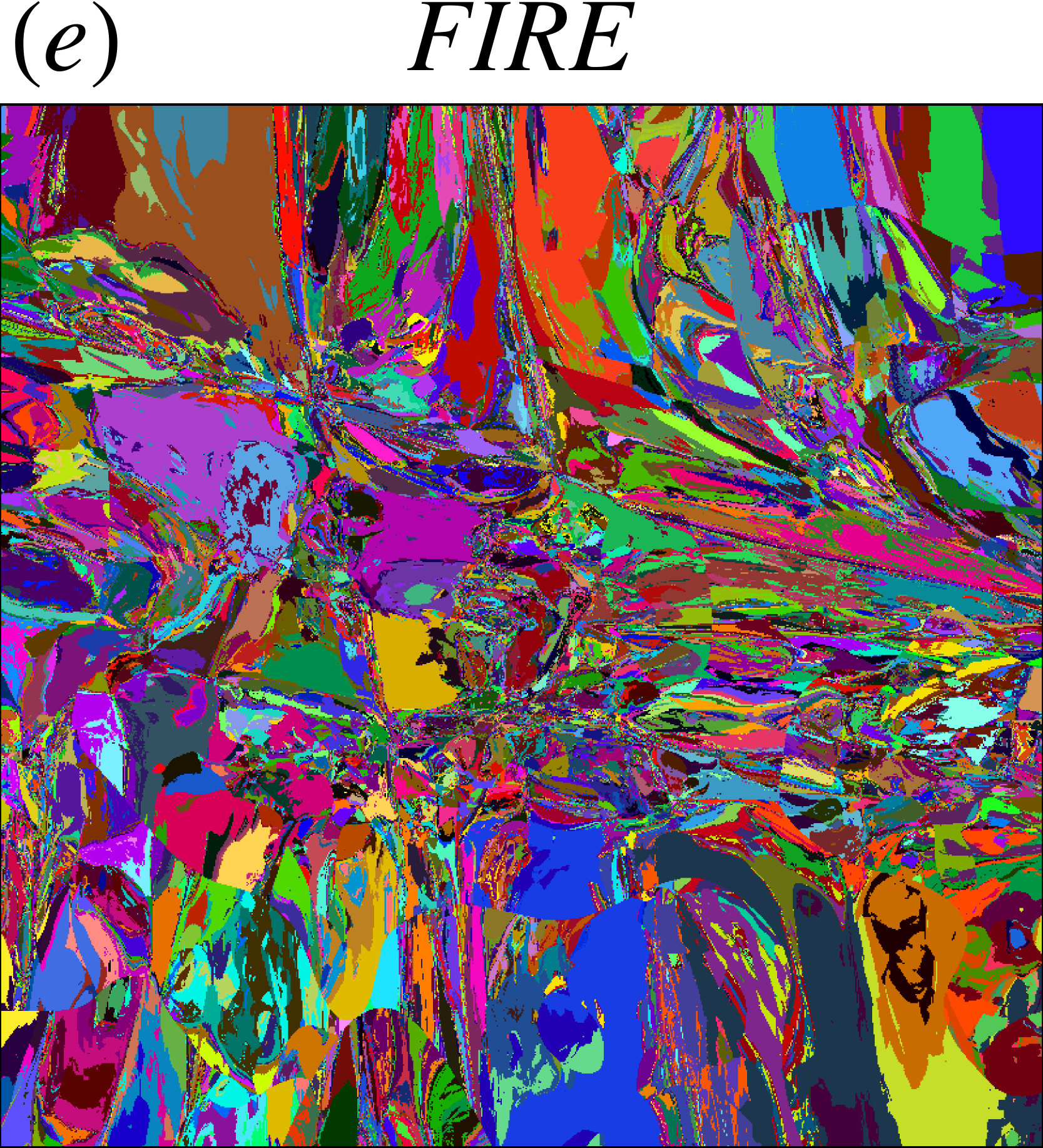}
    \includegraphics[width = 0.24\columnwidth]{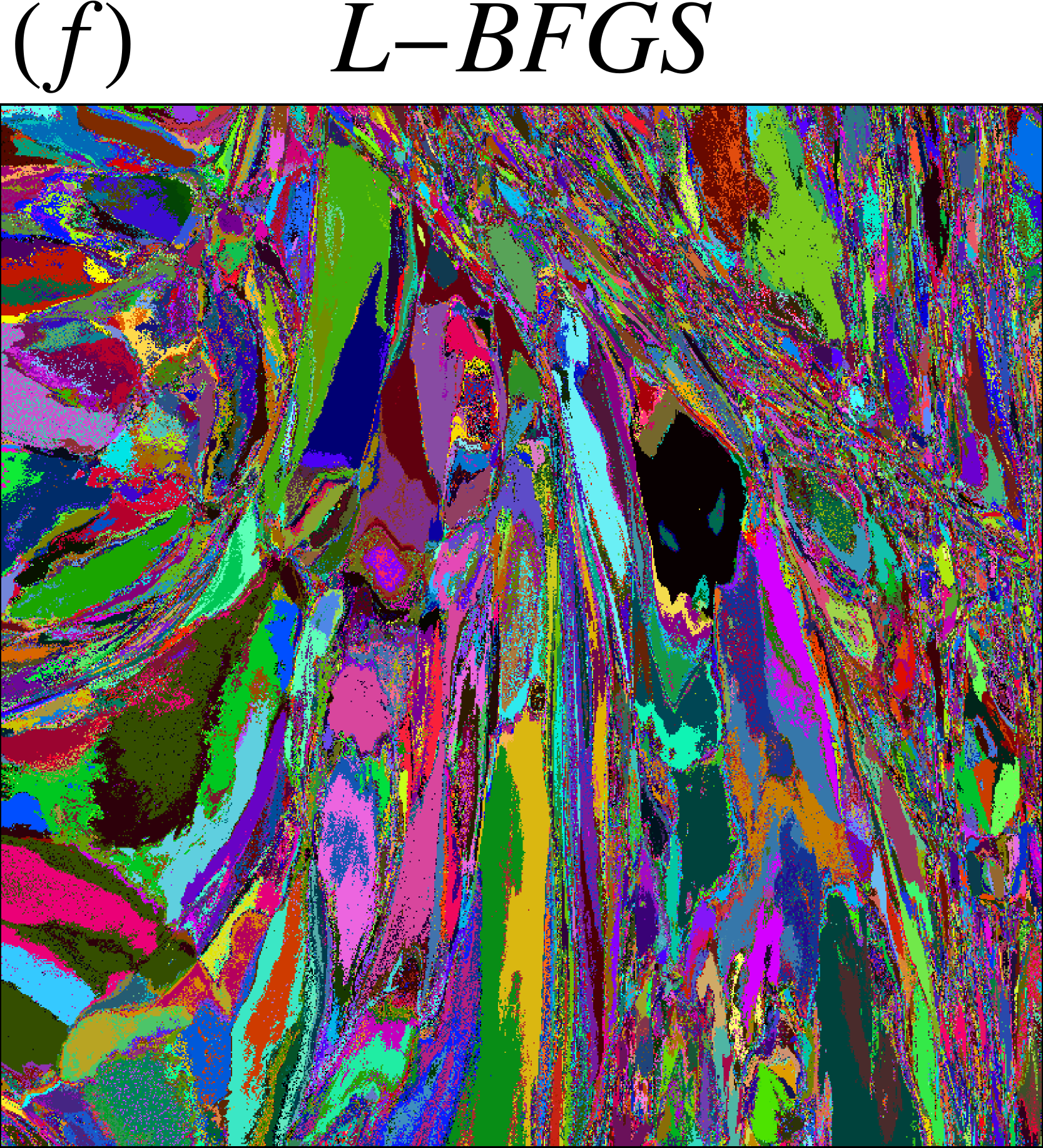}
    \caption{\textbf{Optimizers: fast but inaccurate.}
    $(a)$ Average accuracy of algorithms, computed over $10^4$ random points.
    Error bars are Clopper-Pearson $95\%$ confidence intervals~\cite{Clopper1934,supp}.
    Dashed lines are exponential fits, long-dashed line stretched exponential fits.
    $(b)$ Corresponding average computation times.
    Error bars are Student-T $95\%$ confidence intervals~\cite{supp}.
    $(c)-(e)$ $800\times 800$-pixel slices of configuration space for $N=128$ particles for $(c)$ CVODE, 
    $(d)$ GD, $(e)$ FIRE, and $(f)$ L-BFGS.
    }
    \label{fig:AccuracyTimevsN}
\end{figure}
Quantitatively, in Fig.~\ref{fig:AccuracyTimevsN}, we consider independent random points drawn from the configuration spaces of collections of Hertzian disks with $N \in \left[ 8; 4096\right]$.
Panel $(a)$ shows the accuracy, \textit{i.e.} the fraction of points that are mapped to the right minimum, against $N$ across methods, with the ground truth given by low-tolerance CVODE ($\texttt{rtol} = 10^{-14}$, see SM Sec.~1~\cite{supp}).
The CVODE values are obtained using a looser tolerance, $\texttt{rtol} = 10^{-13}$, showing that it retains high accuracy.
While FIRE and L-BFGS are relatively accurate for $D \lesssim 10$~\cite{Asenjo2013}, their accuracies fall exponentially with $N$ (dashed lines).
For systems with $N \gtrsim N_0 \approx64$, optimizers practically \textit{never} map a point in configurational space to the right basin of attraction.
Furthermore, $N_0$ goes down as $\phi$ decreases, see SM Sec.~3.F~\cite{supp}.
We also check in SM Sec.~3.D~\cite{supp} that the implementation of L-BFGS provided in the broadly used GMIN library~\cite{GMIN} yields qualitatively similar results for accuracy.
We also test an adaptive Gradient Descent algorithm~\cite{Charbonneau2023,supp} with 2 choices of the parameter $\epsilon$ that sets the stepsize.
If $\epsilon$ is fixed (GD), accuracy also falls (like a stretched exponential), and approaches zero for $N > 128$ for the $\epsilon$ used in Ref.~\cite{Charbonneau2023}.
If now $\epsilon$ is set at each $N$ to achieve $98\%$ accuracy (GD98), it comes at a significant time cost.
This is shown in panel $(b)$, where we plot the corresponding wall times against $N$.
While CVODE is slower than FIRE and L-BFGS ($\mathcal{O}(N^{5/2})$ vs. $\mathcal{O}(N^{3/2})$), it achieves computation times much smaller than GD98 (and comparable to GD ones) with better accuracy.
In Fig.~\ref{fig:AccuracyTimevsN}$(c)-(f)$, we show slices obtained like Fig.~\ref{fig:LandscapeSlices}, but this time at $N = 128$, where the accuracies of FIRE and L-BFGS are essentially zero.
These slices reveal a much starker contrast between methods than Fig.~\ref{fig:LandscapeSlices}, as FIRE and L-BFGS turn the whole landscape into an unrecognizable collection of confetti-like, largely disconnected basins.
In fact, \textit{not a single pixel} of the slices obtained with these optimizers falls into the right basin.
The GD slice still looks reasonably smooth, but displays large regions of incorrectly tagged basins.

These results have far-reaching consequences.
Save from the few studies that considered very small systems (\textit{e.g.}~\cite{Xu2005, Gao2006, Xu2011, Ashwin2012, Asenjo2013}), the vast majority of past works likely misattributed every single basin.
Furthermore, using an accurate GD requires unreasonable computation times for this problem.

Interestingly, the distribution of minimum energies found starting from uniform random initial condition, as well as estimates of $\phi_J$, are also altered, see Apps.~\ref{app:energies} and~\ref{app:jamming-point}.
In short, optimizers bias the mean energy downwards by an amount that grows subextensively with $N$.

\textit{Generality of the accuracy results --}
We now briefly discuss the generality of the inaccuracy uncovered in Figs.~\ref{fig:LandscapeSlices} and~\ref{fig:AccuracyTimevsN}.
To do so, we perform two additional numerical experiments.

\begin{figure}
    \centering
    \includegraphics[height=0.32\columnwidth]{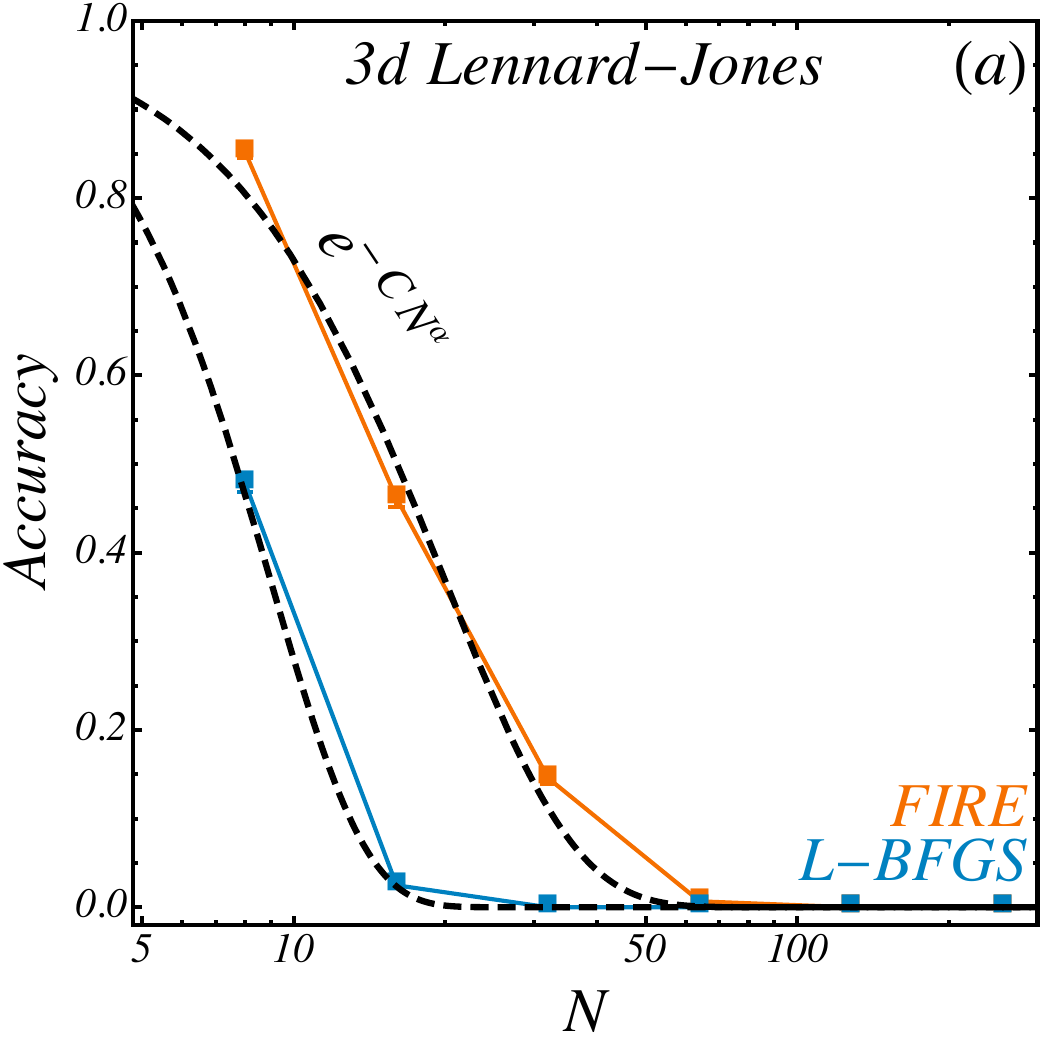}
    \includegraphics[height=0.32\columnwidth]{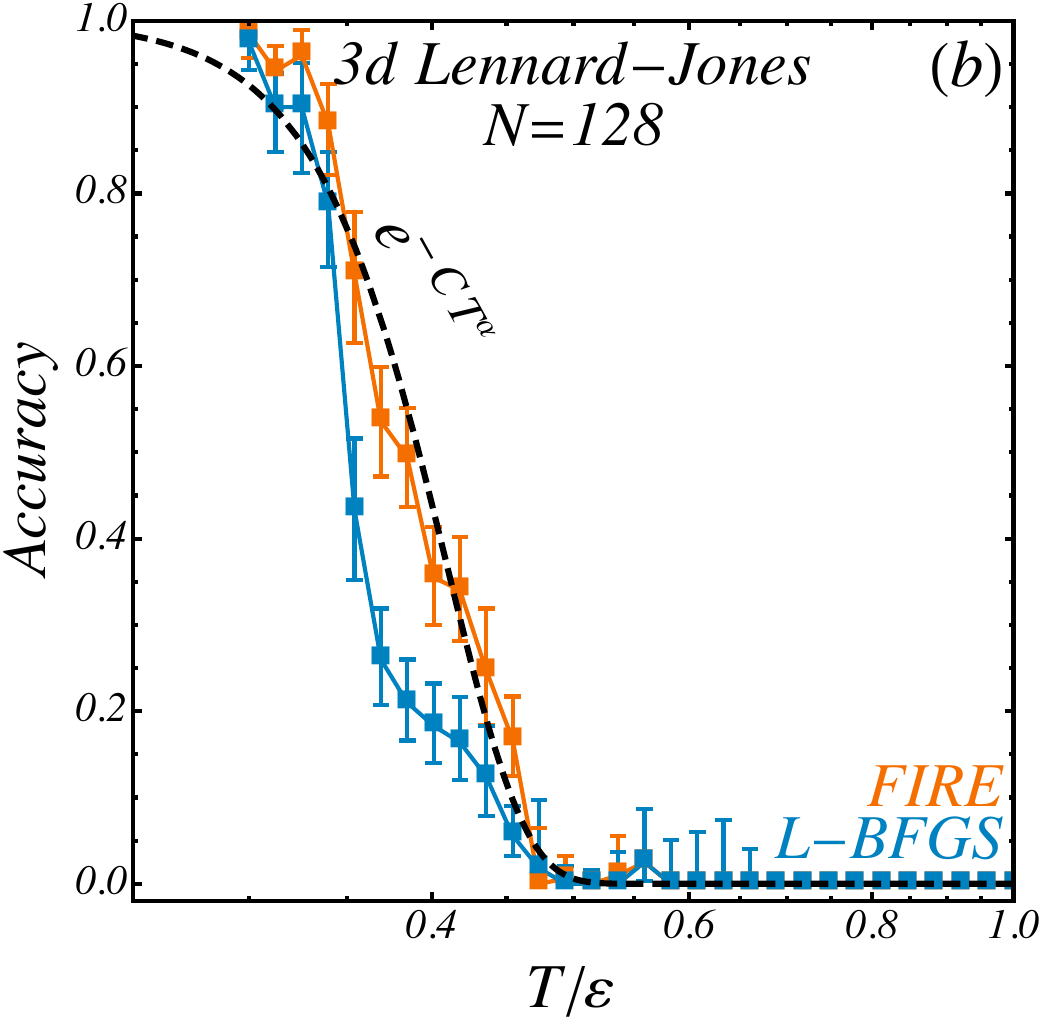}
    \includegraphics[height=0.32\columnwidth]{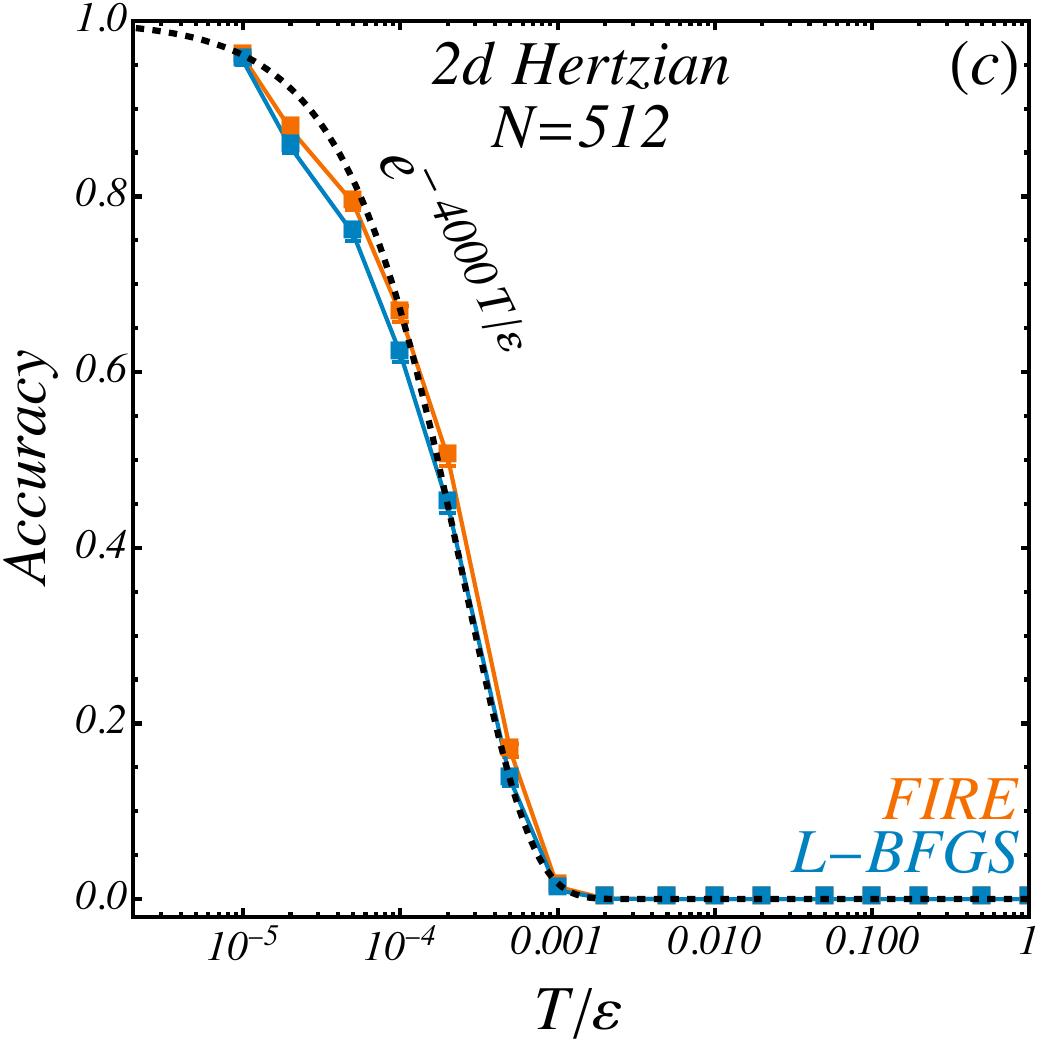} \\
    \includegraphics[width=0.32\columnwidth]{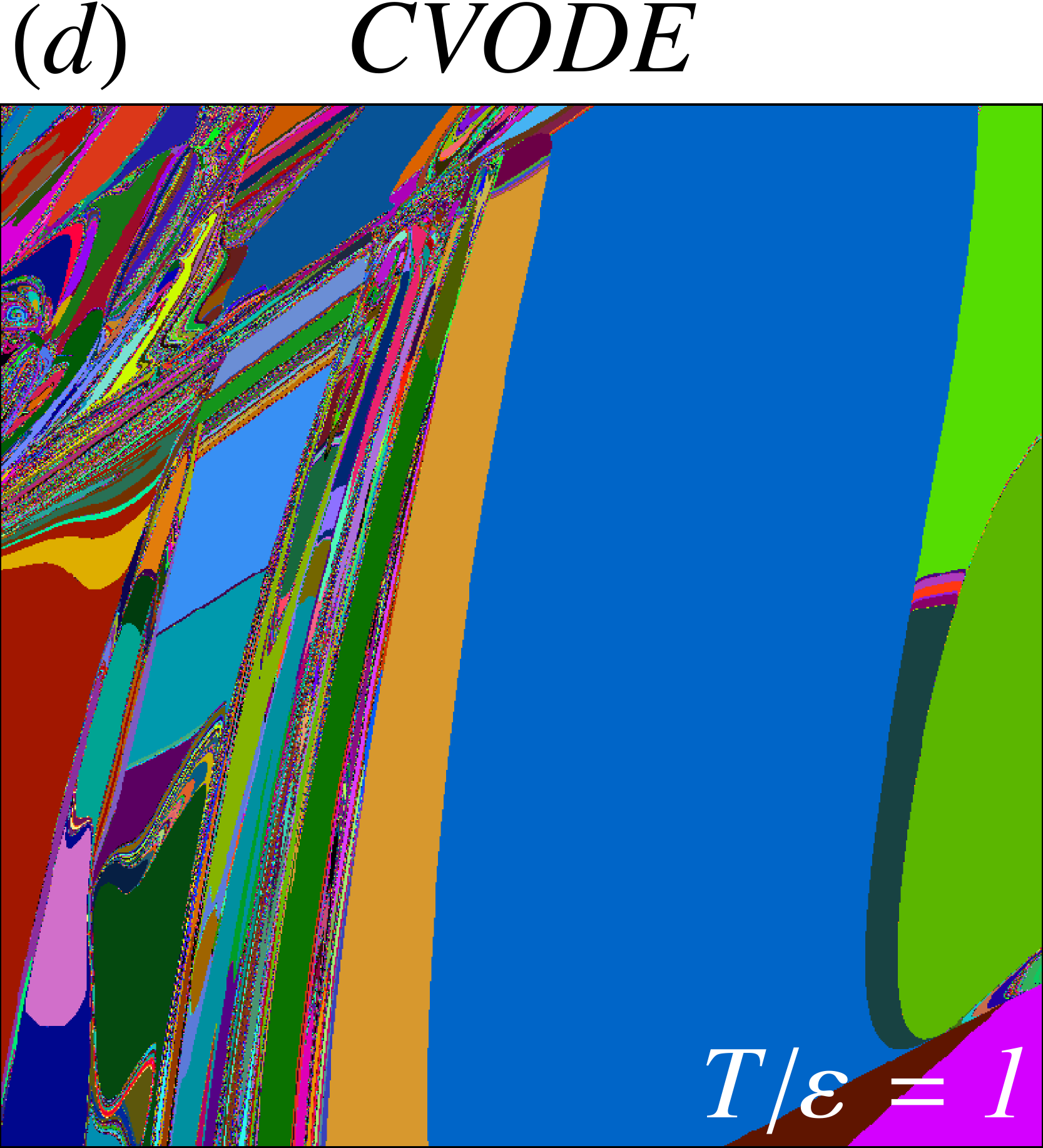}
    \includegraphics[width=0.32\columnwidth]{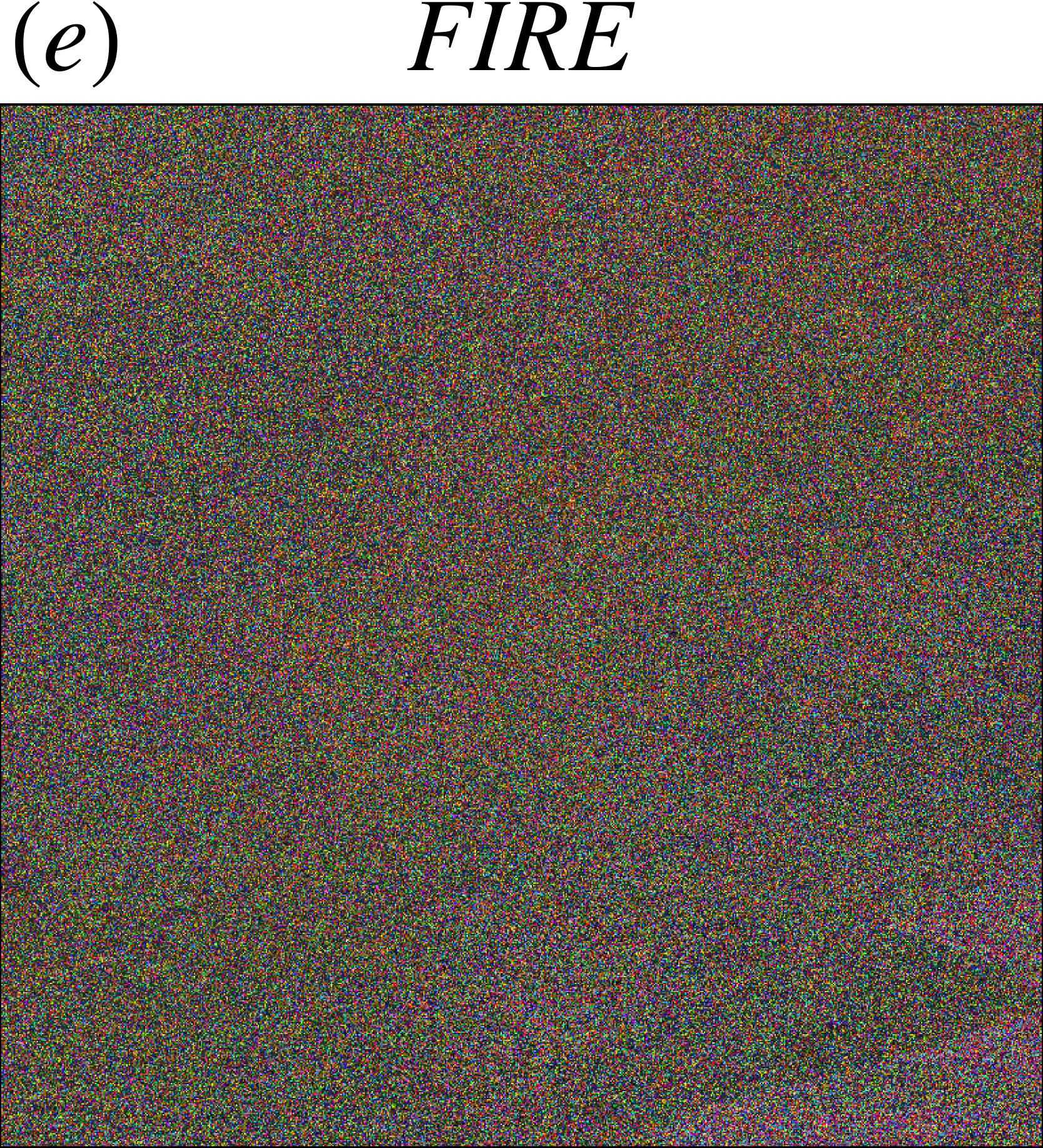}
    \includegraphics[width=0.32\columnwidth]{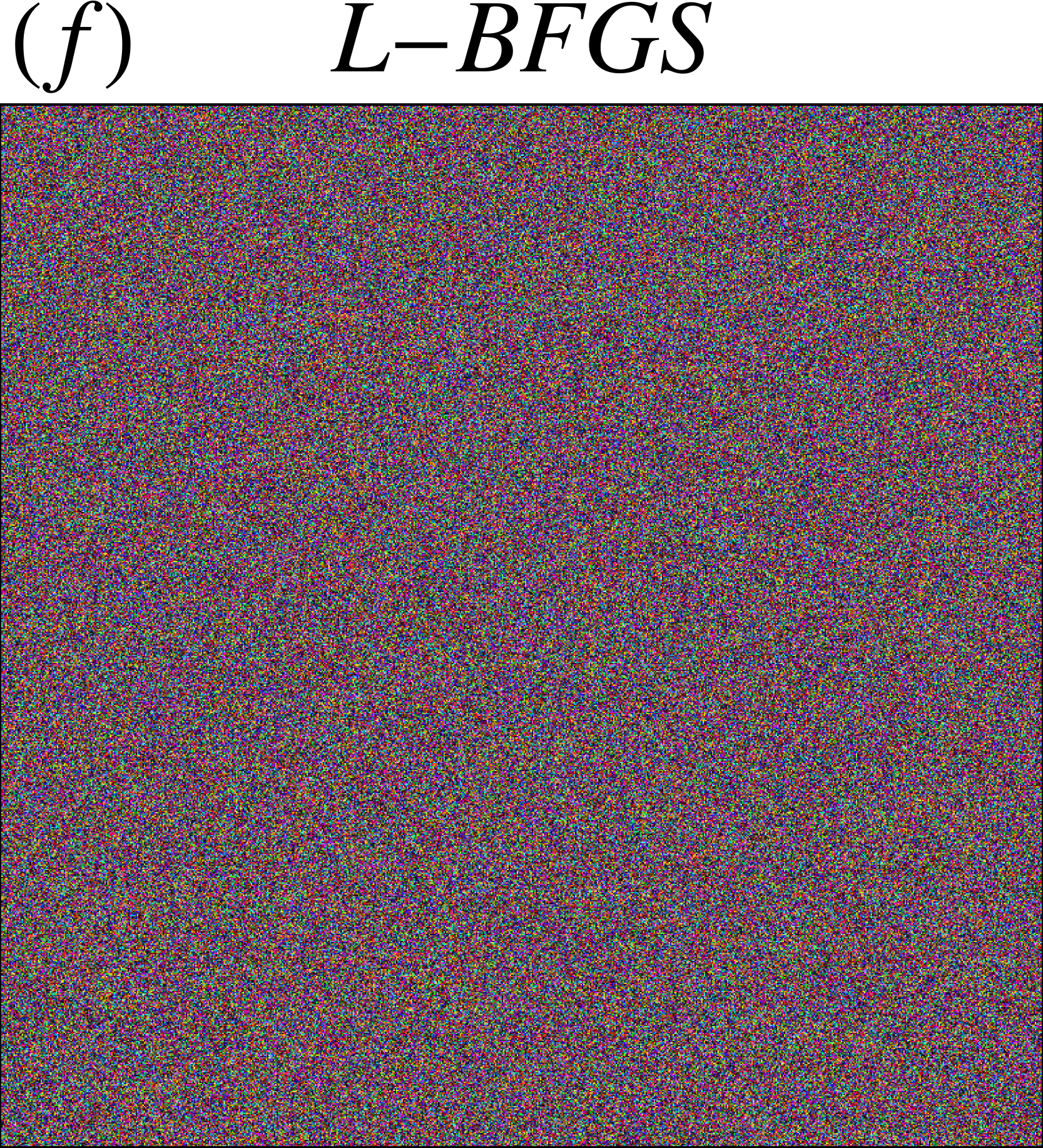} \\
    \includegraphics[width=0.32\columnwidth]{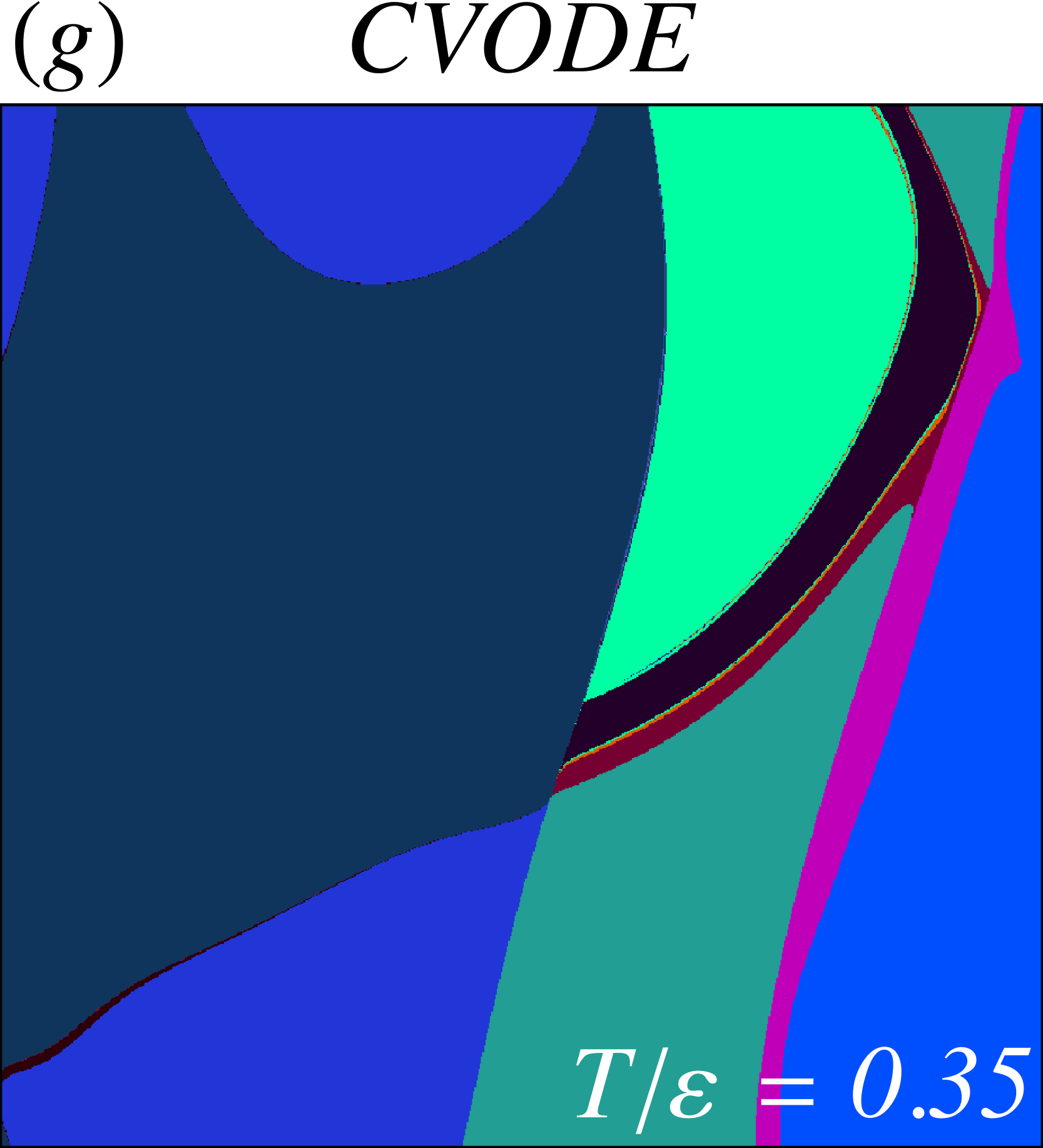}
    \includegraphics[width=0.32\columnwidth]{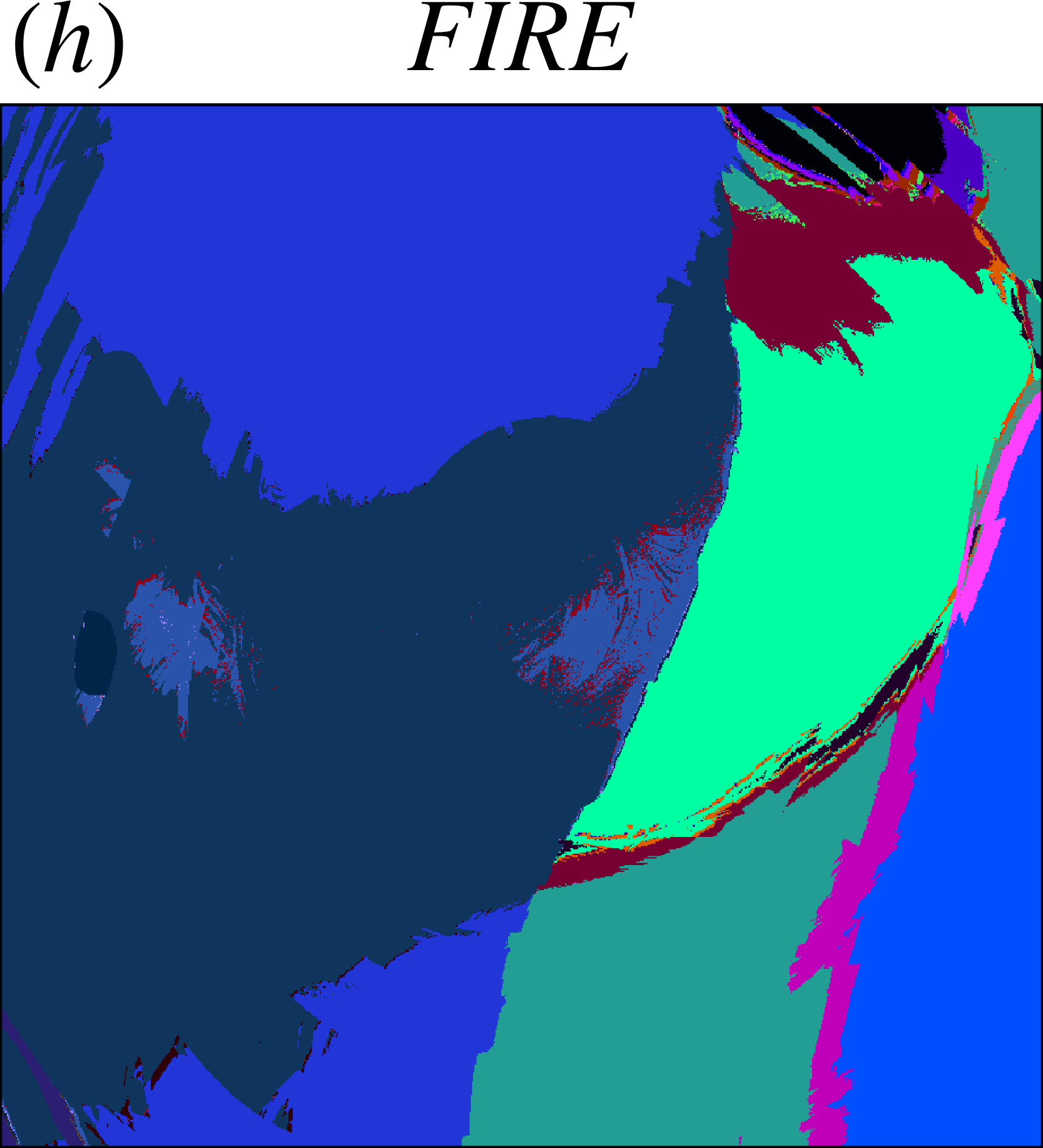}
    \includegraphics[width=0.32\columnwidth]{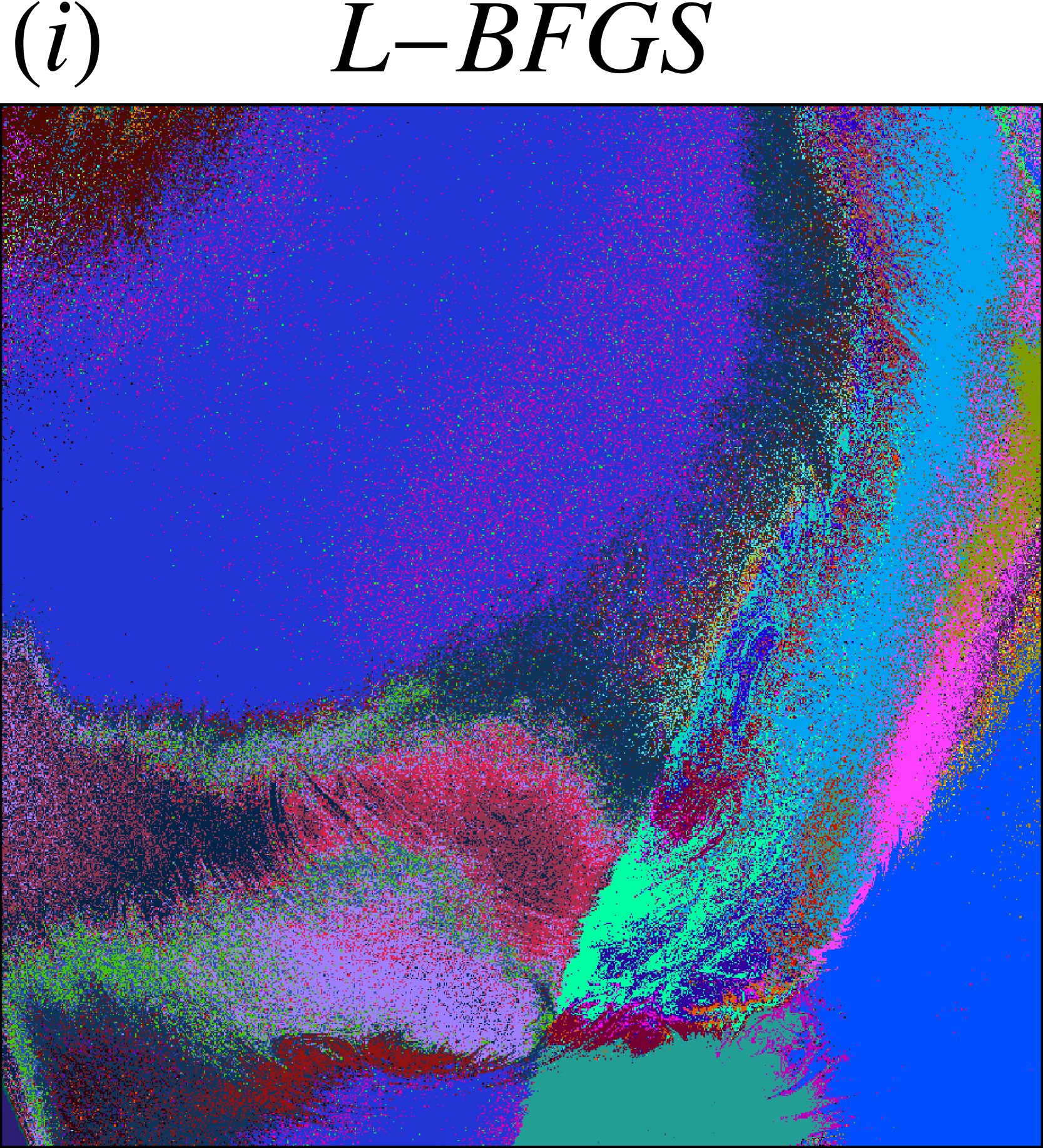}
    \caption{\textbf{Generality of the inaccuracy.}
    $(a)$ Average accuracy of algorithms for the $3d$ Lennard-Jones fluid, computed over $10^4$ configurations equilibrated at $T/\varepsilon = 1$ for each $N$.
    Like in Fig.~\ref{fig:AccuracyTimevsN}, accuracy is the fraction of configurations mapped to the same minimum as the CVODE reference, here run at $\texttt{rtol} = \texttt{atol} = 10^{-10}$ (see SM Sec.~2.F~\cite{supp}).
    $(b)-(c)$ Accuracy obtained when quenching from initial conditions equilibrated at temperature $T$, plotted against the ratio of temperature to the energy scaling factor, $T/\varepsilon$, for $(b)$ a $3d$ Lennard-Jones system with $N=128$ and $(c)$ our model of Hertzian disks, with $N = 512$.
    Dashed black lines indicate exponential fits, and long-dashed black lines stretched-exponential fits, shown as guides to the eye.
    Error bars are Clopper-Pearson $95\%$ confidence intervals~\cite{Clopper1934,supp}.
    $(d)-(f)$ $800 \times 800$ Randomly oriented slice of configuration space around a sample at $T/\varepsilon =1$ for the $3d$ Lennard-Jones fluid with $N = 128$, using $(d)$ CVODE, $(e)$ FIRE, and $(f)$ L-BFGS.
    $(g)-(i)$ Similar slices around a sample at $T/\varepsilon =0.35$, using $(g)$ CVODE, $(h)$ FIRE, and $(i)$ L-BFGS.
    }
    \label{fig:LJ_FiniteT}
\end{figure}

First, instead of our bidisperse system of Hertzian disks, we consider a system of monodisperse Lennard-Jones spheres, interacting via the potential
\begin{align}
    U_{\mathrm{LJ}}(r) = 4\varepsilon \left[ \left(\frac{\sigma}{r}\right)^{12} -\left(\frac{\sigma}{r}\right)^{6} \right]
\end{align}
in 3 dimensions.
This system is very different from the main model studied in this paper: it is attracto-repulsive, long-ranged, with steeply diverging core repulsion, and is defined in $3$ dimensions.
More precisely, we simulate $N$ particles interacting through the full, untruncated potential, without periodic boundaries; to prevent evaporation, the system is enclosed in a spherical container of radius $R = 2 N^{1/3} \sigma$, which keeps the overall density fixed across sizes (see SM Sec.~2.F~\cite{supp}).
Yet, as shown in Fig.~\ref{fig:LJ_FiniteT}$(a)$, when we evaluate the accuracy of basin attribution for L-BFGS and FIRE over $10^4$ configurations per $N$, equilibrated at temperature $T/\varepsilon = 1$, we recover results similar to those of Fig.~\ref{fig:AccuracyTimevsN}: the accuracy decays rapidly already at moderate $N$, on the order of $10$.
Here, we use equilibrated rather than fully random initial points because uniformly drawn configurations can place pairs of particles arbitrarily close to one another, causing the diverging core repulsion to generate energy scales that lead to floating-point precision issues and prevent identification of the true steepest-descent trajectory.
Sampling at $T/\varepsilon = 1$ avoids such overlaps.
Note that this difference in protocols does not hinder the comparison between the two models: for Hertzian disks, the accuracy measured from samples equilibrated at $T/\varepsilon \approx 1$ is already indistinguishable from that obtained from uniformly random points, see Fig.~\ref{fig:LJ_FiniteT}$(c)$.

Second, one may be concerned that our results for accuracy strongly depend on the temperature at which initial points are sampled---uniformly at random in all of configuration space for Hertzian disks (equivalent to infinite temperature), or at $T/\varepsilon = 1$ for the Lennard-Jones fluid.
To address this point, we prepare equilibrium samples of both a $3d$ Lennard-Jones model with $N=128$ and a $2d$ Hertzian model with $N=512$ across $T$, using Monte Carlo simulations (see SM Sec.~2.F~\cite{supp}), and use these collections of samples as starting points to measure accuracy.
The results are shown in Fig.~\ref{fig:LJ_FiniteT}$(b)$ for the Lennard-Jones model and in Fig.~\ref{fig:LJ_FiniteT}$(c)$ for the Hertzian system.
We show that the accuracy of both FIRE and L-BFGS remains high only at low temperature, then falls abruptly to near-zero values above a characteristic temperature. This characteristic temperature is very low for Hertzian disks, so that only starting points extremely close to a minimum converge to the correct one under these minimization algorithms, consistent with past work performing similar measurements~\cite{Nishikawa2022}.
The threshold temperature appears higher in absolute terms for the Lennard-Jones fluid. We speculate that this may be related to the structure of the landscape for such potentials, which are known to display only a small number of distinct minima at low energies for certain values of N~\cite{Doye1999,Wales2018}. As a result, the equilibrium distribution at low T in the Lennard-Jones fluid would concentrate around a relatively small number of minima, compared to the more rugged Hertzian case.

To further illustrate that the kind of inaccuracy discussed above is similar across systems and temperatures, Fig.~\ref{fig:LJ_FiniteT}$(d)$--$(i)$ shows slices of the landscape of a $3d$ Lennard-Jones fluid with $N=128$, analogous to Fig.~\ref{fig:AccuracyTimevsN}$(c)$--$(f)$, here obtained with CVODE, FIRE, and L-BFGS.
To capture the effect of a finite temperature, the slices in Fig.~\ref{fig:LJ_FiniteT} $(d)-(f)$ were centered on a sample equilibrated at $T/ \varepsilon = 1$, so that the center of the slice is in a region where optimizers display near-zero accuracy, while in Fig.~\ref{fig:LJ_FiniteT} $(g)-(i)$ the center point is a sample at $T/ \varepsilon = 0.35$, where the optimizers display moderate accuracy.
The contrast between the high- and low-temperature regions is stark: at $T/ \varepsilon = 1$, not a single pixel obtained by the optimizers agrees with the reference slice, while at $T/ \varepsilon = 0.35$ the slices are much more correlated.
Yet, even in this region of moderate accuracy, using optimizers strongly alters basin identification, with the FIRE slice displaying an overall translation of basins and jagged edges reminiscent of Fig.~\ref{fig:LandscapeSlices}, and the L-BFGS slice showing confetti-like features reminiscent of Fig.~\ref{fig:AccuracyTimevsN}$(e)-(f)$.

In summary, we expect that inaccuracy in the identification of basins when using optimizers becomes significant unless initial points are sampled close to minima. This connects well with past work~\cite{Nishikawa2022,Bautista2026}, which argued that the whole procedure of identifying the basin of a point in configuration space was so sensitive to the precise method used that it was essentially impossible to identify the correct basins. We argue, however, that this issue is resolved by the use of adequate numerical methods, and that, in particular, the landscape is not ``chaotic''~\cite{Nishikawa2022}.

\begin{figure}[b!]
    \centering
    \includegraphics[height = 0.46\columnwidth]{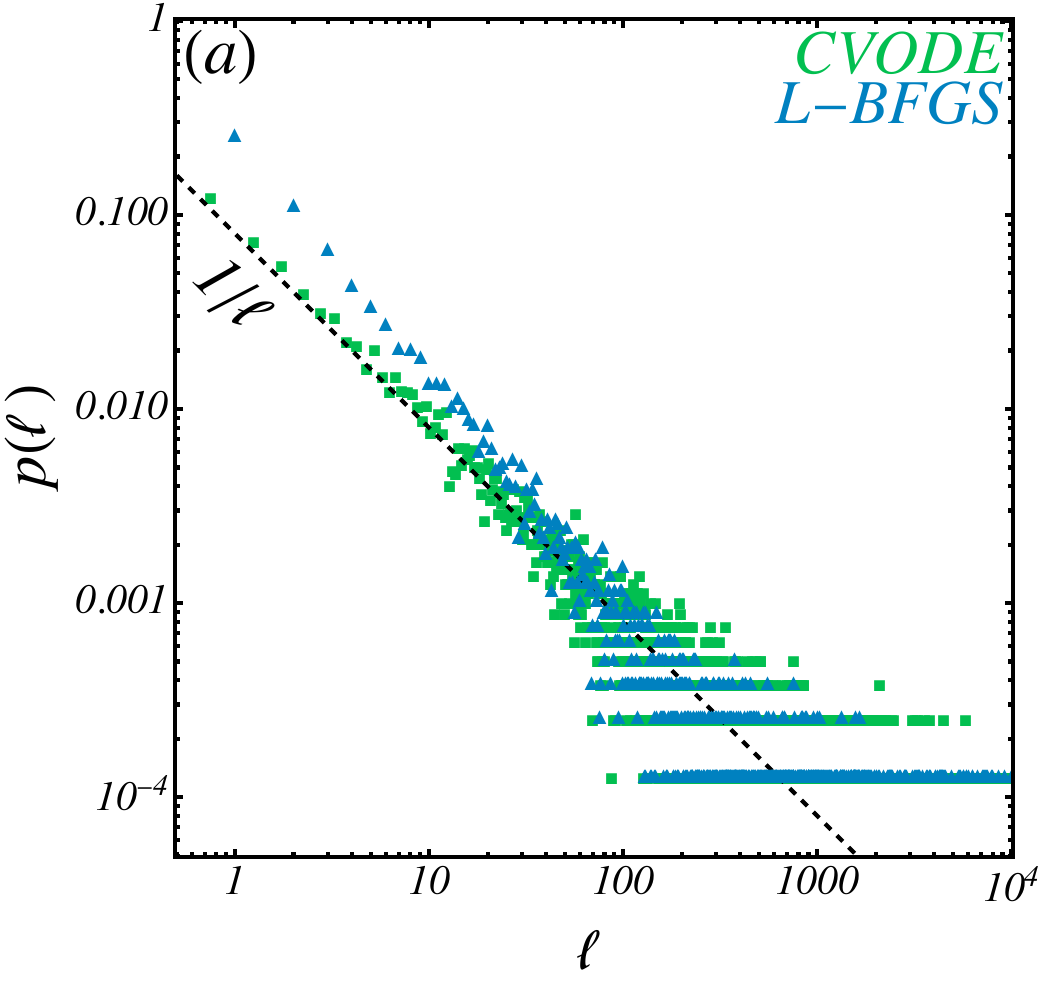} 
    \includegraphics[height= 0.46\columnwidth]{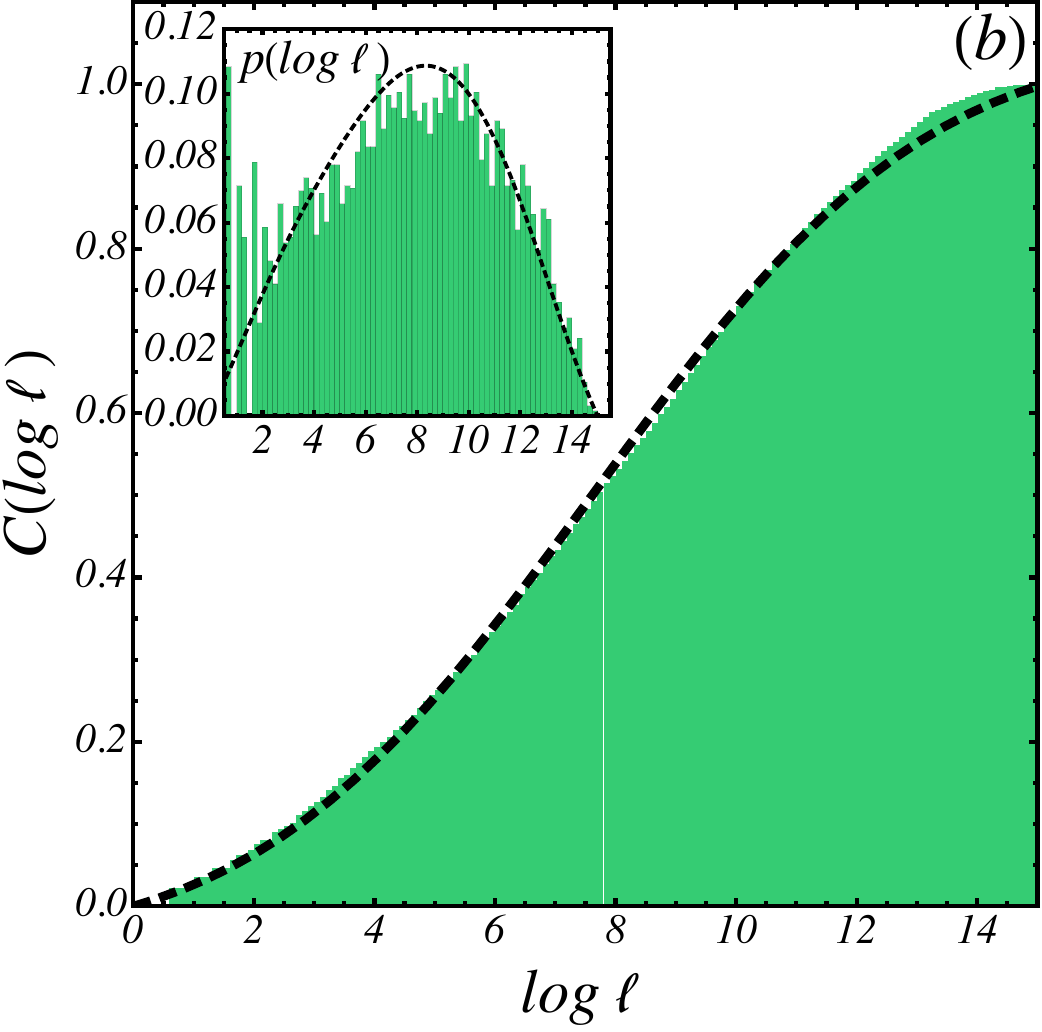}
    \caption{\textbf{Linear intersects of basins.}
    $(a)$ Intersection lengths distributions obtained with L-BFGS (blue triangles) and CVODE (green squares) over $10$ lines of $10^6$ pixels for $N = 16$, in log scales.
    A dashed line indicates $1/\ell$ behavior.
    $(b)$ CDF of the distribution of $\log \ell$ obtained with CVODE by zooming $100\times$ on each basin boundary found from panel $(a)$.
    The dashed black line is a truncated Gaussian fit.
    Inset: Corresponding histogram of the pdf, the dashed line shows a kernel regression.
    }
    \label{fig:LinesCompare}
\end{figure}

\textit{Low-dimensional geometric features --}
To gauge the effects of inaccuracy in basin identification, we study the geometry of basins~\cite{Martiniani2016a}.
First, we consider a simple low-dimensional quantity.
By analogy with Fig.~\ref{fig:LandscapeSlices}, which showed a random $2d$ slice of configuration space, we draw random $1d$ lines in configuration space (by drawing 1 point uniformly in configuration space then a random direction, see SM Sec.~2.A~\cite{supp}).
We consider a number $n_s$ of regularly spaced points on each of these random lines, and use each of them as initial conditions for minimization.
Like in Fig.~\ref{fig:LandscapeSlices}, we then tag each pixel by the basin it falls into and measure the (discretized) intersection lengths $\ell_{ij}$ of segment $i$ with each distinct basin $j$ that it crossed.
The distribution $p(\ell)$ across a collection of random segments is shown in Fig.~\ref{fig:LinesCompare}$(a)$ for CVODE and L-BFGS.
We report that $p(\ell)$, while it retains a power-law-looking decay across methods, looks different between CVODE and L-BFGS measurements.
Indeed, L-BFGS overestimates the amount of small basins, and thus the decay exponent of the distribution, as expected from the ``confetti'' picture, Fig.~\ref{fig:AccuracyTimevsN}$(c)-(f)$.

We take advantage of CVODE to investigate the true distribution of line intersections.
Since this distribution is very broad, an enormous amount of regularly-spaced points would be needed to avoid being resolution limited.
To bypass this limitation, at every boundary between two basins, we produce new segments with a finer resolution.
The results are shown in Fig.~\ref{fig:LinesCompare}$(b)$.
Since the distribution is broad, we focus on the distribution of the logarithm of lengths, $p(\log \ell)$.
We show it is normal, indicating that lengths are log-normal distributed.
This connects well with prior results on basin volumes, which were argued to be log-normal distributed in soft spheres both from numerical measurements~\cite{Frenkel2013,Asenjo2014,Martiniani2016a} and theoretical arguments~\cite{Paillusson2015}.
Indeed, the volume of the intersection of a basin with an  $n$-dimensional affine space may be approximated by a product of $n$ independent log-normal lengths, which yields a log-normal distribution of volumes.
This log-normal distribution invalidates claims that the distribution of basin intersection lengths is scale-free~\cite{Bautista2023}: a power-law tail with exponent $-1$ is observed due to the asymptotic behavior of log-normal distributions with large variances \cite{Clauset2009}.
Likewise, claims of scale-free distributions of basin \textit{volumes}~\cite{Massen2007,Hagh2024} (at odds with direct numerical measurements~\cite{Frenkel2013,Asenjo2014, Martiniani2016a}) likely stemmed from inadequate sampling of a log-normal distribution, of which only the tail was seen, a common issue with small sample sizes~\cite{Broido2019}.

Additionally, we evaluate the box-counting dimension $d_B$ of basins in $2d$ slices of $N=128$ disks such as those in Fig.~\ref{fig:AccuracyTimevsN}, see App.~\ref{app:BoxCounting}.
The average fractal dimension of basin boundaries is significantly closer to $1$ with CVODE and GD $(d_B \approx 1.1)$ than it is with FIRE and L-BFGS $(d_B \approx 1.3)$, see Appendix.
Furthermore, optimizers predict a growing $d_B$ as $\phi \to \phi_J$, while the CVODE estimate remains roughly constant.
In other words, optimizers do create an illusion of fractality, that grows stronger near jamming.
Interestingly, recent work showed large basins of the Random Lorentz Gas, a minimal single-particle proxy model for jamming, to be fractal~\cite{Folena2025}---begging the question of how this feature vanishes in many-body models.

\begin{figure}
        \includegraphics[height=0.45\columnwidth]{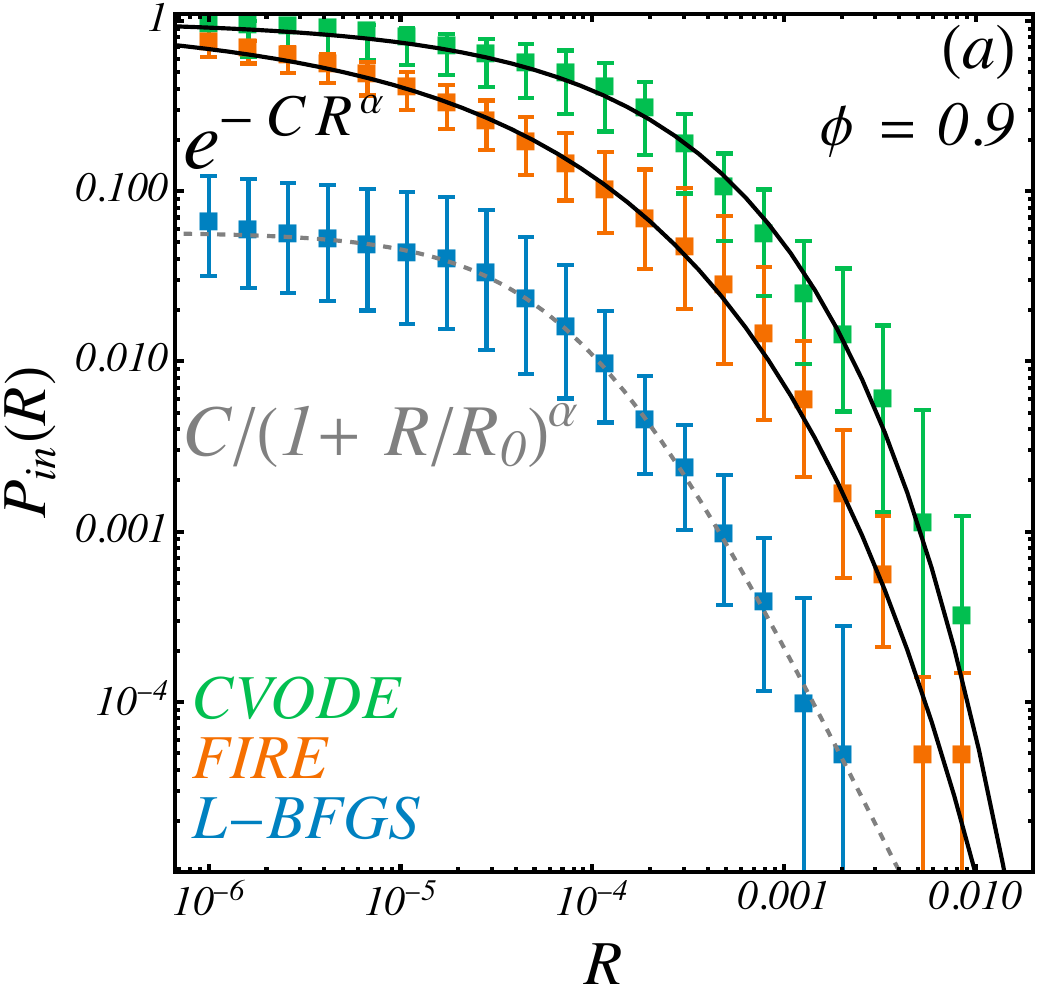}
        \includegraphics[height=0.45\columnwidth]{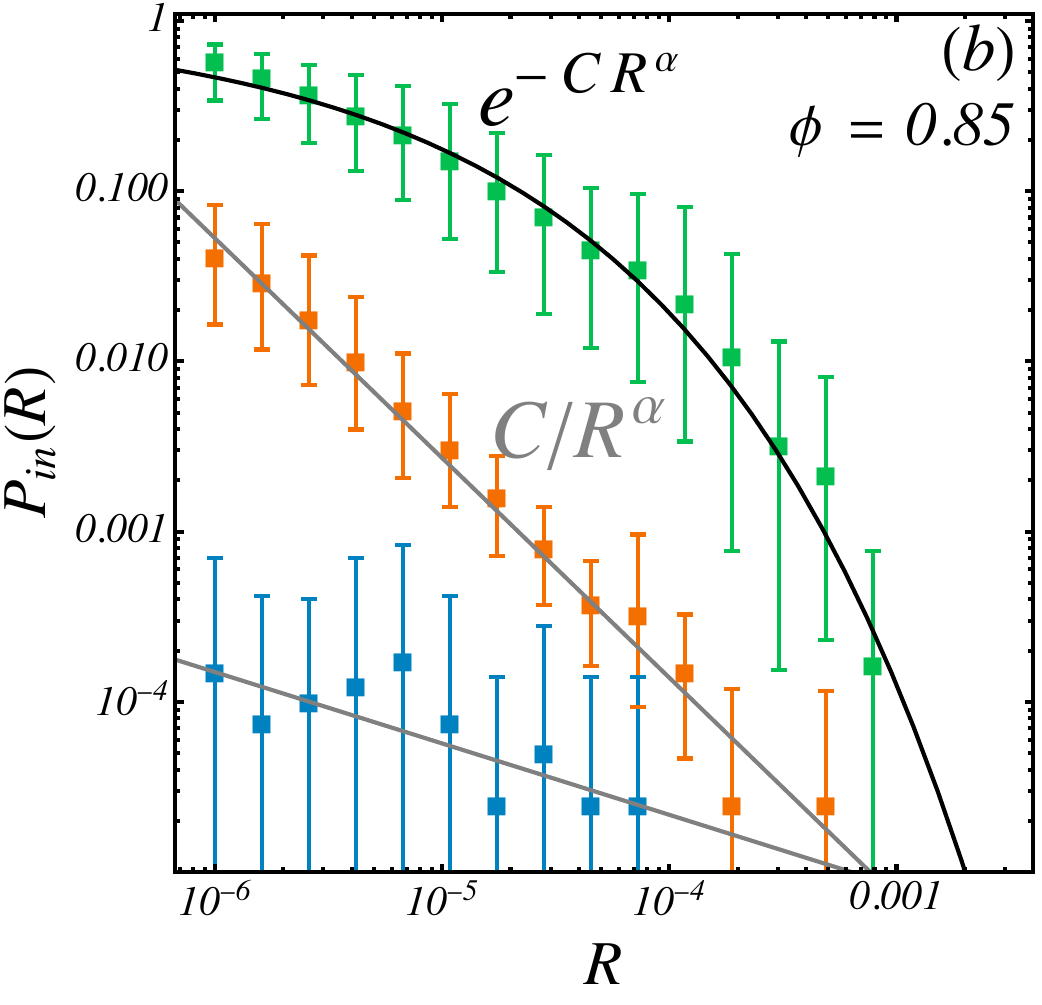} \\
        \includegraphics[height=0.45\columnwidth]{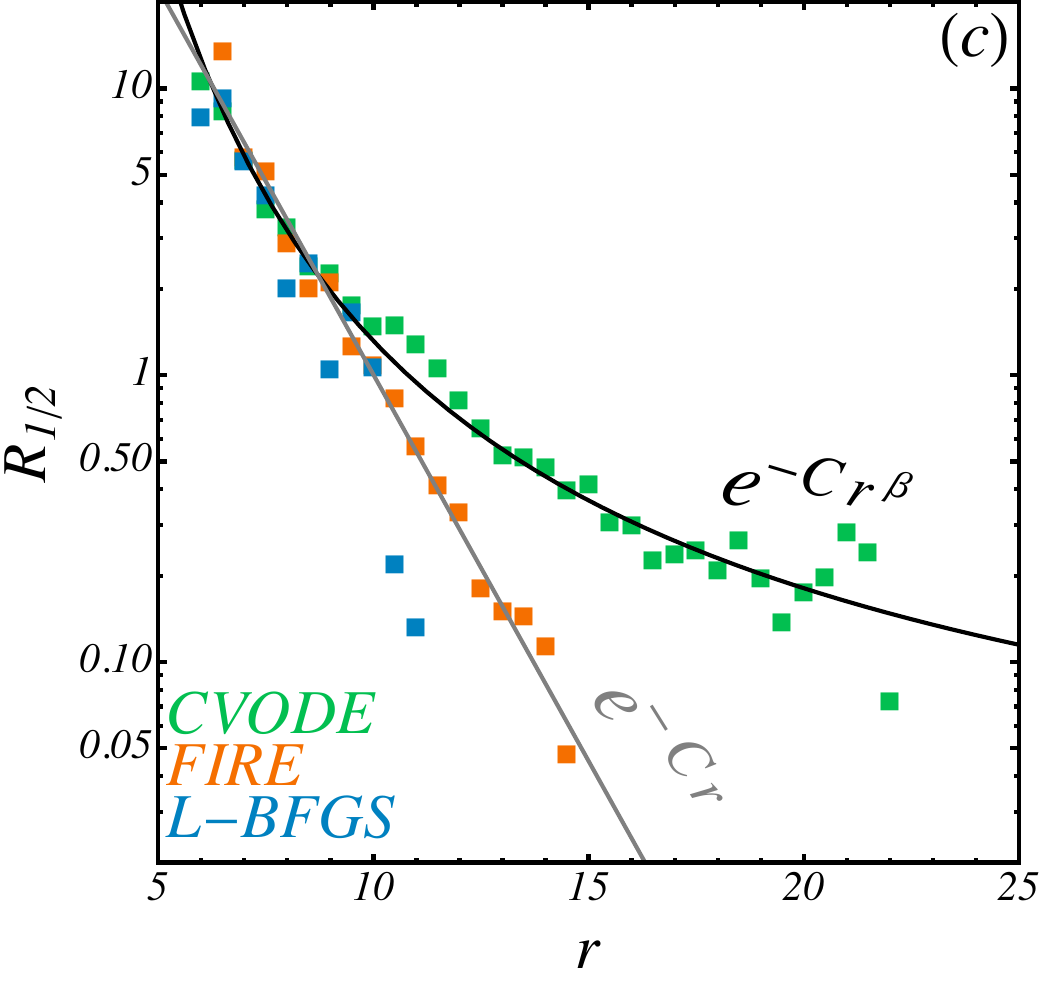}
        \includegraphics[height=0.45\columnwidth]{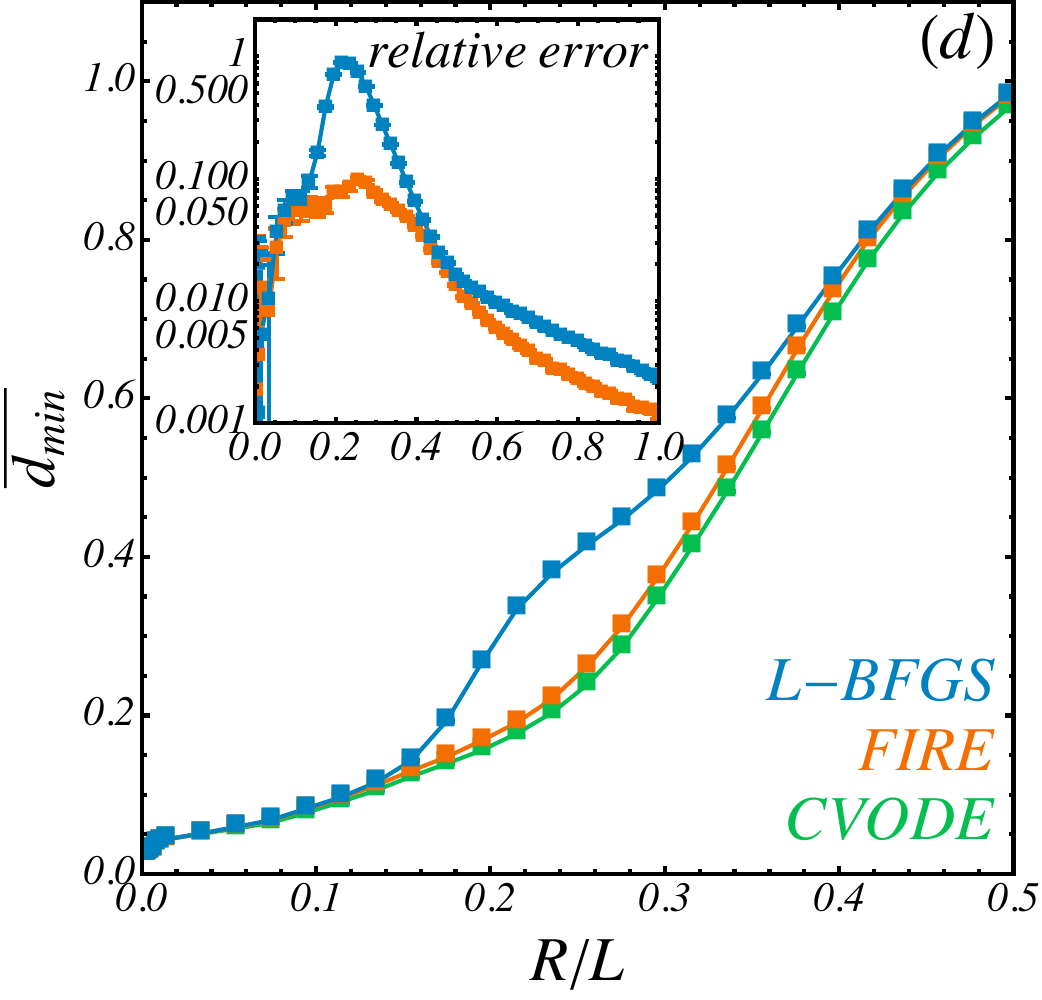}
    \caption{\textbf{Survival}
    $(a)$ Survival from a random point in the landscape of $N=1024$ disks, in log-log scale.
    Solid black lines are stretched-exponential fits for CVODE and FIRE.
    Dashed gray line: saturating power-law fit to L-BFGS.
    Error bars are Clopper-Pearson estimates~\cite{Clopper1934}.
    $(b)$ Same curve at $\phi = 0.85$.
    Gray lines are power-law fits for FIRE and L-BFGS.
    The black line is a stretched-exponential fit to CVODE.
    $(c)$ Half-survival radius $R_{1/2}$ inside a single basin against distance $r$ to its minimum for $N=128$.
    Lines indicate proposed fits, exponential (gray) for FIRE and L-BFGS and stretched-exponential (black) for CVODE.
    $(d)$ Average distance $\overline{d_{min}}$ between neighboring minima using kicks of size $R$, for $N=1024$.
    Inset: log-scale relative error in $\overline{d_{min}}$ for optimizers \textit{vs.} CVODE.
    }
    \label{fig:Survival}
  \end{figure}
\textit{Full-dimensional geometric features --}
We finally focus on full-dimensional geometric measurements.
First, we pick a random point in the landscape, find the basin it belongs to.
Then, we generate points at distance $R$ by uniform hypersphere point-picking, and measure the fraction of points $P_{in}$ that land back at the same minimum, or ``survival'' probability.
The results are shown in Fig.~\ref{fig:Survival}$(a)$.
For $N = 1024$ particles, we use $13$ reference points for CVODE and $43$ points for L-BFGS and FIRE, then around each of them we sample $1000$ points on each of $30$ nested hyperspheres.
Optimizers consistently fall out of the basin at smaller distances, even at high density ($\phi = 0.9$), to the point that L-BFGS falls out even at minute displacements.
In panel $(b)$, we perform the same measurement closer to jamming, $\phi = 0.85$ and find that both optimizers display power-law-looking decays, once again creating an impression of fractality.
However, CVODE reveals a much broader region belonging to the basin, with $P_{in}$ following a stretched exponential $\exp(-C r^\alpha)$.
The basins are thus not scale-free by this measurement either, even close to jamming (additional $\phi$'s in SM Sec.~3.A~\cite{supp}).

To better characterize the shapes of basins, we perform similar survival measurements but this time centering hyperspheres on random samples within a \textit{single basin}.
We measure the distance $R_{1/2}$ at which survival first hits $1/2$ as a function of the distance $r$ between the random sample and the minimum, and thus estimate a typical cross-sectional length of the basin.
Results are shown in Fig.~\ref{fig:Survival}$(c)$ for $N = 128$.
In this figure, we bin radial distance from the minimum into bins with width $0.1$, gather $10$ points per bin, and then sample $1000$ points per hypersphere centered on each of these points across $15$ shell radii (order $10^6$ points per method).
We report exponential decays with optimizers, and stretched exponential behavior for CVODE.
In hypercubes, the cross-section of corners decays exponentially with distance to the center~\cite{ArtsteinAvidan2015}.
Thus, a stretched exponential indicates that basins have ``thicker'' tentacles than cubes, in line with observations that ``tentacles'' contributed to basin volumes up to large distances in sphere packings~\cite{Ashwin2012, Martiniani2016a}, Kuramoto models~\cite{Martiniani2017,Zhang2021,Groisman2026}, and neural networks~\cite{Annesi2023}.

We now study the arrangement of basins, following Ref.~\cite{Dennis2020}.
Starting from $13$ random minima, we define a collection of nested hyperspheres centered on each minimum, and sample $1000$ random points uniformly on each of them. 
We then minimize and keep a record of all new minima that are found.
For each new minimum, we compute a rescaled distance to the initial minimum, $d_{min}$, as defined in Ref.~\cite{Dennis2020} (see SM Sec.~2.D~\cite{supp}), then compute its average $\overline{d_{min}}$ as a measure of the difference between the neighborhoods of basins in each method.
We show in Fig.~\ref{fig:Survival}$(d)$ that for $N = 1024$, $\overline{d_{min}}$ is significantly shifted when using optimizers, with a clear maximal deviation of $10$ to $80\%$ at intermediate hypersphere radii $R$.
Thus, the measured basin organization is clearly affected by the choice of minimization method.

Finally, we perform basin volume measurements, using a Markov-Chain Monte-Carlo (MCMC) method~\cite{Xu2011, Asenjo2014,Martiniani2016a,Martiniani2017a,Casiulis2023} akin to Frenkel-Ladd measurements of free energies~\cite{Frenkel1984} (see SM Sec.~2.E~\cite{supp}).
In short, the method relies on $K$ random walks constrained to remain in the basin of interest, each biased by a different harmonic potential centered on the minimum, and coupled to each other through Replica-Exchange Monte Carlo moves.
The method estimates the free energy of each walker, and in particular that of a free walker $F_0$, which is linked to the volume $V$ of the basin via $F_0 = -\ln V$.
At each MCMC step of each random walk, a full minimization is run to check whether the proposed new position still lies in the basin of interest.
We thus expect this method to be significantly affected by the inaccuracy of optimizers.
We test this hypothesis at $\phi = 0.9$ and across system sizes that remain amenable to using CVODE (recall that the MCMC takes order $10^6$ steps~\cite{Martiniani2017a, Casiulis2023}, and the time of a typical minimization from Fig.~\ref{fig:AccuracyTimevsN}).

\begin{figure}
        \includegraphics[height=0.47\columnwidth]{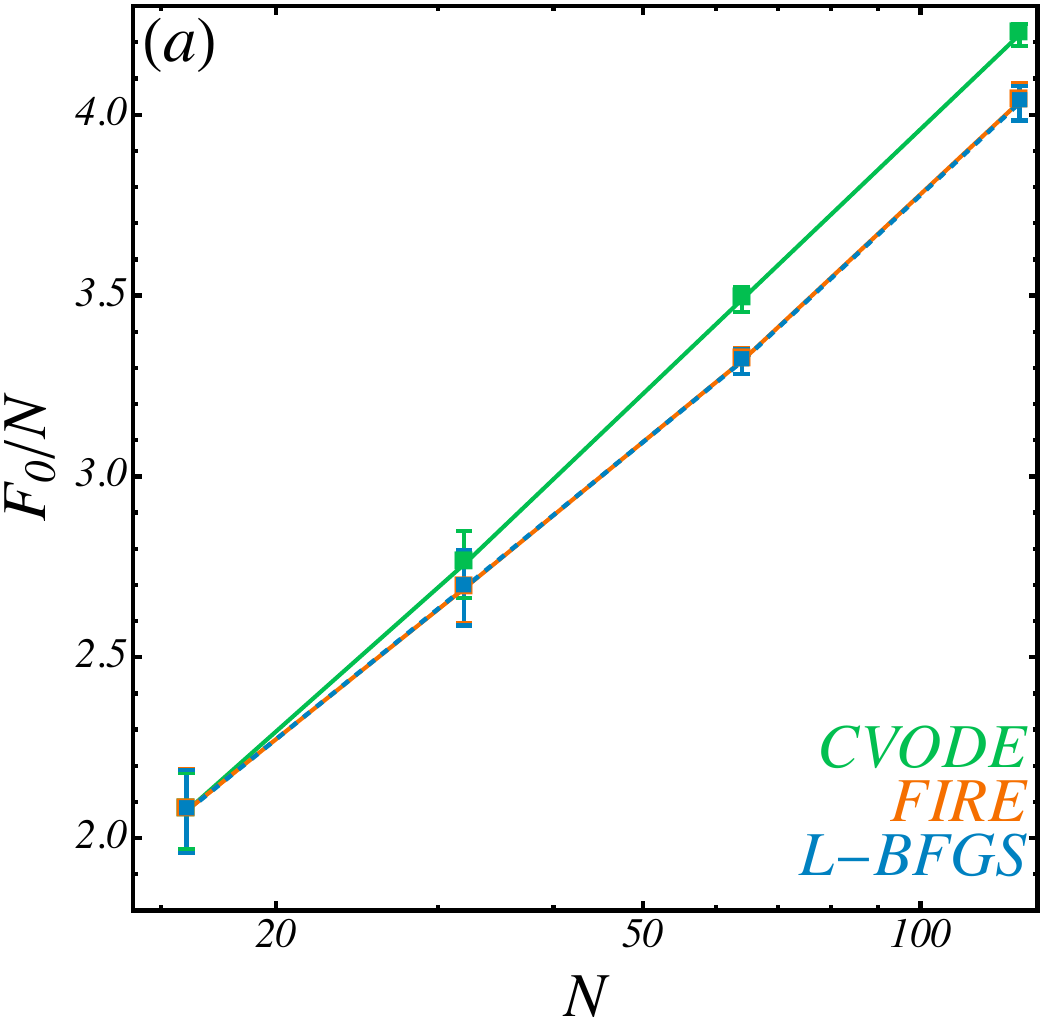}
        \includegraphics[height=0.47\columnwidth]{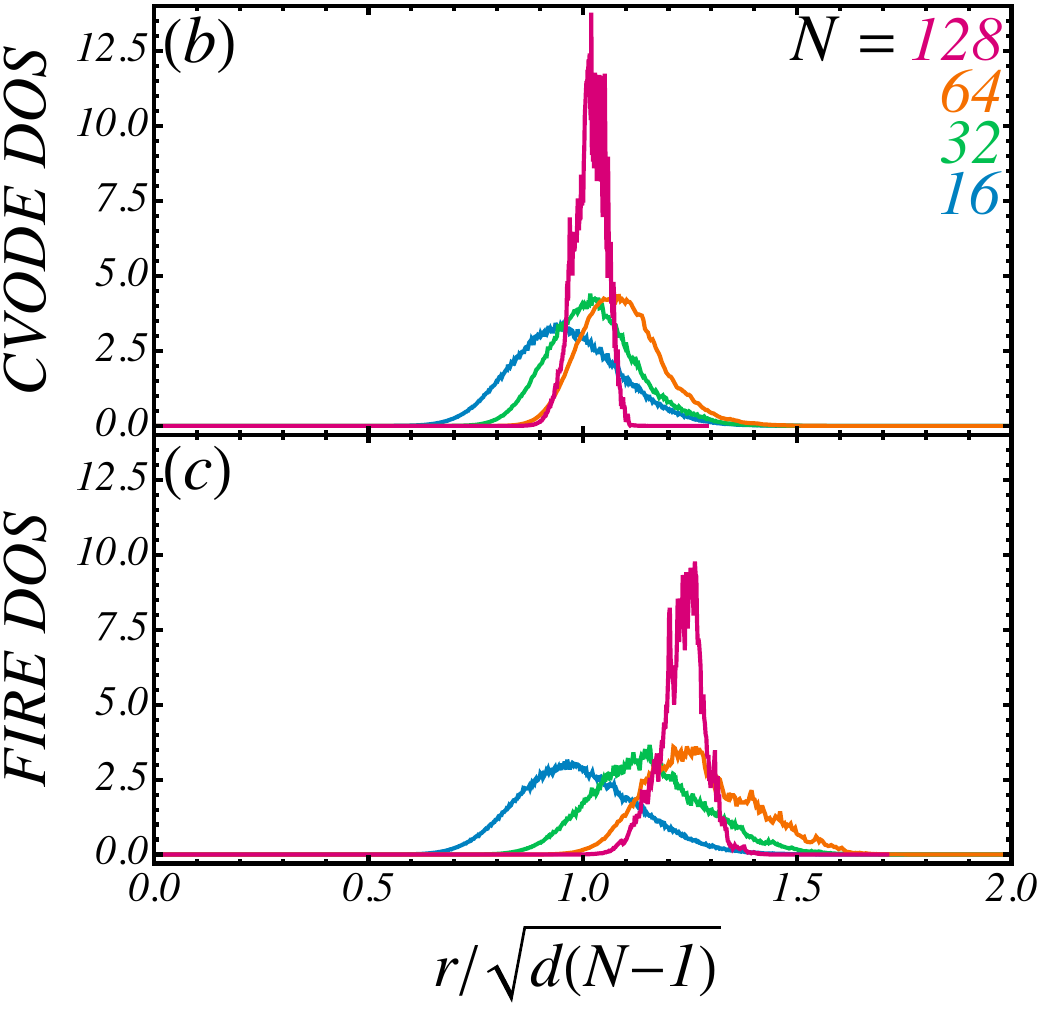} \\
        \includegraphics[width=0.48\columnwidth]{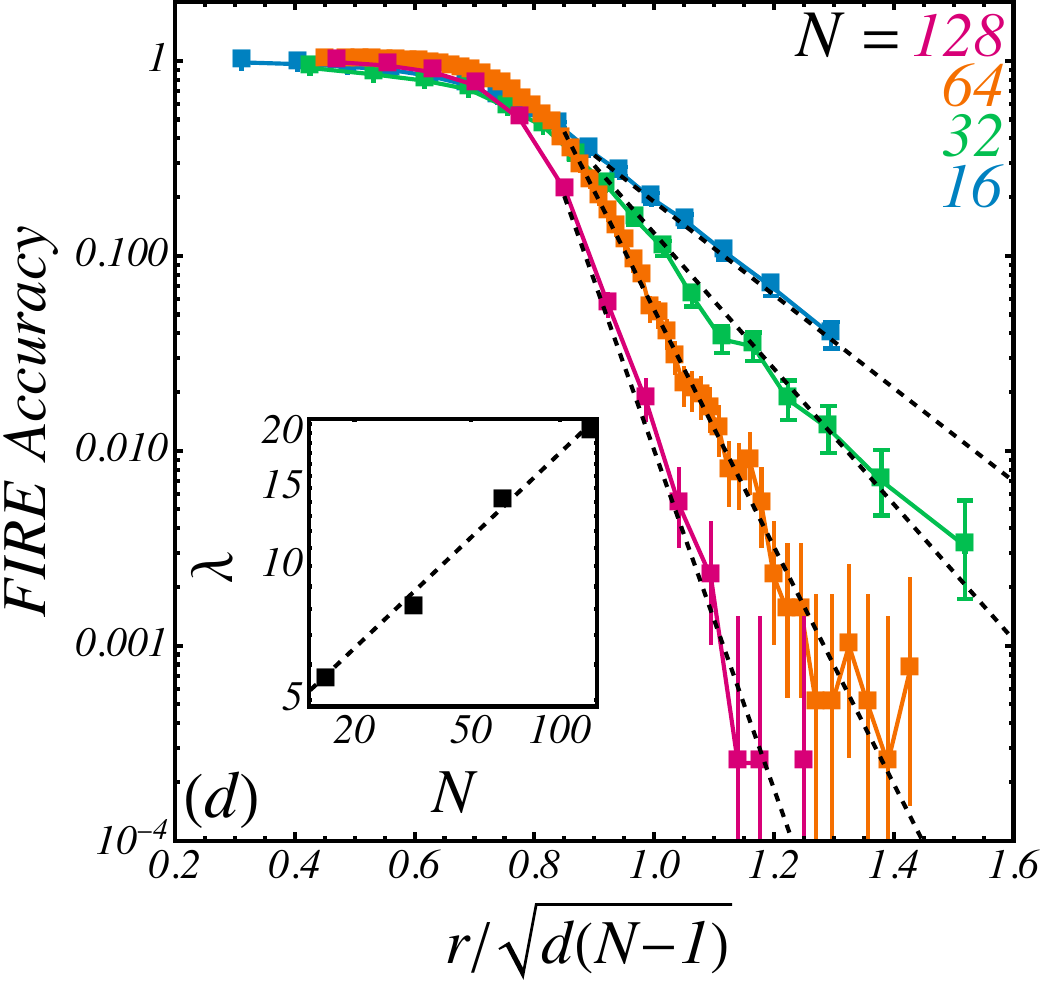}
        \includegraphics[width=0.48\columnwidth]{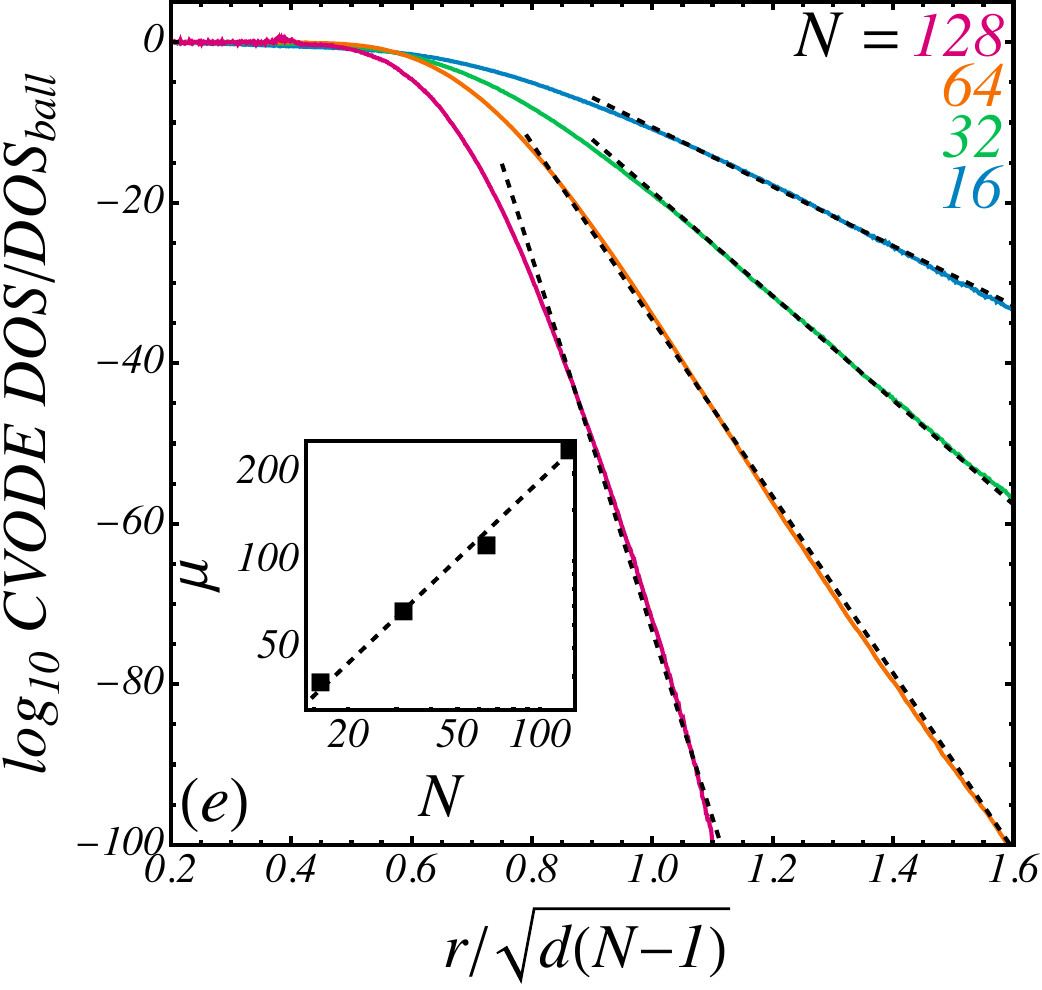}
    \caption{\textbf{Basin volumes.}
    $(a)$ Intensive free energies $F_0/N$ across methods, against $N$, each averaged over the same 5 basins for each $N$.
    $(b)-(c)$ Example densities of states (DOS) for one basin per $N$, for $(b)$ CVODE, and $(c)$ FIRE.
    $(d)$ Averaged accuracy over $\mathcal{O}(10^5)$ FIRE samples used, against their distance to the minimum, in semi-log scale.
    Dashed lines are exponential fits $y = C \exp(-\lambda x)$ of each curve.
    Inset: best decay rate $\lambda$ against $N$ in log-log, with a dashed power-law $\lambda \sim N^{0.6}$.
    $(e)$ $\log_{10}$ of the ratio between the CVODE DOS of the basins in $(b)-(c)$ and that of a hypersphere in $N(d-1)$ dimensions.
    Dashed lines are exponential fits.
    Inset: decay rate $\mu$ against $N$ in log-log, with a dashed power-law $\mu \sim N^{0.9}$.
    }
    \label{fig:Volumes}
  \end{figure}

The results for $F_0/N$ are shown in Fig.~\ref{fig:Volumes}$(a)$.
While FIRE and L-BFGS yield near-indistinguishable values, there is a systematic bias between CVODE and optimizers.
Worse, the systematic bias grows with system size, as FIRE and L-BFGS become more inaccurate.
The volumes measured by optimizers are \textit{larger} than the true volumes, a counter-intuitive result confirmed by the radial densities of states (DOS) reconstructed from samples in example basins, Fig.~\ref{fig:Volumes}$(b)-(c)$.
Using the rescaled radial distance to the minimum $r/ \sqrt{(N-1)d}$ (which keeps the length of a long diagonal of a unit cube constant across dimensions) the CVODE DOS are all maximal around $1$, while the FIRE ones systematically shift to higher values; indeed, FIRE finds samples lying further away from the minimum.
However, in Fig.~\ref{fig:Volumes}$(d)$, we show the accuracy of FIRE samples obtained from the same basins as in $(b)-(c)$ as a function of rescaled radial distance, revealing that the accuracy plummets after $0.7$, with a decay that goes exponentially with distance and gets faster with dimensionality (inset).
In other words, FIRE wrongly tags points outside of basins as belonging to them, likely because inertia facilitates ridge crossing.
In high dimension, a slightly larger spherical shell contributes an enormous volume, so that these erroneous points lead to a systematic volume overestimate in the basins we measured.
This cannot be true of all basins: since optimizers do not create new minima (see App.~\ref{app:energies}) and basins tile configuration space, the average basin volume is fixed by the total volume divided by the number of minima, so that the overestimate of the large basins we sample must be compensated by an underestimate of the smaller basins, Fig.~\ref{fig:AccuracyTimevsN}$(e)-(f)$.
Yet, the overestimate is reproducible in the basins that we measured.
Note that our sampling of basins is itself biased by their size: since basins are selected by picking random points in configuration space, larger basins are sampled with a probability proportional to their volume, so that the basins we measure are larger than average~\cite{Martiniani2017a,Casiulis2023}.
Together, these observations suggest that using optimizers results in a ``rich gets richer'' effect on basin volumes, in which larger basins are made to look even larger while small ones are shrunk even more.
An effect of this bias is that optimizers bias the sampling of minima, as seen in the estimates of energies and of $\phi_J$ (App.~\ref{app:energies},~\ref{app:jamming-point}).
Note that the bias is towards lower energies, which has been shown to be correlated to larger basins in jammed soft particles~\cite{Casiulis2023}, and thus the rich-gets-richer picture is consistent with the observed bias on the energies.

Finally, in Fig.~\ref{fig:Volumes}$(e)$, we show the DOS of the same basins, predicted by CVODE, and divided by that of a sphere~\cite{Martiniani2016a,Casiulis2023}.
This function measures how much sparser than a sphere the basin becomes as $r$ grows.
The accuracy of FIRE samples, Fig.~\ref{fig:Volumes}$(d)$, starts falling at distances slightly larger than the radius of the largest inscribed sphere (where curves start dropping in Fig.~\ref{fig:Volumes}$(e)$), meaning that inaccuracy sets in when entering the basin ``tentacles''~\footnote{Note that the decay exponents of Fig.~\ref{fig:Volumes}$(e)$ (see inset) are much larger than those of panel $(d)$.
Thus, interpreting Fig.~\ref{fig:Volumes}$(e)$ as the success rate of naïve Monte Carlo on a sphere~\cite{Casiulis2023}, the FIRE basin, while inaccurate, remains much more correlated with the true basin than uniform sampling.}.
This also explains the growing disagreement between CVODE and optimizers as $N$ increases: in higher dimensions, the volume of the largest inscribed sphere is vanishingly small!~\cite{Casiulis2023}
These discrepancies call into question the scope of results on the geometry of basins obtained with FIRE~\cite{Martiniani2016a, Martiniani2017,Martiniani2017a}.

\textit{Conclusions --} 
We showed that optimizers are too inaccurate to correctly map minima and basins of attraction, even at low $N$, across two very different liquid models and densities, unless the sampled points lie in the close vicinity of a minimum (\textit{e.g.}, at low temperature).
Proposing CVODE as a viable ODE solver for these problems up to $N \sim 10^3$, and focusing on the paradigmatic example of Hertzian disks, we demonstrated that optimizers consistently attribute points to the wrong basins, and thus fail at capturing simple geometric features of basins.
While we concentrated our efforts on Hertzian disks, qualitatively similar results on $3d$ Lennard-Jones particles suggest that this is a generic feature of high-dimensional configuration spaces of liquids.
Thus, we argue that using appropriate solvers is critical to disentangle the effects of the choice of optimization strategy from that of the structure of the landscape on many-body gradient-type problems, including many dynamical systems~\cite{Wiley2006,Ercsey-Ravasz2011,Zhang2021,Bollt2023, Zhang2024}.
We also showed that basins are not scale-free or fractal objects in the landscape of jammed soft particles, so that their geometry is insufficient to explain the fractality of relaxation paths, and that claims of chaotic behavior~\cite{Nishikawa2022,Bautista2023,Bautista2026} in energy landscapes are generally unfounded.
Likewise, one should be careful about invoking a supposed fractality of basins in machine learning, to justify \textit{e.g.} chaotic-like sensitivity to initial conditions~\cite{Bollt2023,Kwok2025,Ly2025}.
Instead, basins are smooth and their geometric lengths, like volumes, are log-normal distributed at any given $N$ and $\phi$.
Interestingly, log-normal distributions have also been reported in random sequential fragmentation processes~\cite{Baker1992, Ishii1992, Sotolongo-Costa1996, Delannay1996}.
Thus, configuration space splitting into an increasing number of basins of attraction as $\phi \to \phi_J$ might be interpretable as a random fragmentation process, \textit{e.g.} as the high-dimensional crumpling of the potential energy surface.
This geometric interpretation is a promising lead for future research, as relaxation dynamics cannot be explained just from the neighborhood of minima in complex landscapes~\cite{Fournier2025,Charbonneau2025,Pacco2025}.

%\mk{Gardner-like fractality makes optimization better~\cite{Sorkin1991,Mannelli2019}}

\begin{acknowledgments}
\textit{Acknowledgments --}
The authors would like to thank Chris Rackauckas and Yingbo Ma from JuliaHub for help with benchmarking various ODE solvers.
The authors would also like to thank Eric Corwin, Peter K. Morse and R. Cameron Dennis for their comprehensive feedback, as well as David Grier and John Crocker for insightful comments on this work.
P.S., M.C., and S.M. acknowledge the Simons Center for Computational Physical Chemistry for financial support.
This work was supported by the National Science Foundation grant IIS-2226387 and DMR-2443027.
This work was supported in part through the NYU IT High Performance Computing resources, services, and staff expertise.

\textit{Author contributions --} P.S. contributed most of the design and implementation of the simulations, and to the conceptualization, data generation and analysis, and writing of this manuscript. M.C. contributed to the design and implementation of the simulations, and to the conceptualization, data generation and analysis, and writing of this manuscript. S.M. contributed to the conceptualization, data analysis, writing of this manuscript, and funding acquisition. The authors declare no competing interests.
\end{acknowledgments}

% \textit{Contribution statement --}
% P.S. developed the simulation code original to this paper.
% P.S. and M.C. performed the numerical calculations. 
% P.S., M.C. and S.M. conceptualized the work, analyzed the data, and wrote the manuscript.
\bibliographystyle{apsrev4-2}
\bibliography{PostDoc-StefanoMartiniani,supp,supp_extra}

\appendix

\section{Model\label{app:model}}

Throughout the text, we consider a polydisperse system of disks interacting via a Hertzian repulsive potential
\begin{align}
    V_{ij}(r_{ij}) = \left( 1 - \frac{r_{ij}}{R_{i} + R_{j}} \right)^{5/2} \mathbb{1}\left(r \leq R_i + R_j \right). \label{eq:potential}
\end{align}
Here $R_i$ is the radius of particle $i$, $r_{ij}$ is the metric distance between the centers of particles $i$ and $j$, and $\mathbb{1}$ is an indicator function.
Half the radii are positive-normal distributed with mean $\mu_s =1.0$ and standard deviation $\sigma_s = 0.05$, and the other half with $\mu_\ell= 1.4$ and $\sigma_\ell=0.07$.
This choice ensures that particles do not crystallize and that minima of the energy are not connected by permutation symmetry~\cite{Gao2006, Xu2011}.
Due to periodic boundaries, the energy $E = \sum_{i<j} V_{ij}$ is invariant by translation in $d$ directions, so that only $(N-1)d$ degrees of freedom persist.
With these notations, the packing fraction is defined as $\phi  = \pi \sum_{i=1}^{N} R_i^2 / L^2$.

\section{Box-counting dimension of $2d$ cuts\label{app:BoxCounting}}

To complete the data on distributions of $1d$ intersection lengths between random lines and basins presented in the main text, we discuss a $2d$ measure of the fractal dimension of intersections of basins with random $2d$ planes, such as the slices of Fig.~1 of the main text.
To quantify the fractal dimension of basin boundaries, we rely on the Minkowski-Bouligand, or ``box-counting'' dimension~\cite{falconer2013fractal}, $d_B$.
For non-fractal shapes $d_B = 1$, while fractal shapes have $d_B > 1$.
To estimate $d_B$, we isolate a single basin from a $2d$ cut such as the ones in Fig.~1 of the main text, then use the \texttt{porespy} library~\cite{Gostick2019}.
The detailed procedure is described in SM Sec.~2.C~\cite{supp}.

\begin{figure}
    \centering
\includegraphics[width=0.96\columnwidth]{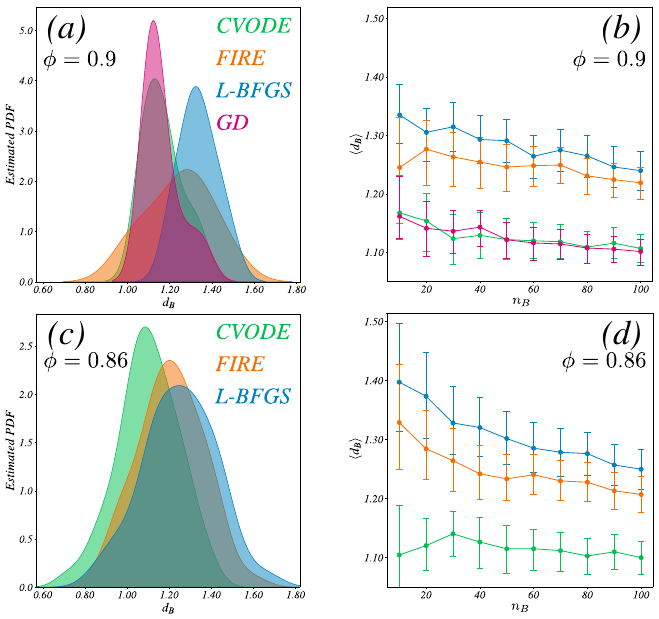}
\caption{\textbf{Box-counting dimension.} $(a)$ Gaussian kernel density estimate of the PDF of the box counting dimension $d_B$ from top $10$ largest basins for $N=128$ and $\phi = 0.9$, and $(b)$ corresponding average box counting dimension estimated from the top $n_B$ largest basins.
In $(c)$, $(d)$, we show the same plots for $N =128$ and $\phi = 0.86$. }
    \label{fig:BoxCounting}
\end{figure}

In Fig.~\ref{fig:BoxCounting}$(a)$, we show Gaussian kernel density estimates (KDE) obtained from the histograms of $d_B$ for each method.
We show that the mode of the distribution is close to unity ($d_B^{mode} \in [1.05;1.15]$) when using CVODE or GD, but is shifted to significantly higher values ($d_B^{mode} \in [1.3; 1.4]$) when using FIRE or L-BFGS.

To estimate the effects of finite resolution, in Fig.~\ref{fig:BoxCounting}$(b)$, we plot the average box counting dimension $\langle d_B \rangle$ calculated for the $n_B$ largest basins, against $n_B$ (error bars are $95\%$ confidence intervals calculated by bootstrap).
We show that as smaller basins are included in the average, the estimated fractal dimension decreases significantly for FIRE and L-BFGS, while it essentially plateaus for ODE solvers.
This is due to the finite resolution of slices: as we include smaller basins, the scale of the pixel size begins to dominate.
As a result, for optimizers, $d_B$ is underestimated when $n_B$ is large.
We repeat the same measurement at $\phi = 0.86$, save for GD which becomes prohibitively expensive near jamming.
The corresponding plots are shown in Fig.~\ref{fig:BoxCounting}$(c)$ and $(d)$ (the slices used at $\phi = 0.86$ are shown in SM Sec.~3.E~\cite{supp}).
The average $d_B$ for $n_B = 10$ are reported in Table~\ref{tab:BoxCounting}.

\begin{table}
  \centering
  \begin{tabular}{|c|c|c|c|c|} \hline $\phi$ & $d_B$ (L-BFGS) & $d_B$ (FIRE) & $d_B$ (CVODE) & $d_B$ (GD) \\ 
    \hline $0.9$ & $1.34 \pm 0.05$ & $1.24 \pm 0.09$ & $1.17 \pm 0.05$ & $1.16 \pm 0.06$ \\
    0.86 & $1.40 \pm 0.09$ & $1.33 \pm 0.09$ & $1.10 \pm 0.08$ & N/A
    \\ \hline \end{tabular}
  \caption{\textbf{Box-counting dimensions.}
    Table of $d_B$ values (with bootstrapped $95\%$ confidence intervals) estimated from the $10$ largest basins in a slice, for each method.
    }
  \label{tab:BoxCounting}
\end{table}

Altogether, basins look significantly more fractal ($d_B \gtrsim 1.25$) when identified with optimizers rather than ODE solvers ($d_B \approx 1.1$).
As the jamming point is approached, the fractal dimension estimated with CVODE slices does not change significantly, while it becomes larger for L-BFGS and FIRE.
Like line measurements, these results suggest that individual basins do not get intrinsically more fractal near jamming---this is just another mirage appearing in optimizer measurements.
In particular, the analogy proposed in past work between basins of attraction and Apollonian gaskets~\cite{Massen2007} quoted a fractal dimension close to $1.3$, which is close to what optimizers report but significantly larger than the CVODE values.

\section{Distribution of Energies at Minima\label{app:energies}}

\begin{figure}
    \centering
    \includegraphics[height=0.45\columnwidth]{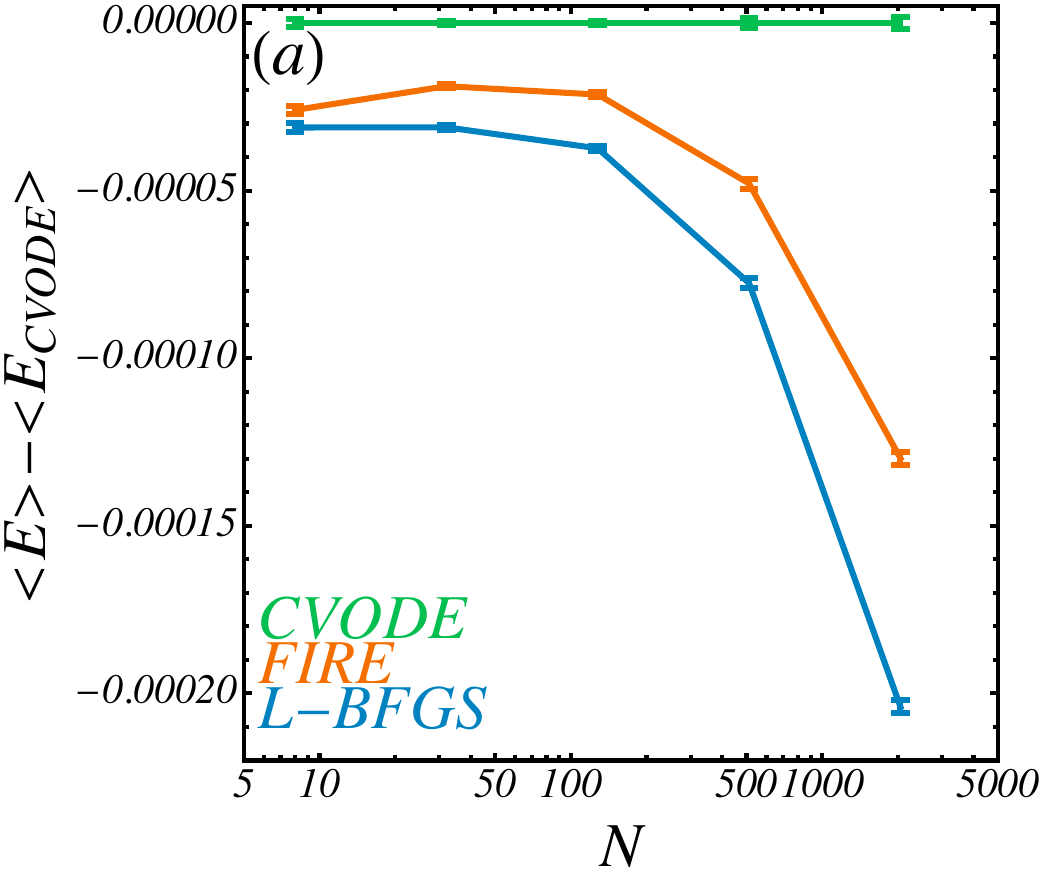} 
    \includegraphics[height=0.45\columnwidth]{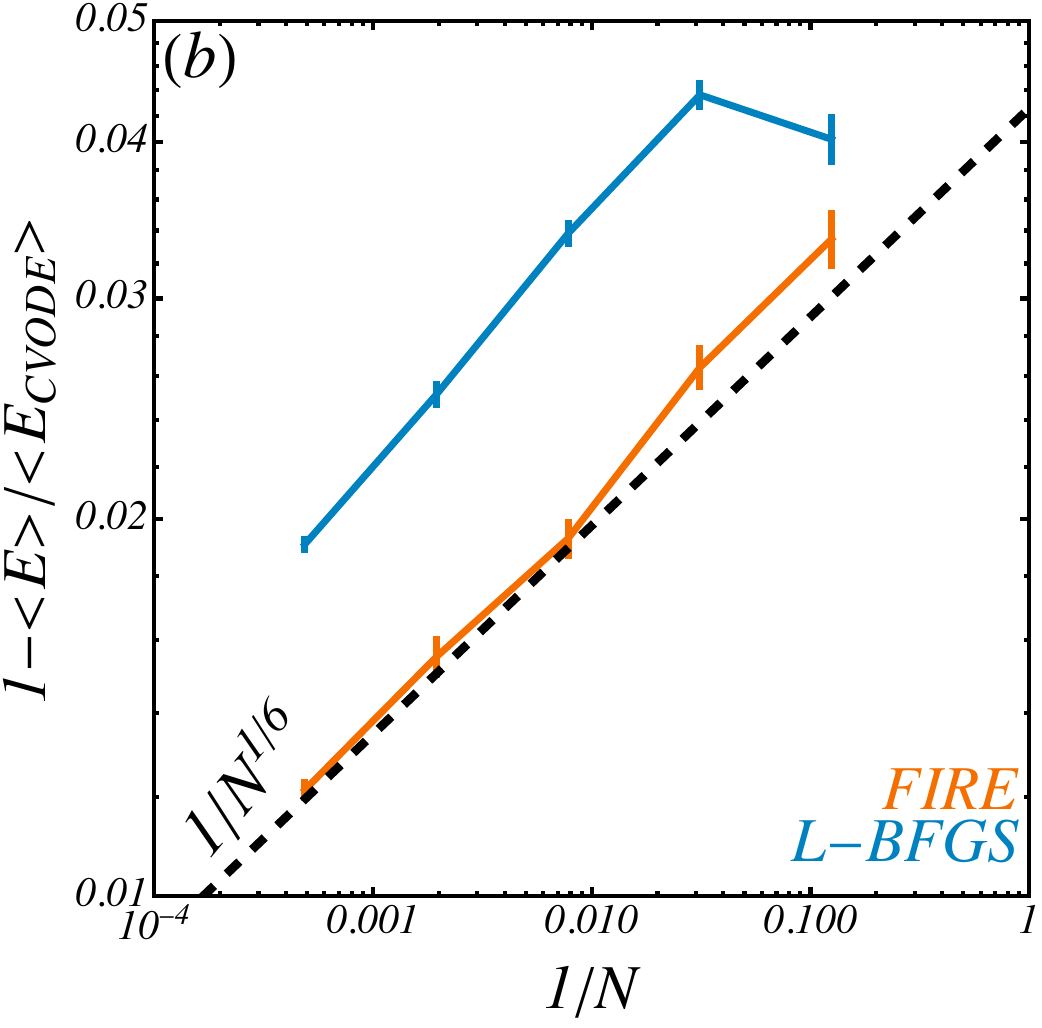} 
    \\
    \includegraphics[height=0.48\columnwidth]{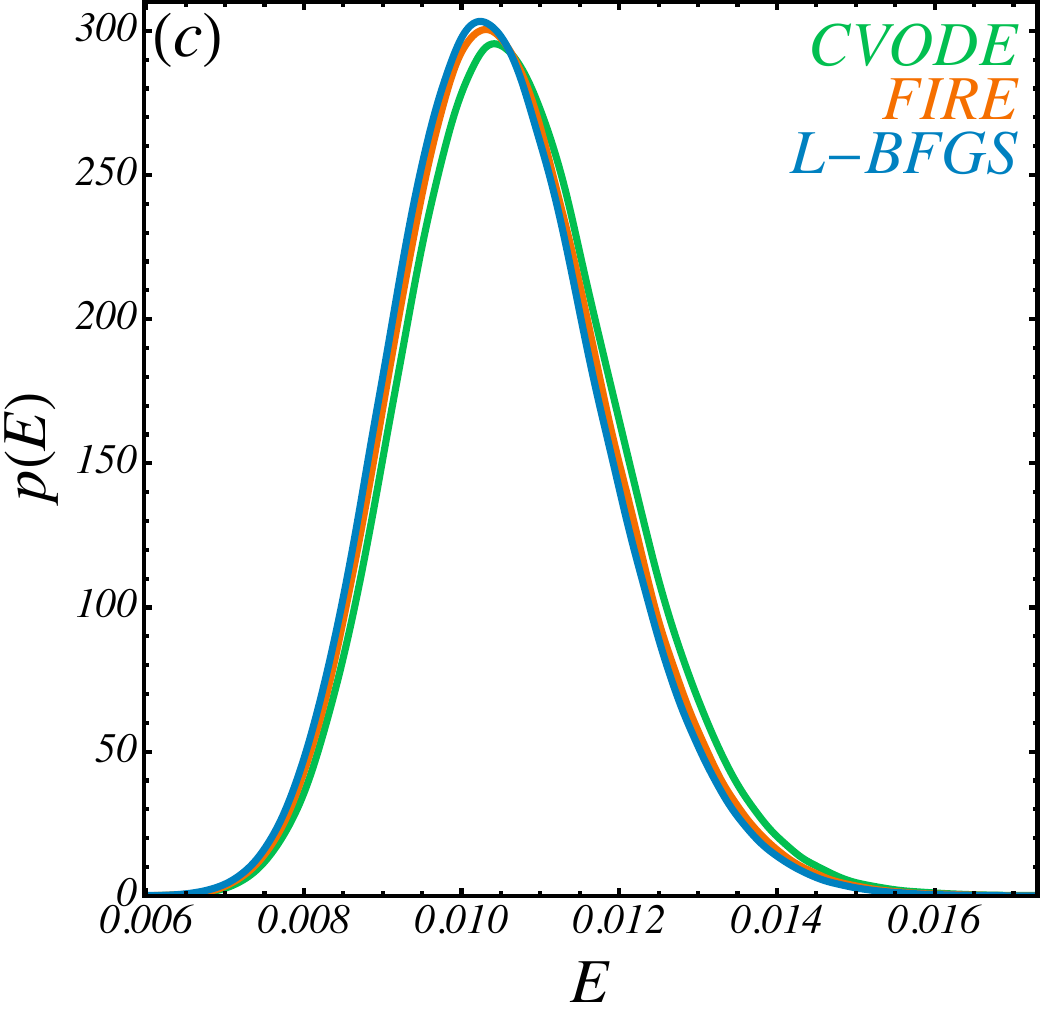} 
    \includegraphics[height=0.48\columnwidth]{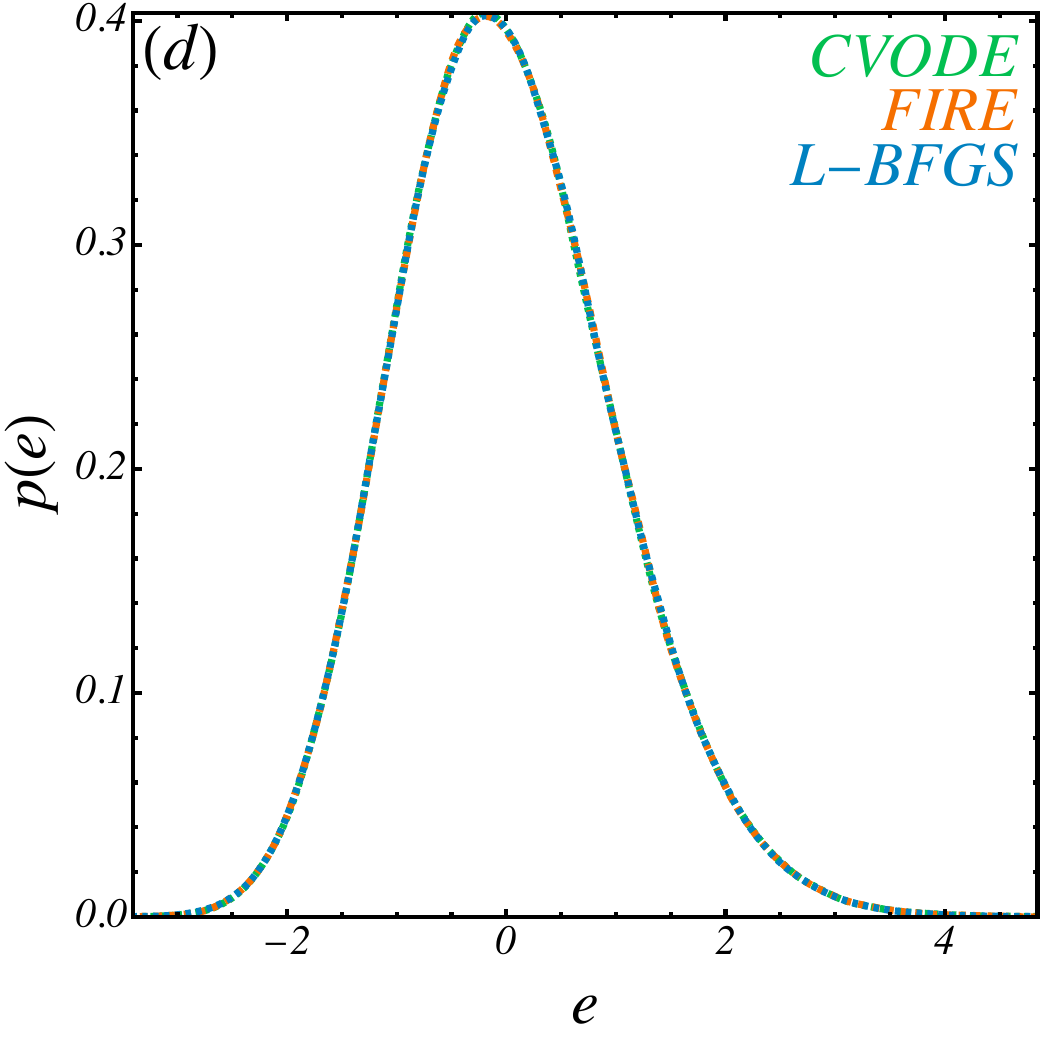} 
    \caption{\textbf{Energies at minima.}
    $(a)$ Differences between mean energies of minima for each method and the mean energies obtained with CVODE against $N$, in log-linear scales.
    The CVODE line is represented to indicate standard error on the CVODE mean.
    $(b)$ Relative error on the energies with respect to CVODE against $1/N$, in log-log scales.
    The dashed black line indicates $1/N^{1/6}$.
    $(c)$ Empirical distribution of energies $E$ at minima for all three methods for $N = 2048$.
    $(d)$ Corresponding empirical distribution of reduced energies $e \equiv (E - \langle E \rangle)/\sigma_E$, in dashed lines.
    Throughout the figure, we encode CVODE by green, FIRE by orange, and L-BFGS by blue.
    Error bars on the mean are obtained by bootstrapping over 1000 subsamples.
    }
    \label{fig:Energies}
\end{figure}

Here we describe the effect of the minimization method on the distribution of energies at minima, mentioned in the main text.
For each method, and for $N \in \left\{8, 32, 128, 512, 2048\right\}$, we use the same $10^6$ initial conditions for optimization, uniformly drawn at random in configuration space, at $\phi = 0.9$.
We then collect the energies at the minima, and study their distributions.
In Fig.~\ref{fig:Energies}$(a)$, we plot the differences between the mean total energies obtained with FIRE/L-BFGS and CVODE, as a function of $N$.
We report a growing bias towards lower energies in both FIRE and L-BFGS as $N$ grows, meaning that the choice of method does affect the distribution of energies.
It is worth noting that FIRE and L-BFGS do not create any new minima compared to CVODE, so that this difference is solely due to how often a given minimum is found---or, in other words, to the distribution of basin volumes being altered.
To investigate the effect of this phenomenon further, we plot the relative errors on mean energy per particle against $1/N$ in Fig.~\ref{fig:Energies}$(b)$.
We show that the relative error in fact decays with $N$, meaning that the bias of Fig.~\ref{fig:Energies}$(a)$ is subextensive.
However, as indicated by a dashed line, the error vanishes algebraically slowly and with a small power, roughly as $N^{-1/6}$ for both FIRE and L-BFGS.
While with our choice of potential the numerical value of the error remains rather small because the prefactor of the algebraic decay is small, another potential could display equally slow convergence with $N$ but with larger prefactors. 
Finally, we assess the effects of the minimization method on the full distributions of energies.
In Fig.~\ref{fig:Energies}$(c)$, we plot the empirical distributions obtained for total energies at minima for $N = 2048$ particles.
The distributions obtained through FIRE and L-BFGS display a systematic bias towards lower energies throughout.
However, as shown in Fig.~\ref{fig:Energies}$(d)$, the \textit{shape} of the distribution is only very weakly modified, as empirical distributions of the reduced energies at minima $e \equiv (E - \langle E \rangle)/\sigma_E$, with $\sigma_E$ the empirical standard deviation, overlap almost perfectly.

\section{Jamming point}
\label{app:jamming-point}
\begin{figure}[htbp]
  \centering
          \includegraphics[height=0.47\columnwidth]{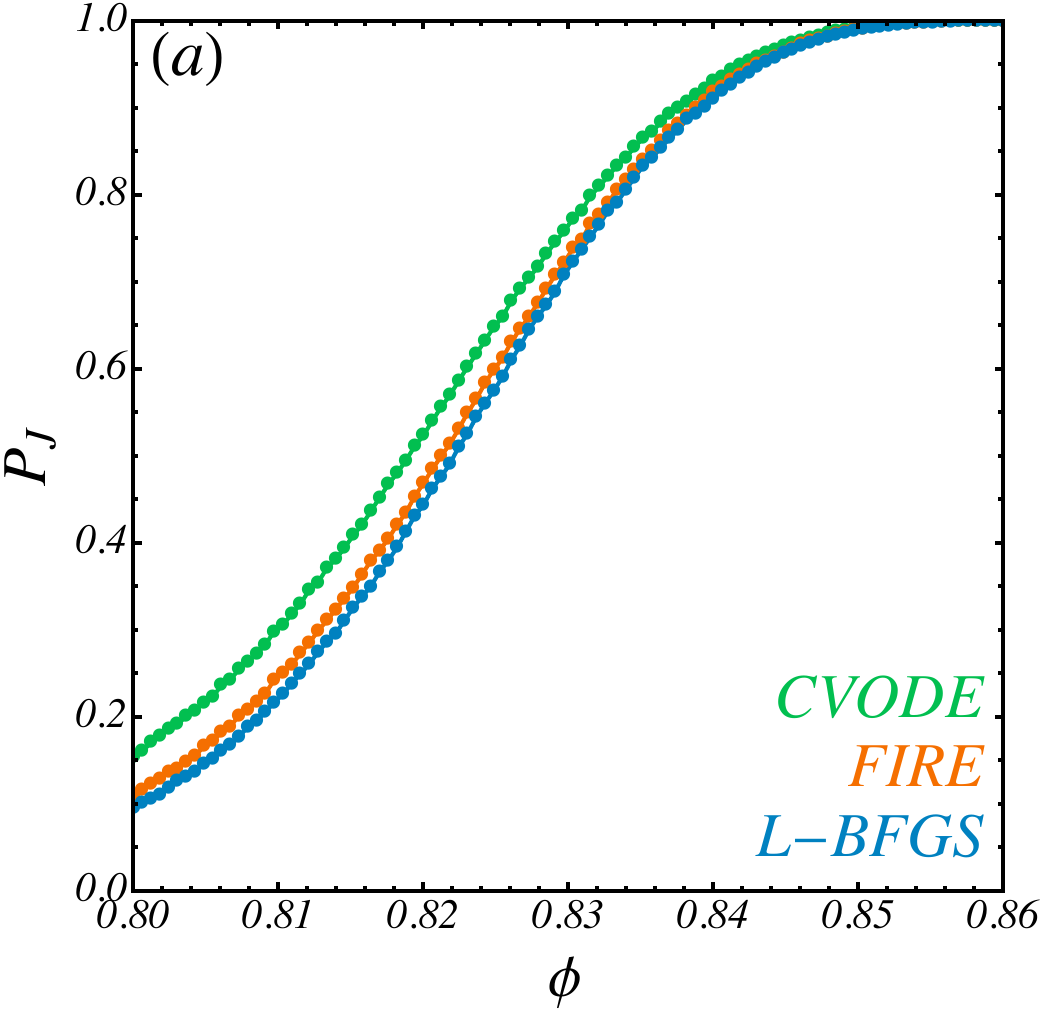}
          \includegraphics[height=0.47\columnwidth]{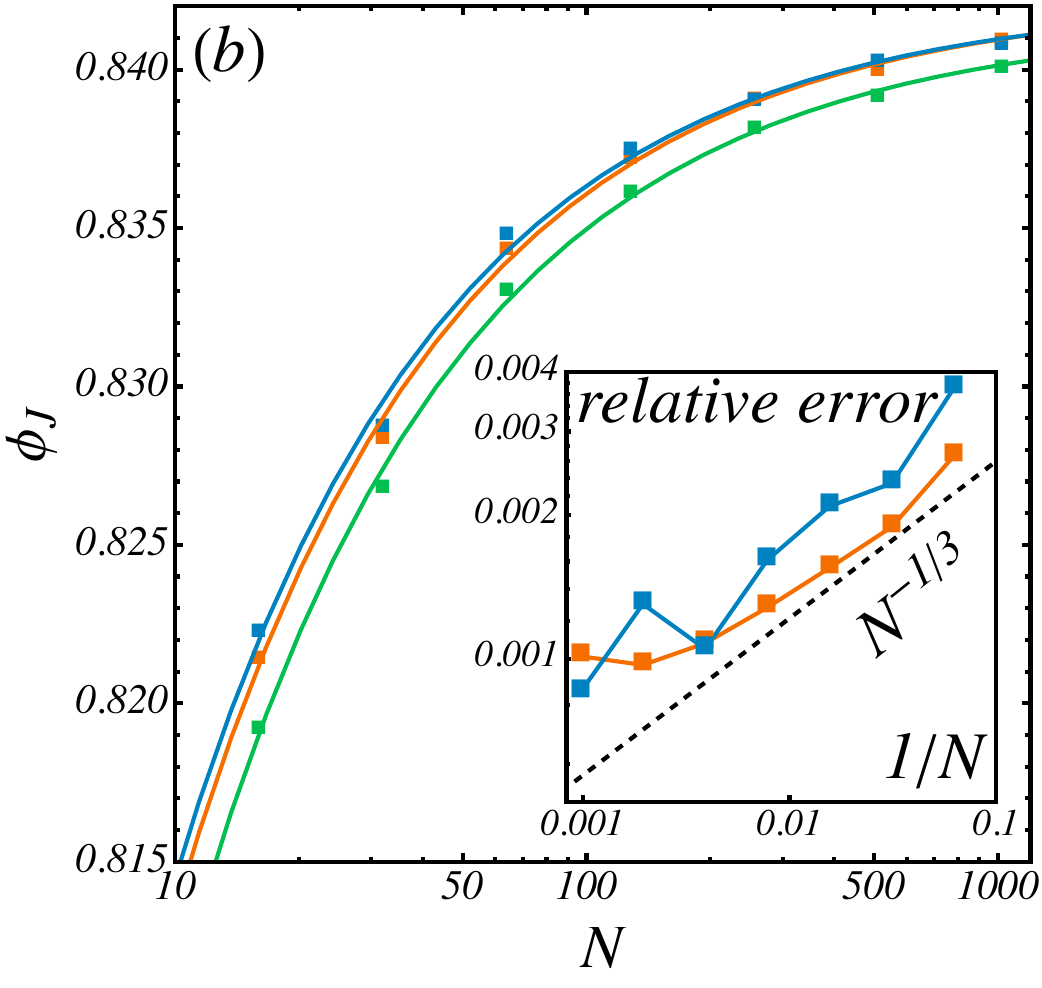}
          
  \caption{\textbf{Jamming Point}
  $(a)$ $P_J$ against $\phi$ for $N = 16$ across methods.
  Student-T $95\%$ confidence intervals are smaller than symbols.
  $(b)$ $\phi_J$ against $N$ across methods.
  Solid lines are critical power-law fits.
  Inset: relative error on $\phi_J$ compared to CVODE against $N$. Dashed black line: $N^{1/3}$ scaling (guide for the eyes).} 
  \label{fig:jammingmain}
\end{figure}

We study the effect of optimizers on the location of the jamming point.
We randomly sample uniform initial conditions in configuration space per packing fraction across a range of $\phi$, and report the fraction $P_J$ of them that falls into a minimum with all methods.
The results are shown in Fig.~\ref{fig:jammingmain}$(a)$ for $N = 16$ and $10^5$ random points.
We show that there is a noticeable shift in the curve when switching to optimizers.
We perform this measurement across $N$, using $10^3$ points per $N$ and density (see SM~\cite{supp} for full curves) and measure the packing fraction $\phi_J$ with $P_J = 1/2$.
The result is shown in Fig.~\ref{fig:jammingmain}$(b)$.
Curves roughly follow power laws of the form $\phi_J(\infty) - \phi_J(N) \propto 1/N^{\theta}$ with $\phi_J \approx 0.842$ and $ \theta \approx 0.67$ as previously reported~\cite{OHern2003,Vagberg2011}.
Like for energies, a deviation between optimizers and CVODE remains noticeable at large $N$.
In the inset, we show the evolution of the relative difference of $\phi_J$ between optimizers and CVODE, showing a trend slower than $N^{-1/3}$.
In short, optimizers introduce a bias on sampled minima that biases finite-size estimates of the jamming density, with a slowly decaying error.

\end{document}

\title{Supplementary Information for\\``The Basins of Attraction of Soft Sphere Packings Are Not Fractal''}

\author{Praharsh Suryadevara}
\affiliation{\nyuphysics}

\author{Mathias Casiulis}
\affiliation{\nyuphysics}
\affiliation{\nyusimons}

\author{Stefano Martiniani}
\email{sm7683@nyu.edu}
\affiliation{\nyuphysics}
\affiliation{\nyusimons}
\affiliation{\nyuchemistry}
\affiliation{\nyucns}
\affiliation{\nyucourant}

\maketitle

%% Adds the main heading for the SI text. Comment out this line if you do not have any supporting information text.

% Custom notations for Figs and Eqs
\renewcommand{\figurename}{FIG.}
\renewcommand{\thefigure}{S\arabic{figure}}
\renewcommand\theequation{S\arabic{equation}}
\renewcommand{\thetable}{S\arabic{table}}
\renewcommand{\arraystretch}{1.2}

\begin{center}
{\bf Supporting Information}    
\end{center}

\section{Minimization and basin identification strategies\label{sec:Minimizers}}

In the main text, we present comparisons between different numerical strategies to map out the basins of attractions of minima in the potential energy landscape of soft repulsive spheres.
Here, we first describe the minimization algorithms we used in more detail and justify our choice of ODE solver.
Then, we discuss how we determine that a minimization has converged, and how we match minima found from different minimizations.
Finally, we provide full tables of parameters for optimizers and ODE solvers used in this work.
The analysis is available in the basinerror repository~\cite{basinerror}.

\subsection{Basin Mapping strategies \label{sec:BasinMappingStrategies}}
The basins of attraction of the potential energy landscape are defined through steepest descent dynamics.
If $\bm{X} = (x_1, y_1, \ldots, x_{N}, y_{N})$ represents a position in the $(N-1)d$ dimensional space over which the energy landscape $E(\mathbf{X}; \mathbf{R})$ is defined, and $\bm{R} = (R_1, R_2, \ldots, R_N)$ is the vector of all particle radii (here assumed fixed), the steepest descent dynamics are  given by the ordinary differential equation (ODE)
\begin{align}
    \dot{\bm{X}}(t) = - \bm{\nabla}_{\bm{X}} E(\bm{X}(t); \bm{R}) \label{eq:GradientDescent},
\end{align}
with the initial condition $\bm{X}(t=0) = \bm{X}_0$ specifying which point to start from.
A basin is then uniquely identified by the infinite-time limit value of the solution to these dynamics, $\bm{X}_\infty = \lim_{t\to \infty}\bm{X}(t)$.
While there are other fixed points in the landscape (saddles, maxima), they are unstable, so that their ``basin'' is reduced to a point, and sampling random initial points $\bm{X}_0$ in configuration space almost surely leads to a minimum in the infinite time limit.
Thus, the basin identification problem reduces to that of computing the map $\bm{X}_0 \mapsto \bm{X}_\infty$, that transforms the configuration space of repulsive spheres into a finite set of attractors.

The ODEs of Eq.~\ref{eq:GradientDescent} may be solved numerically by introducing discrete time steps.
The most naïve approach to integrate it is the forward Euler scheme, which has been the dominant approach to identify basins.
The Forward Euler approach relies on the discretization,
\begin{align}
    \bm{X}(t+dt) \approx \bm{X}(t) - \mathrm{d}t \bm{\nabla}_{\bm{X}} E(\bm{X}(t); \bm{R}), \label{eq:ForwardEuler}
\end{align}
which is unstable on stiff problems such as the one we consider, leading to large errors, unless $\mathrm{d}t$ is made very small~\cite{hairer1996solving}.
This algorithm is often simply referred to as ``gradient descent'', or even ``steepest descent'', in the jamming and glass communities.
In order to increase the accuracy of the method, it is beneficial to introduce adaptive time-stepping.
That is why, in the main text, our ``GD'' method (see Fig. 2 of the main text) relies on the adaptive time-stepping proposed in the context of soft-sphere packings in Ref.~\cite{Charbonneau2023}, where $\mathrm{d}t$ is controlled by a parameter $\epsilon$ based on the cosine similarity between successive gradients,
\begin{align}
    \hat{\bm{g}}(t) \cdot \hat{\bm{g}}(t+\mathrm{d}t) > 1 - \epsilon, \label{eq:GradientCosine}
\end{align}
where $\hat{\bm{g}}(t) = \bm{\nabla}_{\bm{X}} E(\bm{X}(t); \bm{R}) / | \bm{\nabla}_{\bm{X}} E(\bm{X}(t); \bm{R}) |$.
In practice, at each step, a new location is proposed according to Eq.~\ref{eq:ForwardEuler} and, if criterion~\ref{eq:GradientCosine} is met, the step is accepted.
Otherwise it is rejected, $\mathrm{d}t$ is decreased, and another candidate location is proposed.
Additionally, if the cosine similarity condition is satisfied for $n$ successive steps, then the stepsize is multiplied by $n$.
Recent work suggests~\cite{Charbonneau2023} that this adaptive forward Euler algorithm has a better time-accuracy trade-off than other described gradient descent variations in the literature \cite{Asenjo2013, stanifer2022avalanche}.
Yet, the inherent instability of the forward-Euler scheme means that the lowest $\mathrm{d}t$ required for this algorithm to converge to the true mapping between initial points and minima is very small, and as such it has a prohibitive computational cost in most cases.

As a result, many past works have opted for algorithms that do not solve the dynamics~\ref{eq:GradientDescent}, but instead rely on proxy dynamics.
In the context of soft sphere packings, careful studies have shown good accuracy for basin mapping in spite of the proxy dynamics, \textit{e.g.} Ref.~\cite{Asenjo2013}.
However, these studies were restricted to very small systems ($d(N-1) < 5$) and, as we show in this work, the accuracy of these methods plummets as $N$ grows.
Below, we describe the two main methods used in the literature.

On the one hand, some works, \textit{e.g.} Refs.~\cite{Xu2011, zhao2011new} have opted for quasi-Newton methods such as L-BFGS~\cite{Nocedal1999}, that use information about the second derivative of the energy to accelerate convergence to the minimum.
In a nutshell, these methods rely on a second-order Taylor expansion of the energy around any point $\bm{X}_0$,
\begin{align}
    E(\bm{X}_0 + \delta \bm{X}, \bm{R}) \approx E(\bm{X}_0; \bm{R}) +  \bm{\nabla}_{\bm{X}} E(\bm{X}_0; \bm{R})  \cdot \delta \bm{X} + \frac{1}{2} \delta \bm{X}^T \cdot \overline{\overline{\bm{H}}}(\bm{X}_0; \bm{R}) \cdot \delta \bm{X}, 
\end{align}
where $\overline{\overline{\bm{H}}}$ is the Hessian matrix of the energy, and $\delta \bm{X}$ is a small displacement vector in configuration space, such that $||\delta \bm{X}||^2$ is small compared to the scales of variation of the gradient and Hessian.
One may then compute the derivative of this expression with respect to $\delta \bm{X}$ to find an optimal update,
\begin{align}
    \bm{\nabla}_{\bm{X}} E(\bm{X}_0 + \delta \bm{X}; \bm{R}) \approx \bm{\nabla}_{\bm{X}} E(\bm{X}_0; \bm{R}) + \overline{\overline{\bm{H}}}(\bm{X}_0; \bm{R}) \cdot \delta \bm{X}.
\end{align}
The optimal update corresponds to the case $\bm{\nabla}_{\bm{X}} E(\bm{X}_0 + \delta \bm{X}; \bm{R}) = 0$, that would converge to the minimum in a single step in a quadratic energy landscape.
To achieve this, the ``Newton'' step is then given by
\begin{align}
    \delta \bm{X}_{N} = - \overline{\overline{\bm{H}}}^{-1}(\bm{X}_0; \bm{R}) \cdot \bm{\nabla}_{\bm{X}} E(\bm{X}_0; \bm{R}).
\end{align}
A quasi-Newton method is one that replaces the inverse of the Hessian, which is costly to compute, by a cheaper numerical evaluation.
The issue with such methods is that they are guaranteed to map a point to the steepest descent minimum only if the energy landscape is strictly convex.
As a result, this method is in general likely to introduce \textit{some} error in the mapping between initial points and minima defined through steepest descent.

Many other works have relied on momentum-based methods, \textit{e.g.} Refs.~\cite{Martiniani2016, Martiniani2016a, Martiniani2017a,Dennis2020,Hagh2024}, most notably FIRE~\cite{Bitzek2006} where a small amount of inertia is added to the dynamics.
Instead of the gradient descent equation, Eq.~\ref{eq:GradientDescent}, FIRE solves the second-order ODE~\cite{Bitzek2006}
\begin{align}
    m \ddot{\bm{X}}(t) + \gamma(t) \left(\dot{\bm{X}}(t) + |\dot{\bm{X}}(t)| \frac{\bm{\nabla}_{\bm{X}} E(\bm{X}(t); \bm{R})}{|\bm{\nabla}_{\bm{X}} E(\bm{X}(t); \bm{R})|} \right) = - \bm{\nabla}_{\bm{X}} E(\bm{X}(t); \bm{R}),
\end{align}
with $m$ a mass and $\gamma(t)$ a damping coefficient.
The effect of the term proportional to $\gamma$ is to re-orient the velocity onto the steepest descent direction, but with a delay.
In FIRE, $\gamma(t)$ is dynamically updated so as to avoid uphill motion~\cite{Bitzek2006}.
Although the addition of momentum accelerates convergence to a minimum, the dynamics are very different from those of steepest descent. We use a strictly downhill variant of FIRE with limited stepsize~\cite{Asenjo2014}.

\subsection{Choice of ODE solver}\label{sec:choice-ode-solver}

\begin{figure*}[htbp]
  \centering
    \includegraphics[width=\textwidth]{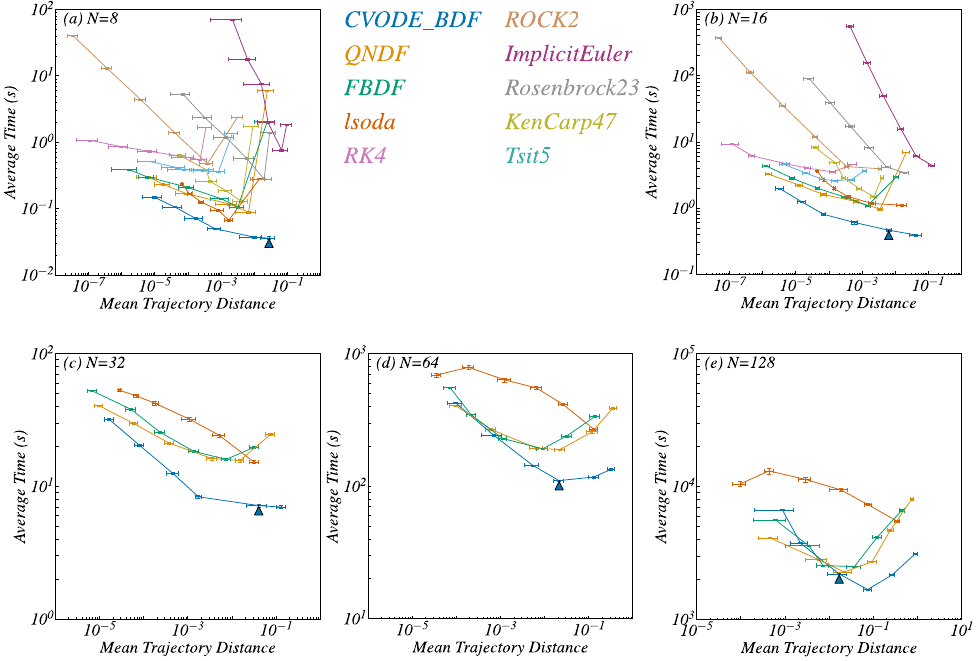}
  \caption{\textbf{Benchmarking ODE solvers.} $(a)$--$(e)$ Plots of the Mean Trajectory Distance versus Average Computation Time for different numbers of particles $N = 8$, $16$, $32$, $64$, and $128$, respectively.
  Error bars represent the standard error of the mean (SEM) for both the Mean Trajectory Distance and Average Computation Time.
  Points highlighted by filled triangles correspond to the maximum tolerance values used among all calculations.}
  \label{fig:WorkPrecisionDiagrams}
\end{figure*}

We investigate which ODE solver offers the best possible time-for-error trade-off.
Due to the stiffness of the problem, we focus on adaptive implicit solvers.
We test a number of candidate algorithms on the energy landscape of Hertzian disk packings, and report a benchmark in Fig.~\ref{fig:WorkPrecisionDiagrams}.
The benchmark consists of work-precision diagrams computed for $N = 8$, $16$, $32$, $64$, $128$, with the same size distribution as in the main text.
For each system size, we spawn 100 random starting points $\bm{X}_0$, drawn uniformly in the periodic square box $[0;L]^2$, then solve the steepest descent dynamics, Eq.~\ref{eq:GradientDescent}, using a wide range of ODE solvers available in the Julia \textit{DifferentialEquations.jl} package\cite{Rackauckas2017}, including \textit{lsoda}, a Fortran ODE solver library that switches between implicit and explicit methods based on a stiffness criterion.
For $N = 32$, $64$, and $128$, we restrict ourselves to more performant methods to avoid prohibitive computation times.
The error in adaptive ODE solvers is managed via a relative local error tolerance, $\texttt{rtol}$, and an absolute error tolerance $\texttt{atol}$, that ensure at each step that the trajectory does not stray away from the true ODE solution more than a specified amount (using a relative and an absolute distance, respectively).
The absolute error tolerance $\texttt{atol}$ sets the error bound when the coordinates are close to the origin, where the relative indicator $\texttt{rtol}$ is ill-defined.
Here, each ODE solver is run independently for a variety of values of the relative tolerance parameter $\texttt{rtol}$, and for a large enough integration time that the distance of the final point to the minimum is less than $10^{-2}$, in this case $t_{\text{stop}} = 10000$.
We ensure that the relative performance of ODE solvers is insensitive to the stopping time.
We then measure for each computed trajectory: 1) the time the computation took (in seconds), and 2) the maximal distance a particle deviates from the true steepest descent path over the whole trajectory,
\begin{align}
    d^{\text{solver}}_{\max}(\bm{X}_0, \texttt{rtol}) \equiv \max_{0\leq t\leq t_{\text{stop}}}\left\|\bm{X}_{\text{solver}}(t; \bm{X}_0,\texttt{rtol}) - \bm{X}_{\text{ref}}(t ; \bm{X}_0) \right\|,
\end{align}
where $\bm{X}_{\text{solver}}(t; \bm{X}_0,\texttt{rtol})$ is the trajectory found by the solver from the initial position $\bm{X}_0$ and at a given value of $\texttt{rtol}$, and $\bm{X}_{\text{ref}}$ is the steepest descent trajectory.
We then construct a Mean Trajectory Distance, defined as the average over random initializations $\bm{X}_0$ of $d^{\text{solver}}_{\max}$.
In practice, we estimate the Mean Trajectory Distance using 100 evenly spaced points along the trajectory.
The reference trajectory $\bm{X}_{\text{ref}}$ is obtained using \textit{CVODE\_BDF} with stringent tolerances (\texttt{rtol} = \texttt{atol} = $10^{-12}$).

The results show that among the wide range of parameters we use, \textit{CVODE\_BDF}~\cite{Hindmarsh2005, Gardner2022} (an adaptive-step implicit method that is part of the SUNDIALS suite) outperforms other ODE solvers, with \textit{QNDF} and \textit{FBDF} also showing competitive performance as system size increases.
We emphasize that all ODE solvers converge to the same trajectory as \texttt{rtol} is decreased, and our calculations are always performed in the regime where our identified basin is independent of precise solver parameters.
We therefore adopt CVODE as our reference ODE solver in the rest of the paper.

\subsection{Performance on other systems}

To rule out the possibility that CVODE is optimal only for our choice of interaction potential, we perform the same benchmark as in Sec~\ref{sec:choice-ode-solver} on $3$ other systems, spawning $100$ random starting points uniformly in configuration space for each of them.

\begin{figure}
  \centering
  \includegraphics[width=\textwidth]{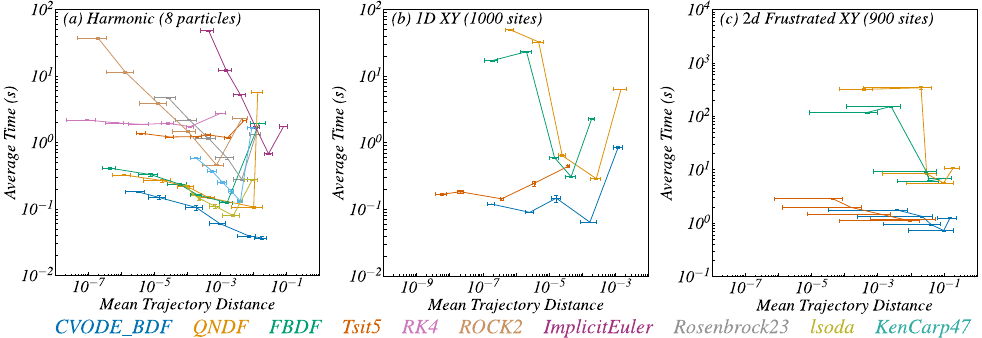}
  \caption{\textbf{Benchmarking ODE solvers on other systems.}  (a) Interacting pairwise potential with harmonic interactions (b) 1D XY model (Kuramoto) (c) Frustrated XY model on a triangular lattice. Error bars represent the standard error of the mean (SEM) for both the Mean Trajectory Distance and Average Computation Time. For the XY model [(b) and (c)] we utilize Jacobian Free GMRES solvers provided by the libraries for all implicit methods (CVODE, QNDF, FBDF) for speed.}
  \label{fig:OtherEnergyLandscapes}
\end{figure}

In Fig.~\ref{fig:OtherEnergyLandscapes}$(a)$, we consider a set of $N = 8$ soft spheres interacting via a harmonic potential,
\begin{align}
    V_{ij}(r_{ij}) = \varepsilon \left( 1 - \frac{r_{ij}}{R_{i} + R_{j}} \right)^2 \mathbb{1}\left(r \leq R_i + R_j \right), \label{eq:HarmonicPotential}
\end{align}
rather than a Hertzian one.
We show that CVODE outperforms other solvers in a way similar to that described in the case of a Hertzian potential, Fig.~\ref{fig:WorkPrecisionDiagrams}.

In Figs.~\ref{fig:OtherEnergyLandscapes}$(b)-(c)$, we consider variations on the XY model~\cite{Kosterlitz1974}, with the energy function
\begin{align}
    E = - \sum\limits_{i = 1}^{N} \sum\limits_{j \in \partial i} J_{ij} \hat{\bm{s}}_i\cdot\hat{\bm{s}}_j, \label{eq:XY}
\end{align}
where $\hat{\bm{s}}_i = (\cos\theta_i, \sin\theta_i)$ is a two-dimensional unit vector parametrized by an angle $\theta_i \in [0; 2\pi)$, $J_{ij}$ is the interaction constant between spins $i$ and $j$ ($J_{ij} > 0 $ is aligning, $J_{ij} < 0$ is anti-aligning), and $\partial i$ is the neighborhood of spin $i$, which depends on the chosen geometry for the problem.
Note that the dynamical system defined by the steepest descent equation for Eq.~\ref{eq:XY} with aligning interactions is also known as the Kuramoto model, whose basins of attraction have been previously investigated~\cite{Wiley2006, Martiniani2017, Zhang2021, Zhang2024}.
In Fig.~\ref{fig:OtherEnergyLandscapes}$(b)$, we choose a $1d$ geometry with nearest-neighbor interactions only, and we set $J_{ij} = +1$ between neighbors, defining a usual ferromagnetic XY model.
In Fig.~\ref{fig:OtherEnergyLandscapes}$(c)$, we choose a $2d$ triangular lattice with $J_{ij} = -1$ interactions between nearest neighbors, thus defining a fully-frustrated XY model~\cite{Villain1977, Villain1977a, Lee1991}, a classical model of deterministic spin glass.
In both cases, Tsit5 \cite{Rackauckas2017,tsitouras2011runge}, a variant of RK4, performs comparably to CVODE.
This implies that the steepest descent equation in the case of the XY model, even in a glassy regime, is not stiff enough that there is a significant performance gap between implicit and explicit methods.
Note that we checked that \textit{lsoda}, the solver used in Refs.~\cite{Zhang2021, Zhang2024}, performs significantly worse than the solvers we show in this benchmark, so that we do not show it in Fig.~\ref{fig:OtherEnergyLandscapes}.

\subsection{Convergence Criterion}
\label{sec:conv-crit}

Throughout this study, a convergence criterion is needed to decide whether an optimizer has run for sufficiently long that it has approached its infinite-time attractor.
Far from jamming (here, empirically, for packing fractions $ \phi \gtrsim 0.86 $), the Hessian eigenvalues are relatively large throughout the landscape, so that we use a simple gradient-based convergence criterion: we end ODE solving and optimizations when $ \|\mathbf{g}\| < 10^{-10} $.
However, as the system gets closer to jamming, the Hessian eigenvalues may become small, as the energy landscape flattens.
As a result, it is possible to have very small forces on particles far from the minimum, so that a purely gradient-based criterion can trigger premature termination of the minimization.
To circumvent this issue, we employ a Newton-step-based convergence criterion that uses a second-order estimate of the distance to the minimum,
\begin{equation}
\delta \mathbf{X}_{\text{min}} \approx \left(\overline{\overline{\bm{H}}}(\bm{X}) + \lambda (\bm{X})\, \overline{\overline{\bm{I}}}\right)^{-1} \mathbf{g}(\bm{X}),  
\end{equation}
where $ \mathbf{g} $ is the gradient, $\overline{\overline{\bm{H}}} $ is the Hessian, $\overline{\overline{\bm{I}}}$ is the identity matrix, and $ \lambda $ is chosen to ensure that $ H + \lambda I $ remains well-conditioned.
In practice, $ \lambda = \max\left(2|\lambda_{\min}|,\;\gamma\,\lambda_{\text{avg}}\right) $, where $ \lambda_{\min} $ and $ \lambda_{\text{avg}} $ are the minimum and average Hessian eigenvalues at $\bm{X}$, respectively, and $ \gamma = 0.1 $.
We consider the dynamics converged when
\begin{equation}
  \label{eq:r}
\|\delta \mathbf{X}_{\text{min}}\| < 10^{-5}.  
\end{equation}
This threshold is three orders of magnitude smaller than the maximal distance between non rattler particles used to declare two minima as different, $ 10^{-2} $.
The reason for using a smaller value is that Newton steps are only a heuristic and may lower accuracy if one does not make its range smaller than every other scale of the landscape near minima.
We check that this value does not introduce errors compared to a very small gradient criterion in the landscape of Hertzian disks, but comes at a lower computational cost.

\subsection{Minima matching procedure}
\label{sec:minima-match-proc}

It is crucial to be able to accurately determine whether $2$ points obtained as large-time limits of numerical dynamics map to one and the same minimum of the energy landscape.
In the context of jammed packings, the procedure is complicated by the presence of rattlers, \textit{i.e.} particles that can move over a finite-measure set of positions without changing the overall energy of the system.
Thus, to match minima, we first identify all rattlers in the system by checking whether they are contained by a convex hull of non-rattler neighbors.
We then check whether the remaining particles and the number of contacts are enough to create a stable backbone, using the fact that we need at least $2 (N_{\text{stable particles}}-1)  + 1$ contacts for a stable backbone.
If there are not enough contacts, we conclude that the configuration is a fluid state, and is simply tagged as fluid.
Otherwise, we conclude that the minimum is jammed.

To identify whether two jammed minima are different, we first check whether they have the same number of rattlers---if not, they are different.
If they do, to account for translational symmetry, we superimpose the first non-rattler particle of each packing onto each other (recall that the set of radii is the same so that there is a clear particle identity).
The two minima are identified as different if the maximal distance across all non-rattler particles with the same radius is greater than $\texttt{dtol} = 10^{-2}$.
In practice, at jamming densities over $0.85$, and across various minima, all matches we find correspond to maximal distances smaller than $10^{-4}$ when we set a gradient tolerance $\text{tol} = 10^{-10}$ in the convergence criterion (see Sec.~\ref{sec:Params}).
Our choice of $\texttt{dtol}$ is the largest value that doesn't lead to appreciable change in the landscape at any value of $(N,\phi)$.

\subsection{Choices of solver parameters\label{sec:Params}}

In this section, we provide complete lists of parameter for all solvers used in the main text---namely, CVODE, FIRE, L-BFGS, and Gradient Descent.
Whenever relevant, we also provide a rationale for these choices.

\subsubsection{CVODE Parameters\label{app:CVODETolerance}}

Many adaptive ODE solvers, including CVODE, adjust their step size by imposing an upper bound on the estimated local error for each step.
As noted in Sec.~\ref{sec:Minimizers}, this bound is specified by two tolerance settings: relative error ($\texttt{rtol}$) and absolute error ($\texttt{atol}$).
However, constraining this local error estimate does not guarantee that the global error will adhere to the same bounds, particularly in systems where small errors can lead to diverging trajectories.
To address this limitation, we set values for tolerances based on an analysis of the accuracy of the mapping $\bm{X}_0 \mapsto \bm{X}_\infty$ between random initial points and their corresponding final minima.

To assess this accuracy, one needs a reference (a ground truth).
Relying on the fact that CVODE is an ODE solver, and thus asymptotically converges to the true steepest-descent trajectories at vanishing $\rtol$, we expect minimal error when $\rtol$ is small enough that the mapping remains unchanged over large random sets of points as we decrease $\rtol$.
Across all our simulation parameters, this is achieved when $\rtol$ is set to $10^{-14}$.

\begin{figure*}[htbp]
    \centering
    % Row 1: N = 8, 8, 16, 16
    \includegraphics[width=0.23\textwidth]{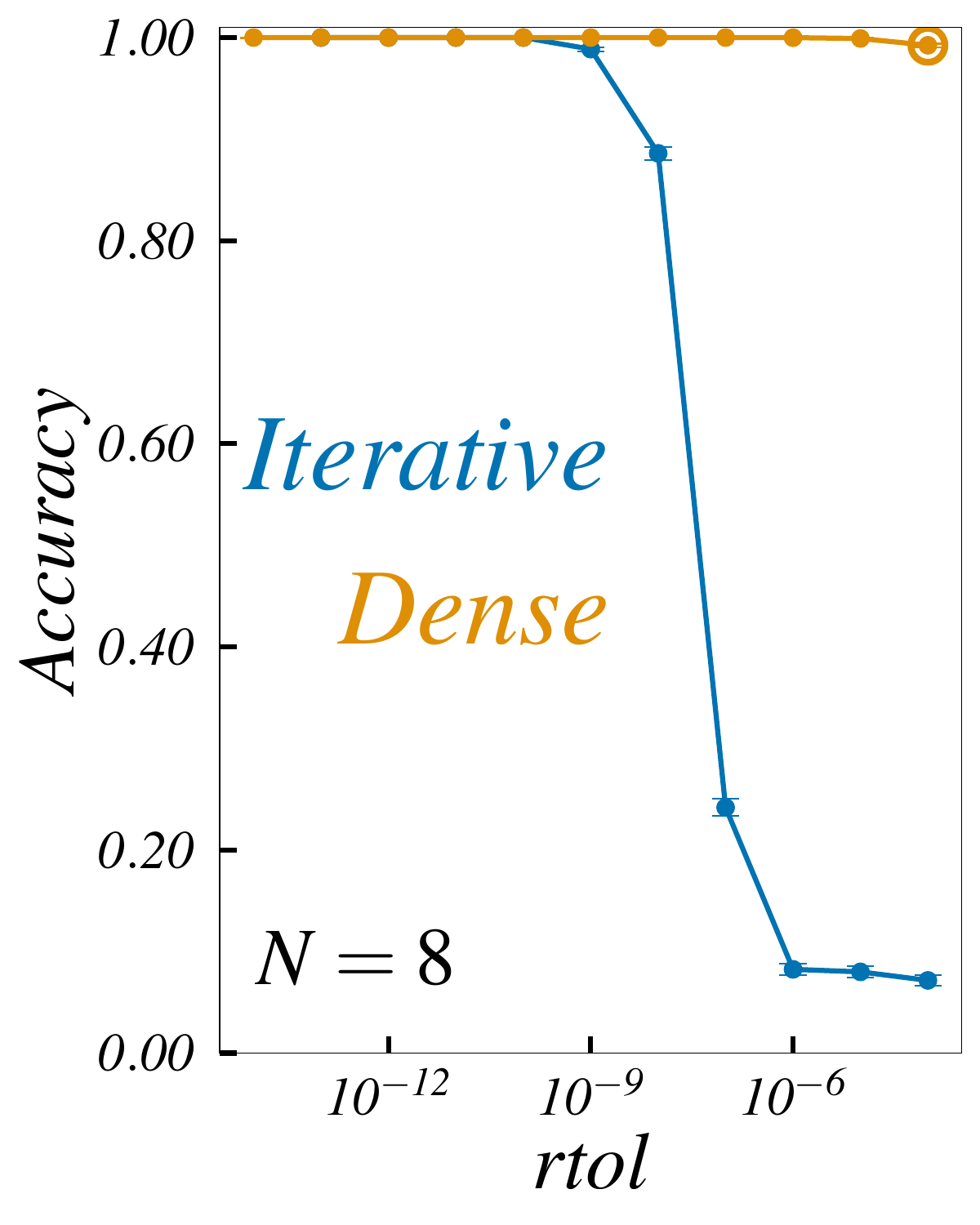}\hfill
    \includegraphics[width=0.23\textwidth]{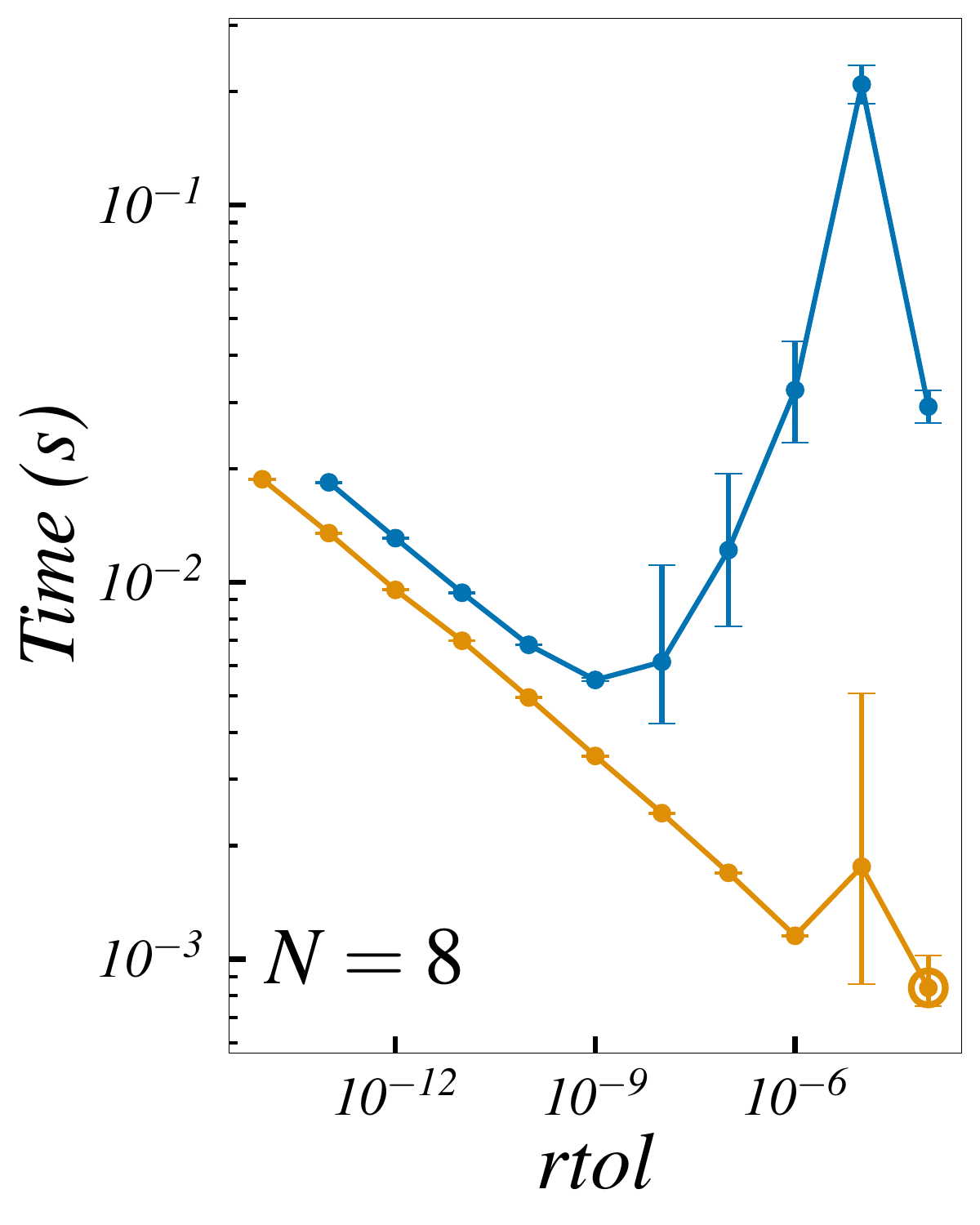}\hfill
    \includegraphics[width=0.23\textwidth]{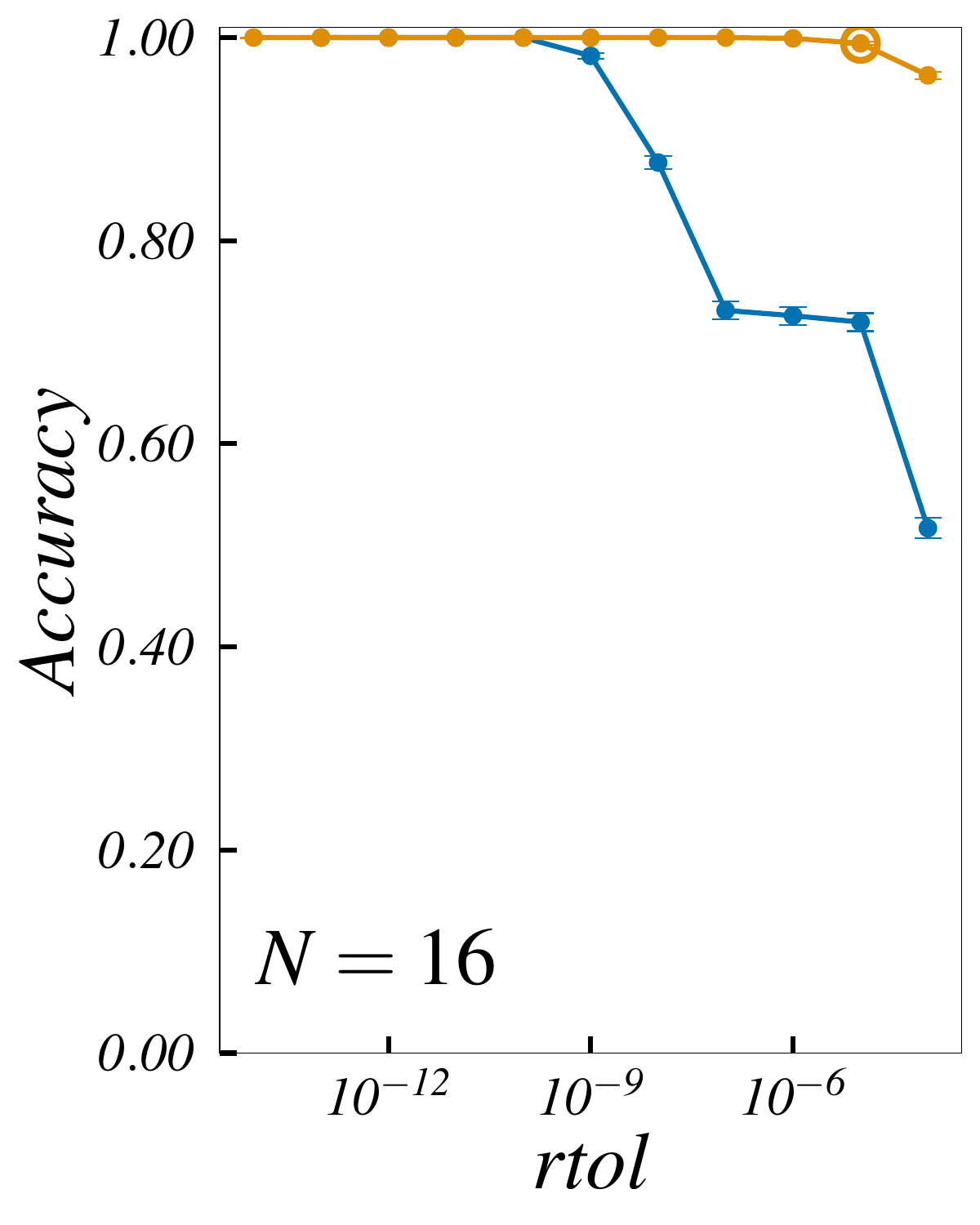}\hfill
    \includegraphics[width=0.23\textwidth]{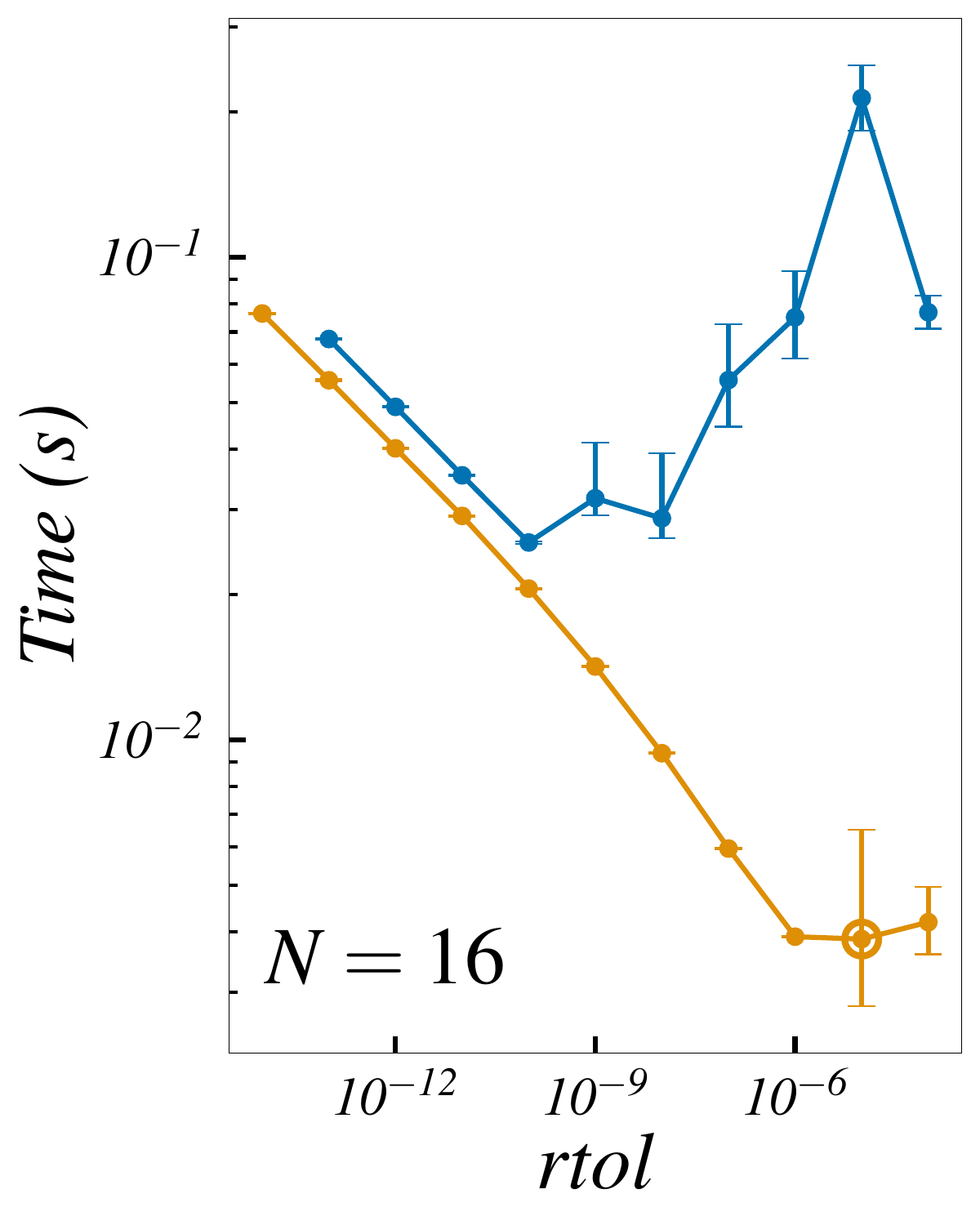}
    
    \vspace{1em}
    % Row 2: N = 32, 32, 64, 64
    \includegraphics[width=0.23\textwidth]{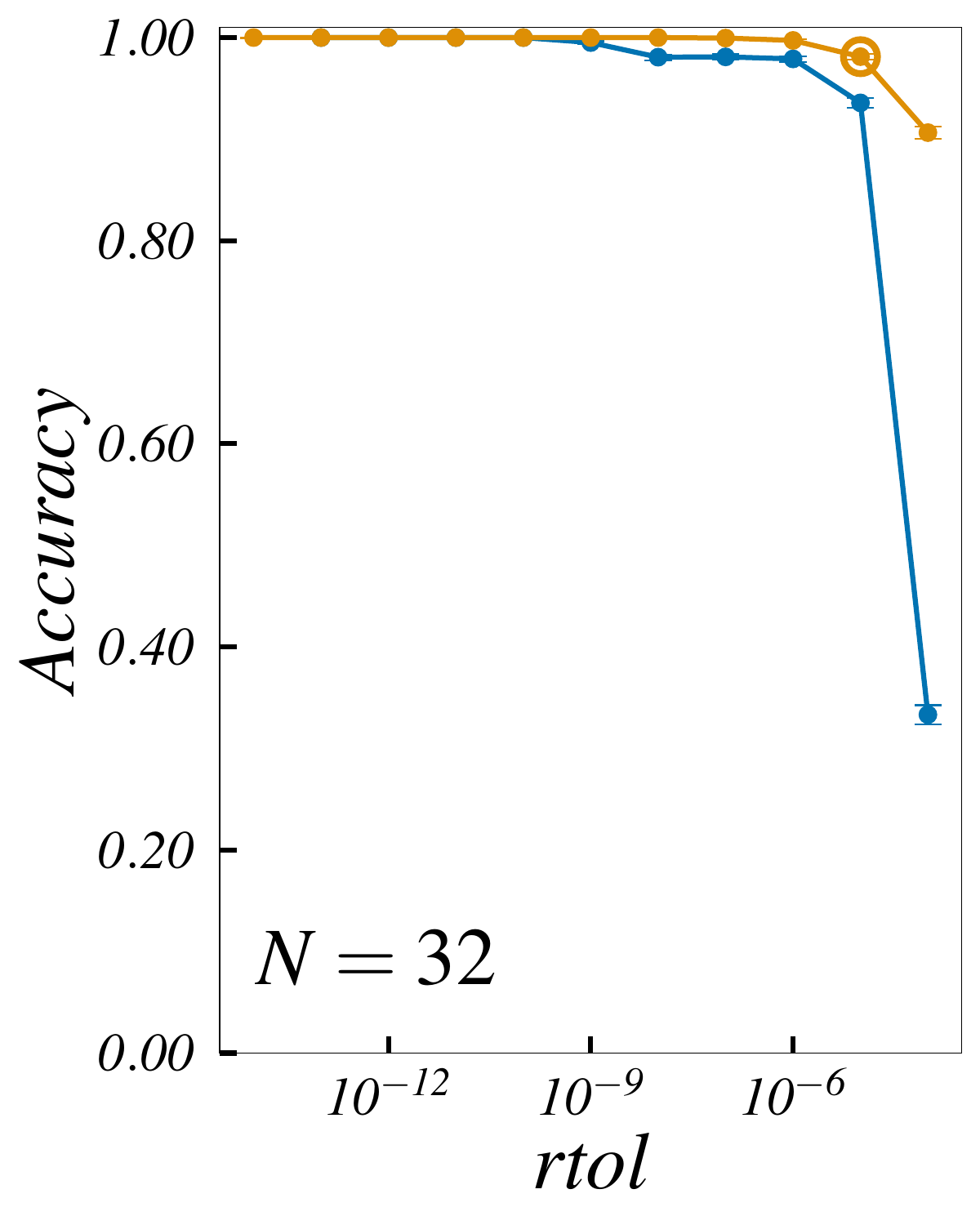}\hfill
    \includegraphics[width=0.23\textwidth]{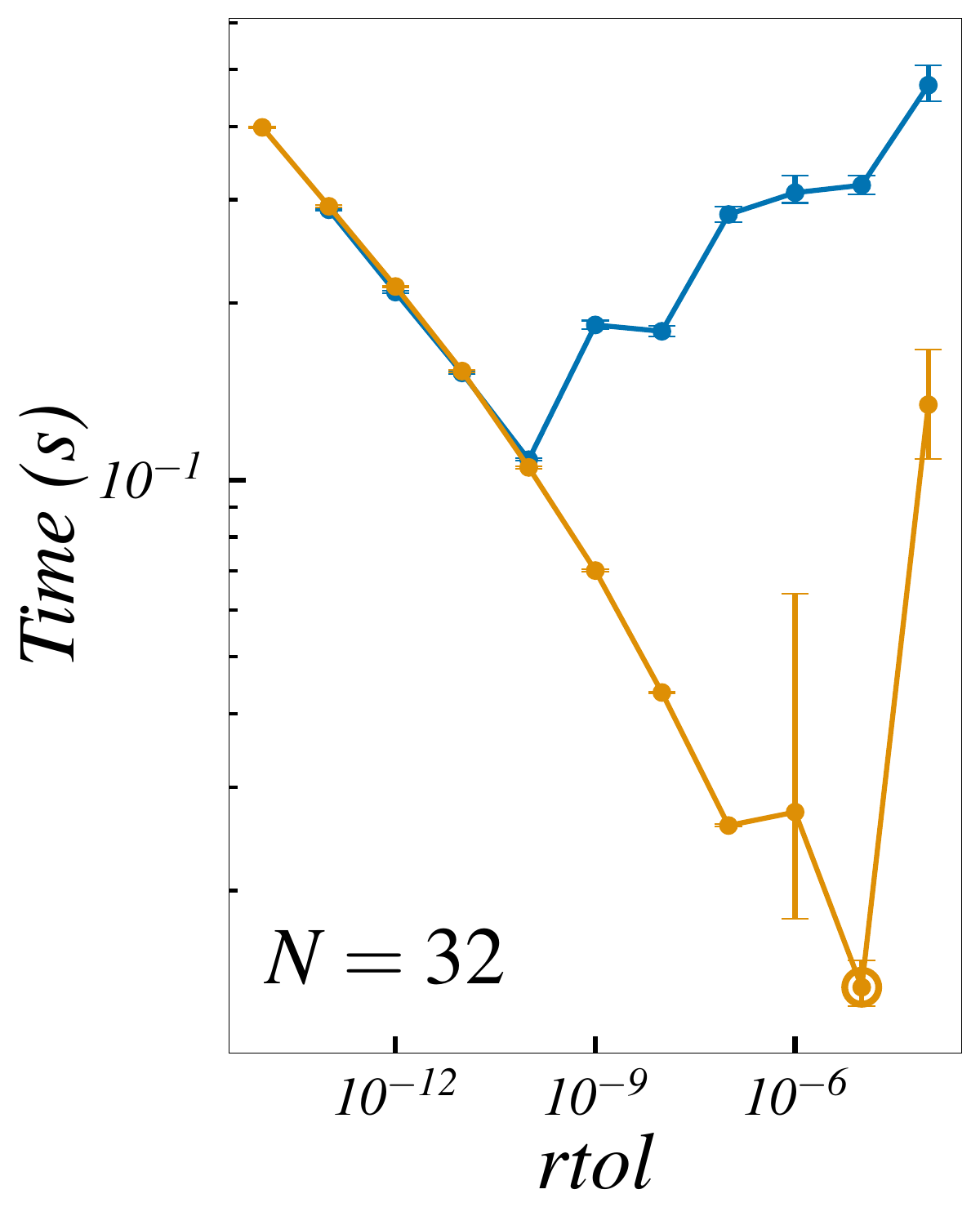}\hfill
    \includegraphics[width=0.23\textwidth]{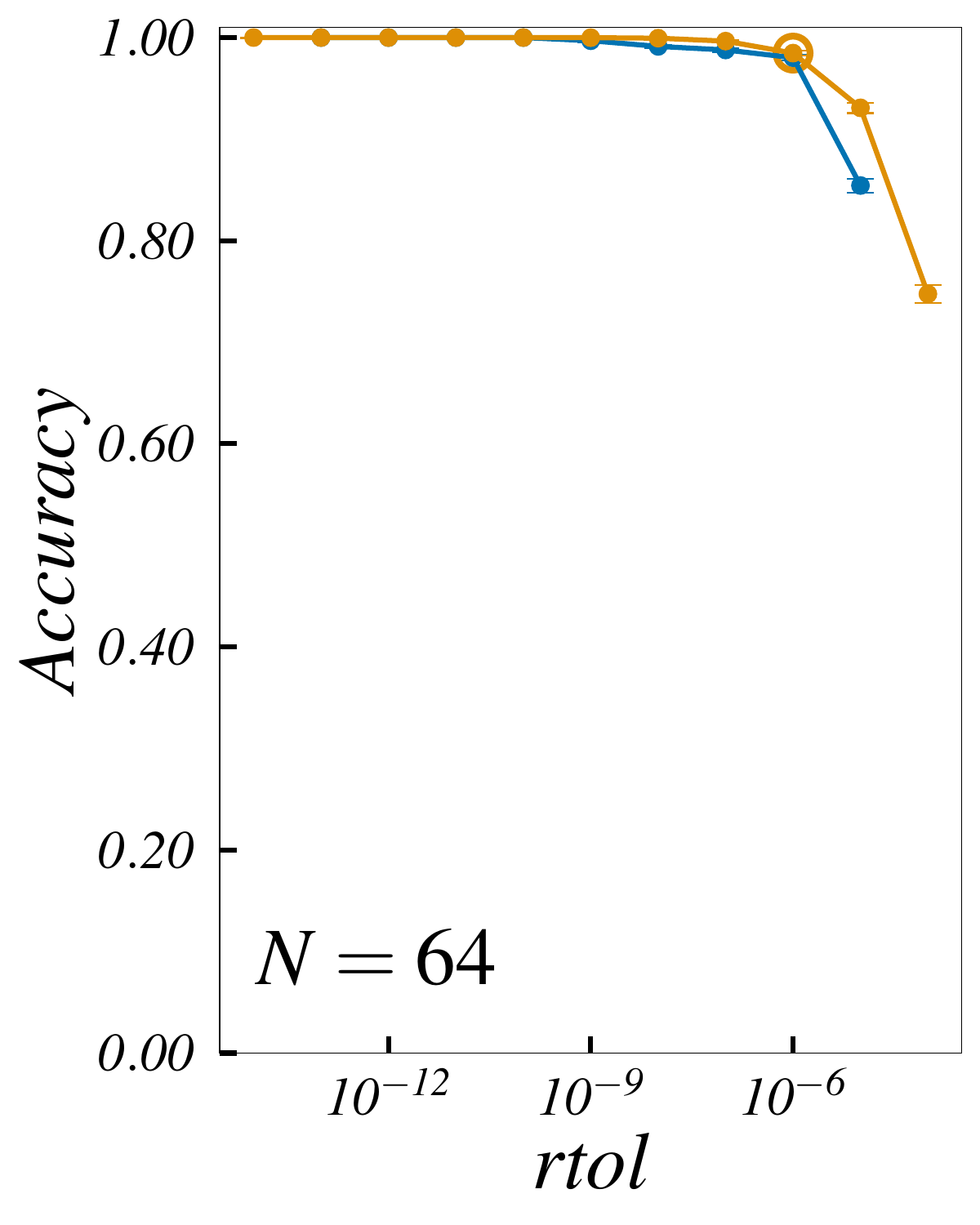}\hfill
    \includegraphics[width=0.23\textwidth]{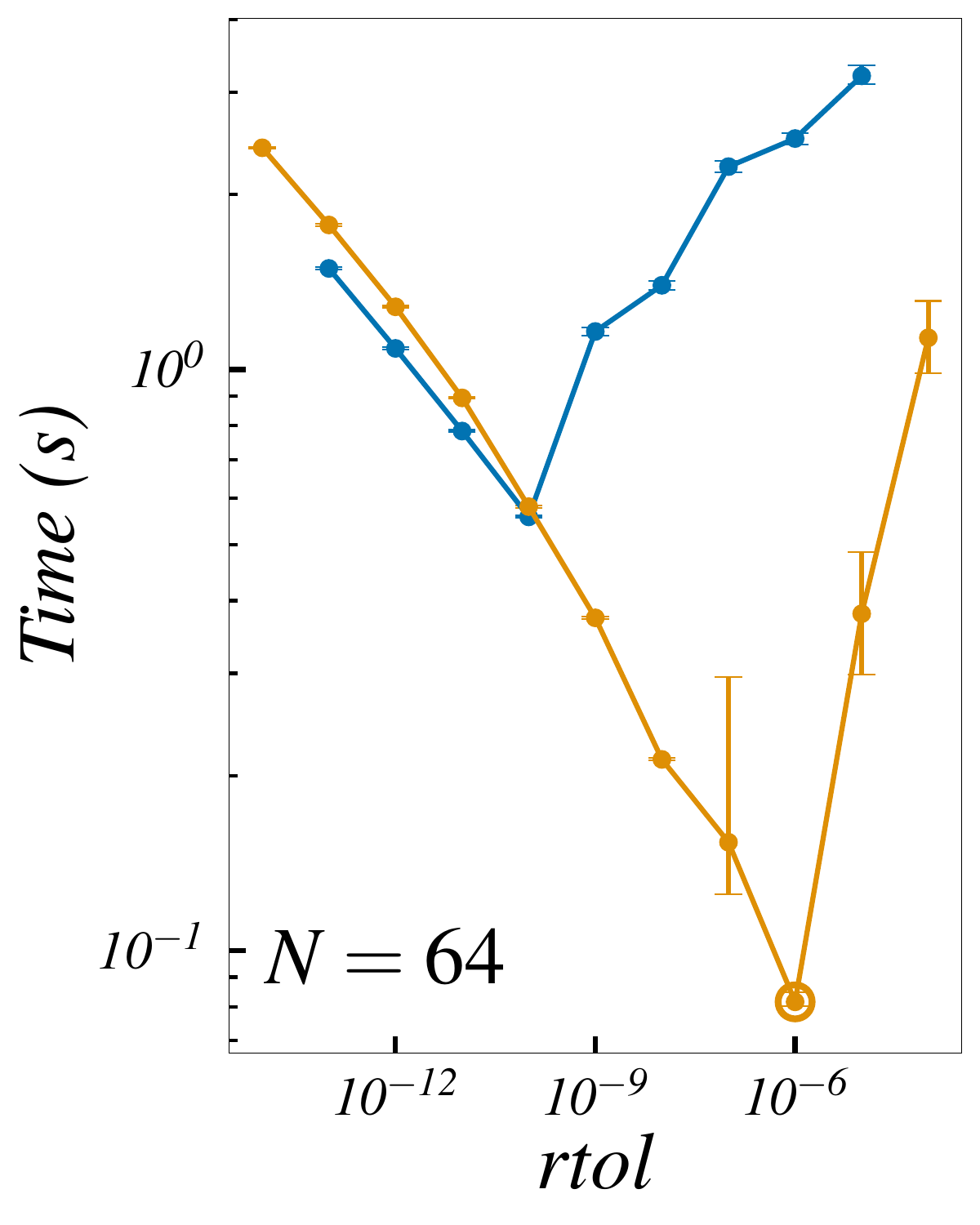}
    
    \vspace{1em}
    % Row 3: N = 128, 128, 256, 256
    \includegraphics[width=0.23\textwidth]{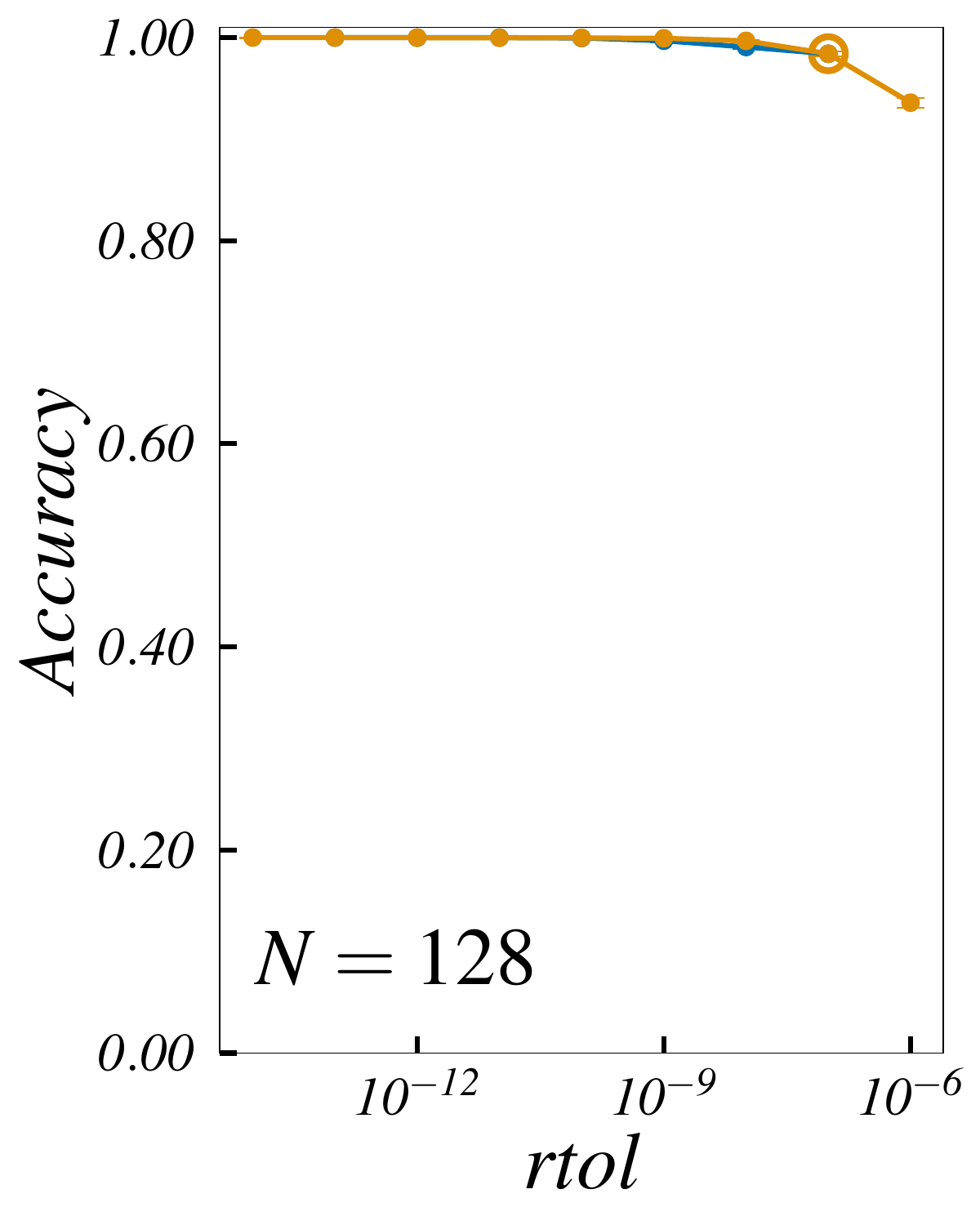}\hfill
    \includegraphics[width=0.23\textwidth]{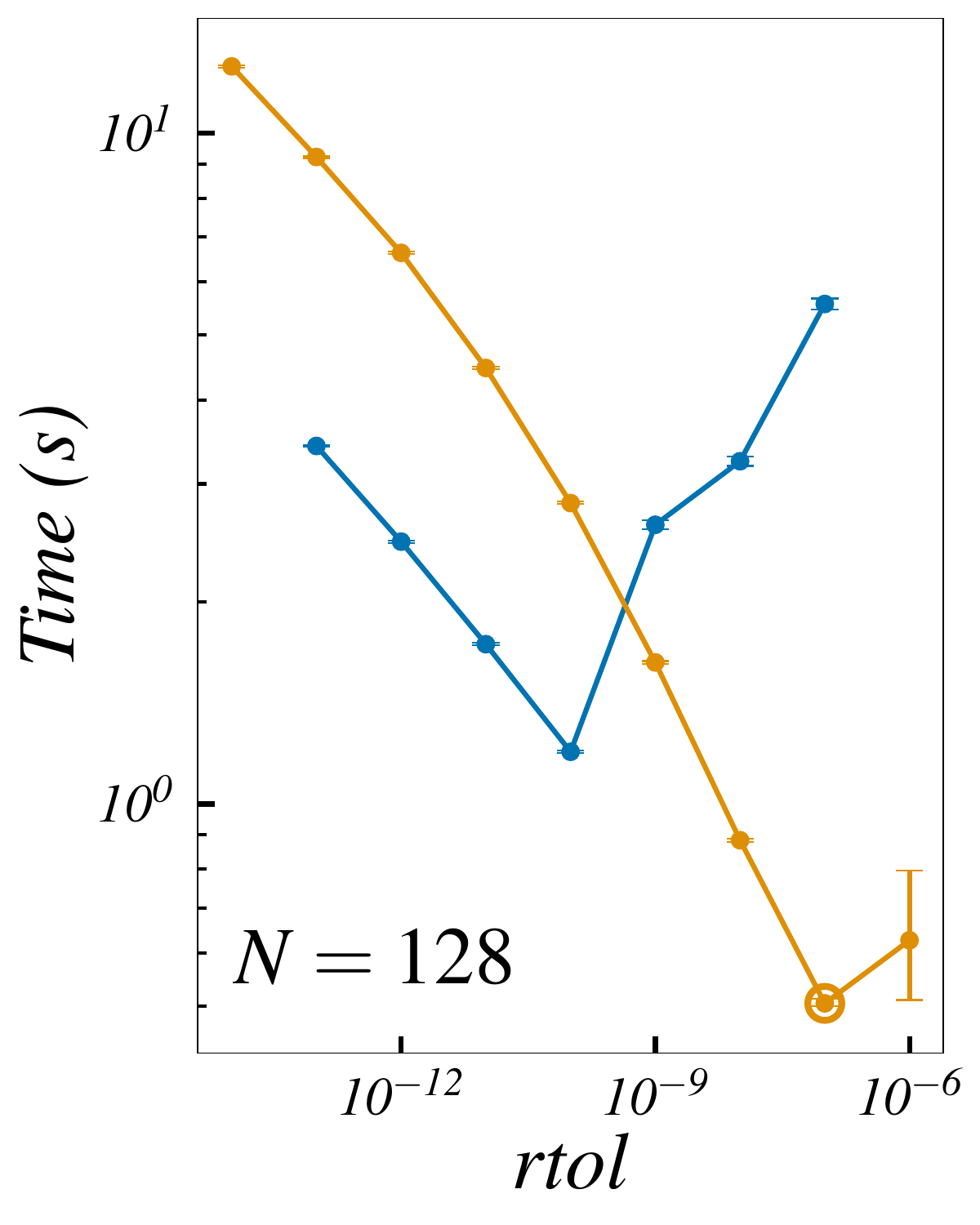}\hfill
    \includegraphics[width=0.23\textwidth]{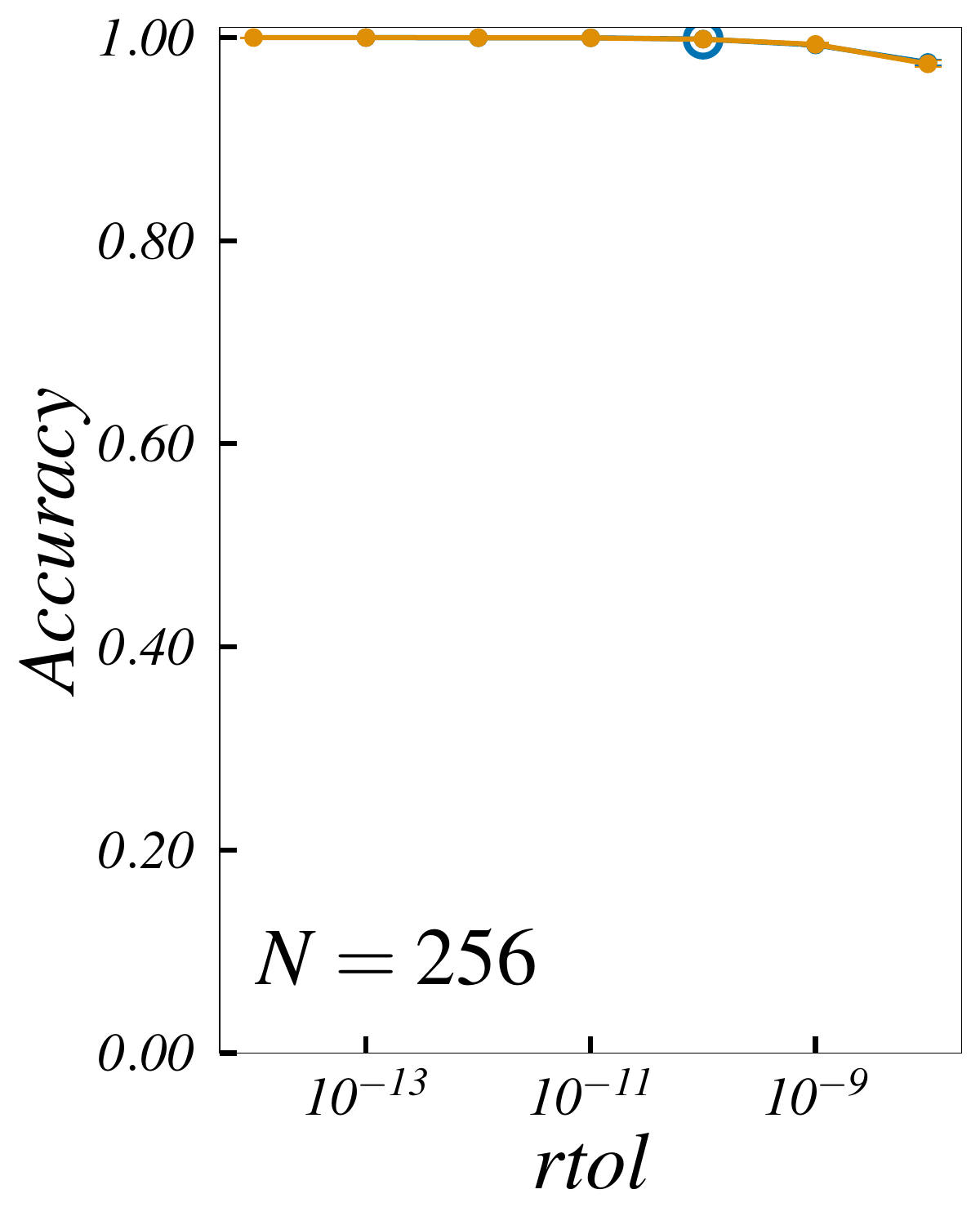}\hfill
    \includegraphics[width=0.23\textwidth]{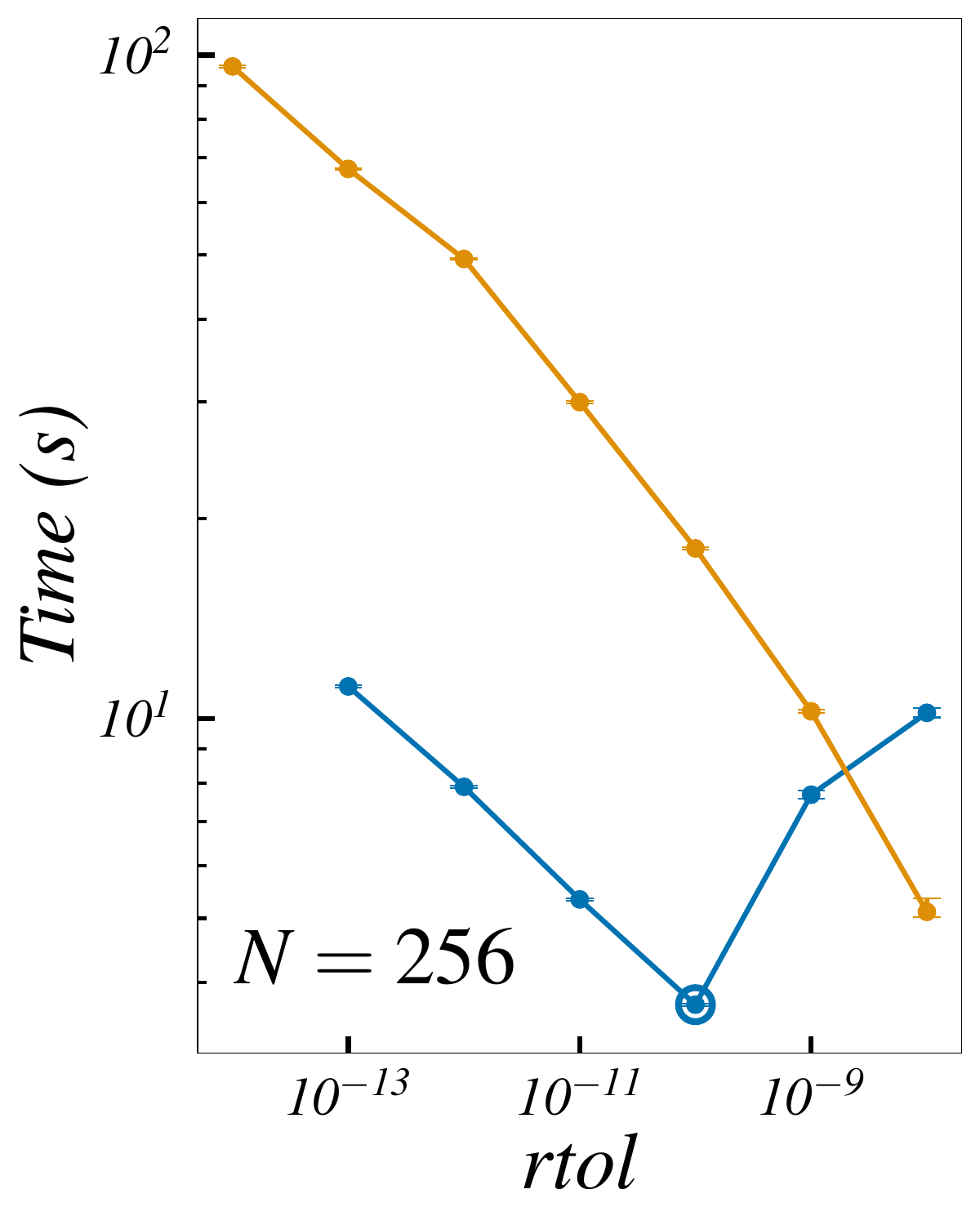}
    
    \vspace{1em}
    % Row 4: N = 512, 512, 1024, 1024
    \includegraphics[width=0.23\textwidth]{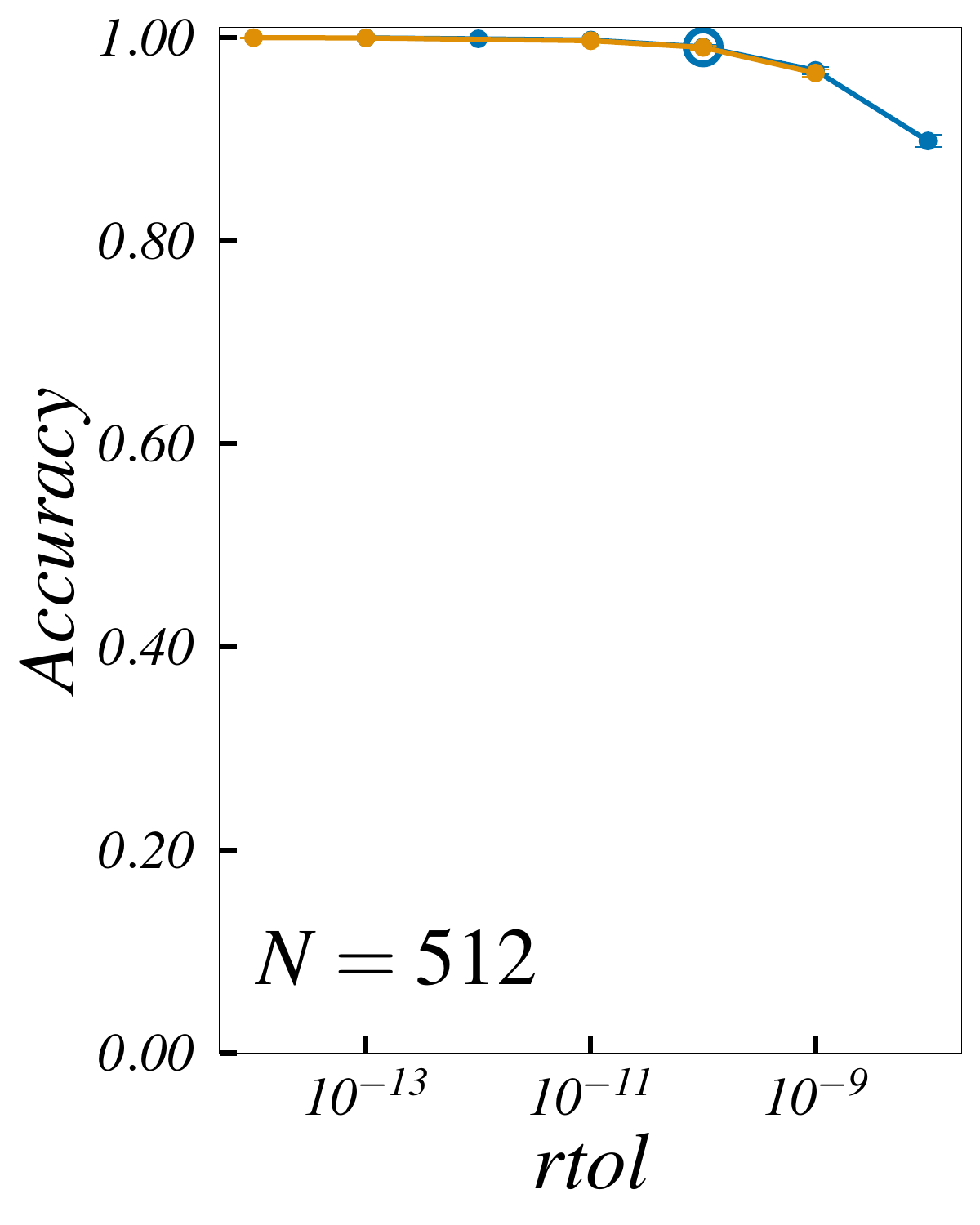}\hfill
    \includegraphics[width=0.23\textwidth]{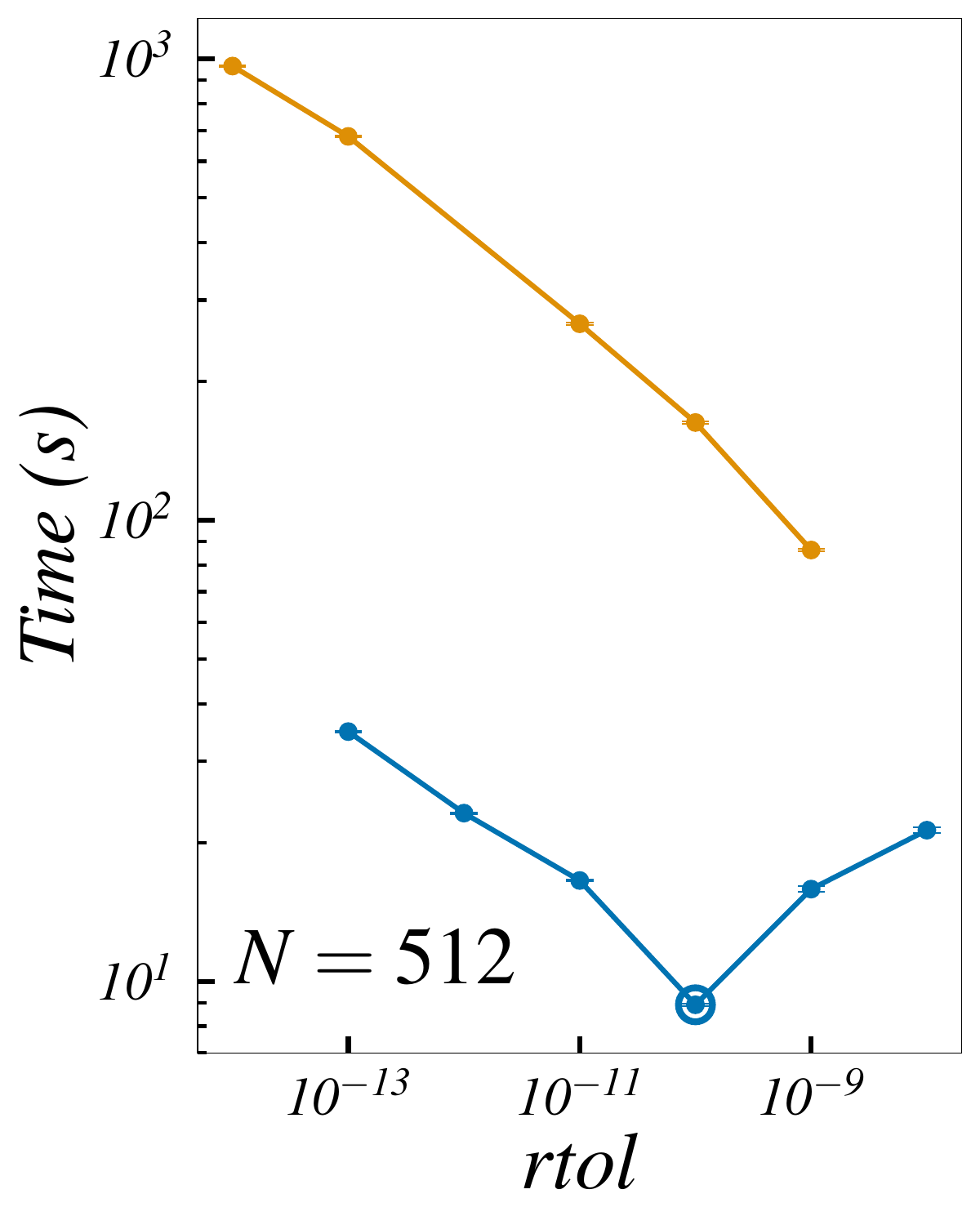}\hfill
    \includegraphics[width=0.23\textwidth]{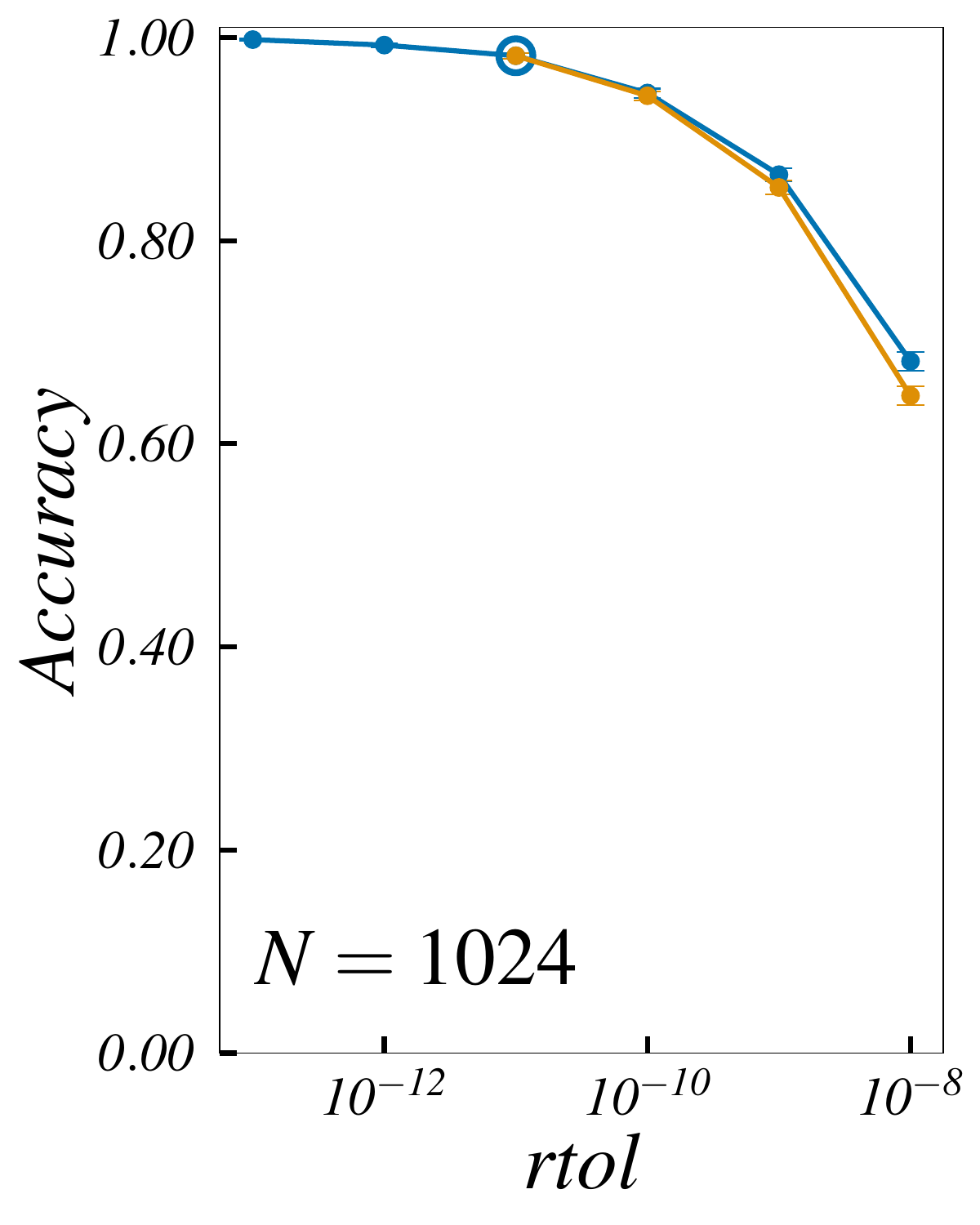}\hfill
    \includegraphics[width=0.23\textwidth]{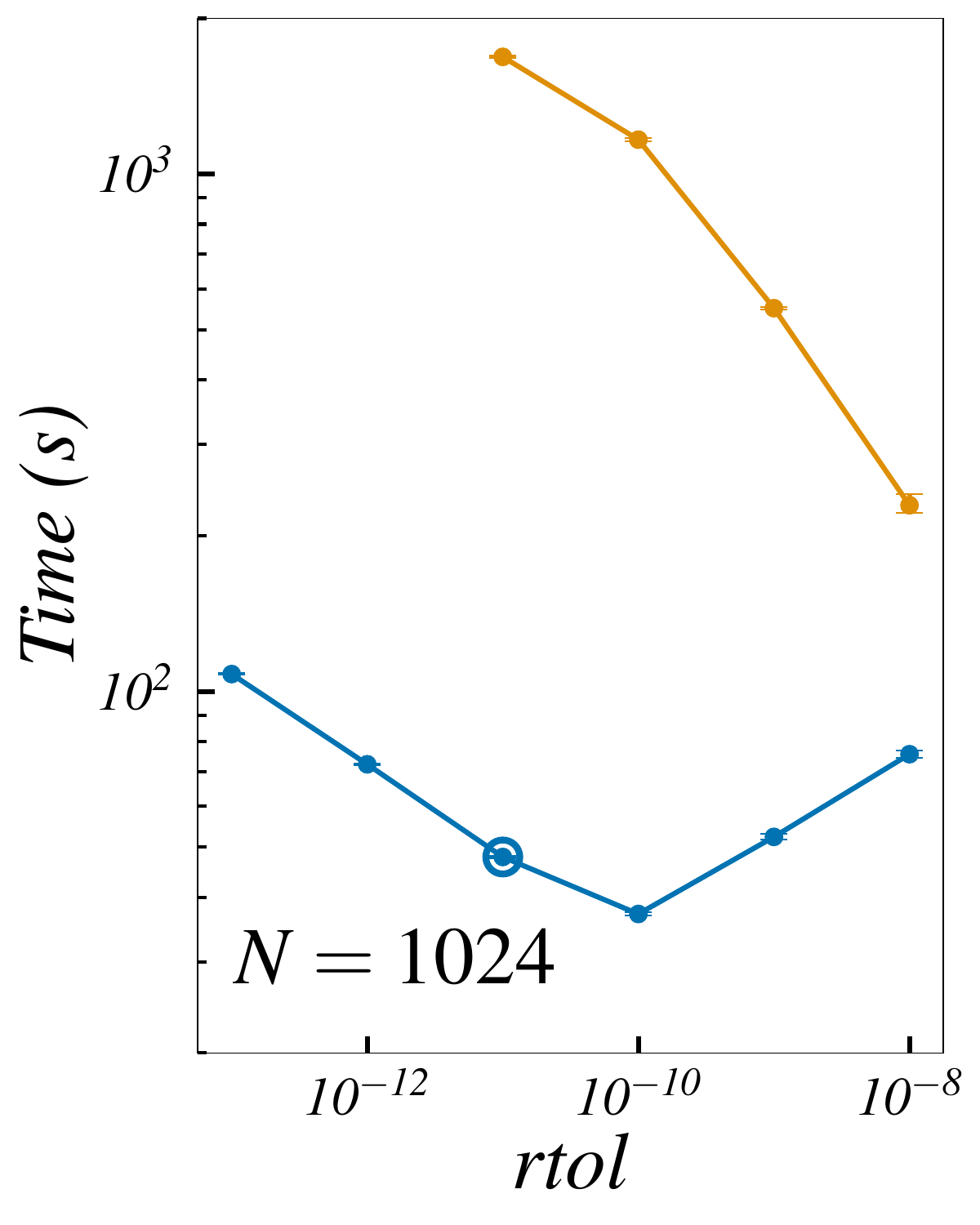}
    \caption{\textbf{CVODE Accuracy and Time vs Tolerance.}
    For each $N$, we vary $\rtol$ and report accuracy and time, averaged across $10^4$ random initial conditions.
    Error bars represent Clopper-Pearson confidence intervals for accuracy and standard error for time.
    Highlighted points correspond to points used in the main text, corresponding to $>98\%$ accuracy.
    The point at the lowest rtol is taken to have accuracy $1.0$}
    \label{fig:rtol_all}
  \end{figure*}

Using this reference, we assess performance at larger tolerances by running CVODE from approximately $10000$ random initial points at each $\rtol$ value.
We define an accuracy for each $\rtol$ as the fraction of initial points that converge to the same minimum as when $\rtol = 10^{-14}$.
Results are shown in Fig.~\ref{fig:rtol_all}, where we plot accuracies and computation times against $\rtol$ at all system sizes for which CVODE is considered in the paper.
We also report performance based on whether we switch on an iterative Newton-Krylov scheme~\cite{Saad1986} provided within CVODE~\cite{Hindmarsh2005,Gardner2022} to solve an implicit time-stepping equation instead of using the dense Hessian.
For each $N$, we highlight the $\rtol$ and choice of scheme (iterative or dense) corresponding to the fastest time with $>98\%$ accuracy at identifying basins accurately.
This \texttt{rtol} is the value we use in practice.

We summarize our choices of \texttt{rtol} and scheme for each $N$ in Table~\ref{tab:cvode-params}, where we also give values for accuracies larger than $95\%$.
Note that the optimal time to attain these accuracies involves switching on the iterative scheme at sizes $N>64$, which leads to a slight change of trend of time scalings at the junction $N = 64$.

\begin{table}[htbp]
  \centering
  \begin{tabular}{|c|c|c|c|} \hline System Size & $\rtol \, (\text{acc} \gtrapprox 95)$ & $\rtol \, (\text{acc} > 98)$ & Iterative \\ 
    \hline 8 & $10^{-4}$ & $10^{-7}$ & False \\
    16 & $10^{-5}$ & $10^{-7}$ & False \\
    32 & $10^{-5}$ & $10^{-7}$ & False \\
    64 & $10^{-6}$ & $10^{-7}$ & False \\
    128 & $10^{-7}$ & $10^{-8}$ & True \\
    256 & $10^{-8}$ & $10^{-9}$ & True \\
    512 & $10^{-9}$ & $10^{-10}$ & True \\ 1024 & $10^{-10}$ & $10^{-11}$ & True
    \\ \hline \end{tabular}
  \caption{\textbf{Tolerances for CVODE.}
    Table of CVODE tolerances we choose.
    For small systems, while a higher $\texttt{rtol}$ is optimal for performance, the resulting numerical noise can prevent the gradient from satisfying the $10^{-10}$ convergence threshold close to a minimum.
    Here iterative means that an iterative Newton-Krylov scheme is used to solve the Newton's equation instead of using the dense Hessian.}
  \label{tab:cvode-params}
\end{table}

Also note that, while the matrix structure for the Hessian is sparse, we are not aware of already available methods that may take advantage of this property, because the sparsity structure constantly changes as particles start or stop interacting through the minimization process.
Finally, from the values in Tab.~\ref{tab:cvode-params}, it is interesting to notice that \texttt{rtol} seemingly scales with $N$ with a power close to $N^{-3}$, that is reminiscent of the timestep scaling with $N$ reported in Ref.~\cite{Bautista2025} -- although we emphasize that \texttt{rtol} is not simply a timestep.

\subsubsection{FIRE, LBFGS, and Gradient Descent parameters}
\label{sec:fire-lbfgs-param}

We employ the FIRE and L-BFGS algorithms as implemented in the pele package~\cite{pele2024}, which adhere to the recommendations by Ref.~\cite{Asenjo2013} to generate accurate basins.
To prevent excessively large step sizes, we cap the maximum step at 
\begin{align}
    \Delta_{\max} = \frac{\mu_s}{10}
\end{align}
with $\mu_s$ the mean radius of the smaller particles in our bidisperse distribution of radii.
In the basin volume calculations, however, we use a different bound to be consistent with parameter choices in Ref.~\cite{Martiniani2017a},
\begin{align}
    \Delta_{\max}^{\text{BV}} = \frac{R_{min}}{4}
\end{align}
with $R_{min}$ the smallest radius.
In practice, for our choice of distributions, we typically have $\Delta_{\max}^{\text{BV}} \approx  2  \Delta_{\max}$.

In our FIRE implementation, we incorporate an additional safeguard: the minimizer is prevented from taking uphill steps by halting and resetting momentum, as in Ref.~\cite{Martiniani2017a}.
The list of FIRE parameters for the pele implementation used in the paper are given in Tab.~\ref{tab:fire_params}.
\texttt{finc} determines the factor to be applied to $\mathrm{d}t$ after \texttt{Nmin} steps, $\mathrm{d}t' = \texttt{finc} \, \mathrm{d}t$, if the velocity is downhill (i.e pointing in the direction of the gradient).
If we encounter an uphill step, then we reduce $\mathrm{d}t$ by multiplying it by \texttt{fdec}, $\mathrm{d}t' = \texttt{fdec} \, \mathrm{d}t$.
The maximum timestep,\texttt{dtmax}, is set to $1$.

\begin{table}[htbp]
  \centering
  \begin{tabular}{|c|c|}
    \hline
    \textbf{Parameter} & \textbf{Value} \\
    \hline
    \texttt{dtstart} & 0.1 \\
    \texttt{dtmax}   & 1 \\
    \texttt{maxstep} & 0.5 \\
    \texttt{Nmin}    & 5 \\
    \texttt{finc}    & 1.1 \\
    \texttt{fdec}    & 0.5 \\
    \texttt{fa}      & 0.99 \\
    \texttt{astart}  & 0.1 \\
    \texttt{stepback} & True \\
    \texttt{tol} & $10^{-10}$ \\
    \hline
  \end{tabular}
  \caption{FIRE algorithm parameters}
  \label{tab:fire_params}
\end{table}

Our L-BFGS parameters are listed in Tab.~\ref{tab:LBFGSparameters}.
Here, \texttt{M} controls how well the algorithm approximates the Hessian, \texttt{maxErise} determines how much the energy can rise due to the method, (in this case $0$), \texttt{H0} determines the initial approximate inverse Hessian.
We use the default parameter $\texttt{M}=1$ as recommended by Ref.~\cite{Asenjo2013}.
\begin{table}[htbp]
  \centering
  \begin{tabular}{|c|c|}
    \hline
    \textbf{Parameter} & \textbf{Value} \\
    \hline
    \texttt{tol} & $10^{-10}$ \\
    \texttt{M} & 1 \\
    \texttt{maxstep} & 0.1 \\
    \texttt{maxErise} & $ 10^{-10}$ \\
    \texttt{H0} & 0.1 \\
    \hline
  \end{tabular}
  \caption{L-BFGS algorithm parameters}
  \label{tab:LBFGSparameters}
\end{table}

Finally, in the variant of Gradient Descent adapted from Ref.~\cite{Charbonneau2023} and described in Sec.~\ref{sec:BasinMappingStrategies}, we use the parameters given in Tab.~\ref{tab:GradientDescentparameters}.
Here $\epsilon$ is a bound on the cosine similarity between successive gradients. $\mathrm{d}t_{\text{initial}}$ is the starting value of the timestep $\mathrm{d}t$.
If $n_{\text{backtrack}}$ successive steps satisfy the cosine similarity condition then we modify the timestep $\mathrm{d}t = n_{\text{backtrack}} \mathrm{d}t$, and if the condition is violated, we reduce the timestep by doing $\mathrm{d}t = \mathrm{d}t/n_{\text{backtrack}}$.

\begin{table}[htbp]
    \centering
    \begin{tabular}{|c|c|}
      \hline
    \textbf{Parameter} & \textbf{Value} \\
    \hline
    $\epsilon$ & $10^{-2}$ \\
   $dt_{\text{initial}}$  & $ 10^{-5}$ \\
      $n_{\text{backtrack}}$ & $5$ \\
      \hline
    \end{tabular}
    \caption{Gradient Descent algorithm parameters}
    \label{tab:GradientDescentparameters}
  \end{table}

\section{Detailed numerical experiments}

In this section, we describe numerical methods used to produce the data discussed in the main text in complete detail.

\subsection{Low-dimensional intersections: $2d$ slices and $1d$ line cuts\label{sec:LinesandSlices}}

To produce $2d$ slices of the energy landscape color-coded by basin, such as Fig.~1 of the main text, we first sample a single configuration $\bm{X}_0$ uniformly in $[0;L]^{Nd}$.
We then randomly sample two orthogonal directions, encoded by two unit vectors $\hat{\bm{n}}_1$ and $\hat{\bm{n}}_2$ such that $\hat{\bm{n}}_1 \cdot \hat{\bm{n}}_2 = 0$ and construct a grid of configurations centered around that random point, \textit{i.e.} a set of points with locations $\bm{X}_{pq} = \bm{X}_0 + pP_x \hat{\bm{n}}_1 + q P_y \hat{\bm{n}}_2$ with $(p,q) \in \mathbb{Z}^2$ such that $p$ spans $N_1$ integers and $q$ spans $N_2$ integers (both with ranges centered on $0$), and $P_x, P_y>0$ are the spacings between two measurement points along the two axes.
The output is translated into a $N_1 \times N_2$ picture, where each pixel is colored by the basin it belongs to.
The list of colors is generated using the Glasbey colormap~\cite{glasbey2007colour} to ensure that the color corresponding to each minimum is perceptually distinct.

In Fig~1 of the main text, we produce pictures of $1920\times 1080$ pixels that correspond to a rectangle in configuration space with side-lengths $12\mu_s\times 6.75\mu_s$ (in units of the mean radius $\mu_s$ of the small disks).
For Fig.~2 of the main text, and all $N=128$ cuts in this document, we produce $800\times 800$ pictures corresponding to squares in configuration space with side-lengths
$0.44\mu_s\times 0.44\mu_s$.
This configuration-space width is chosen so that the $2d$ density of basins on the slices would be fairly similar to that of Fig.~1 of the main text.
In this document, we also produce slices for $N = 8$ (see Fig.~\ref{fig:8_slice}).
They are produced using $800\times 800$ pictures corresponding to squares in configuration space with side-lengths $10\mu_s \times10\mu_s$.

Line intersections are produced in a similar way.
We start by selecting a configuration uniformly, then sample a random direction uniformly, thus defining a line.
On that line, we define a segment symmetrically around the starting configuration, with a total length $L_S \approx 10 \mu_s $ for $N=16$, similar to the sidelengths of one of the $2d$ slices of Fig.~1.
We then divide that segment into $10^6$ regularly spaced points, which we use as initial conditions for optimization, and we tag the basins they fall into.
Having thus constructed a discretized map of basins on the line, we list all pairs of neighboring pixels falling into distinct basins.
At each such pair, we define a new collection of points on the line, with a resolution $10^2$ finer, and linking the 2 neighboring pixels.
We run optimizations at these new points to better resolve the junction between the two basins, possibly uncovering new basins.
Finally, we reconstruct the intersection lengths between the basins and the line using the two resolutions.
Note that, in principle, the zoom being performed selectively at junctions found at the base resolution could introduce biases in the measured lengths.
However, we here use resolutions much finer than in the $2d$ cuts from the main text, as the base resolution before zooming is about $1000$ times finer than that of Fig.~1 of the main text (the zoomed in version is thus $10^5$ times finer than Fig.~1).
We also check that we never see new basins within the bulk of a basin when switching to finer scales.
In other words, empirically, basins are regular enough that we do not expect a measurable effect of zooming everywhere versus zooming only at intersections.
Furthermore, note that for the data shown in Fig.~3 of the main text, the peak in the distribution of log-lengths obtained with CVODE is already visible before the zoom, so that the zoom helps resolve single-pixel intersections but does not artificially introduce a peak into the distribution.

\subsection{Video caption}
The video attached to the article is obtained as a succession of $2d$ slices of the energy landscape similar to the ones shown in Figs. 1 and 2 of the main text, for a system of $N = 16$ Hertzian disks at $\phi = 0.9$.
To obtain motion along a closed trajectory, and thus a looping video, we first define a random $2d$ slice in the usual way, see Sec.~\ref{sec:LinesandSlices}.
That slice is defined by a random point $\bm{X}$ in configuration space, and two orthogonal unit vectors $\hat{\bm{n}}_1$ and $\hat{\bm{n}}_2$.
We then sample a third and fourth random direction $\hat{\bm{e}}_3$ and $\hat{\bm{e}}_4$, and use them to construct the unit vectors $\hat{\bm{n}}_3$ and $\hat{\bm{n}}_4$ such that $\hat{\bm{n}}_i \cdot \hat{\bm{n}}_j = \delta_{ij}$ for $(i,j) \in \{ 1,2,3,4\}^2$.
Using the two vectors perpendicular to the starting slice, we define a circular trajectory in configuration space and generate regularly spaced slices along it, such that the successive slices are always perpendicular to the circle and are spaced by a distance equal to the size of a pixel.
We then match minima across all slices so that any one basin is encoded by the same color across slices.
The resulting collection of slices forms a continuous-looking video that hints at a smooth $3d$ structure of basins.

\subsection{Fractal analysis of slices}
\label{sec:fract-analys-cuts}

In this section, we describe in detail the procedure used to measure bounding-box dimensions of $2d$ slices of basins.
The box-counting dimension is defined as
\begin{equation}
  \label{eq:4}
  d_{B} = \frac{\log(N(\ell_B))}{\log(\ell_B)}.
\end{equation}
where $N(\ell_B)$ is the number of cubic boxes with sidelength $\ell_B$ that are needed to cover a shape.

Starting from a $2d$ slice of configuration space such as Fig.~1 of the main text, we crop the image to the smallest bounding box that encloses the basin boundary, with a padding of $L/16$ to mitigate boundary effects, then convert the picture into a binary map with values $1$ in the basin, $0$ elsewhere.
We then feed the result into the \texttt{porespy} package~\cite{Gostick2019} to count $N(\ell_B)$.
To estimate $d_B$, we perform a linear fit on the $ \log(N(\ell_B)) $ versus $ \log(\ell_B) $ data, excluding the largest scales so as to mitigate finite-size effects due to the finite slice size.
We thus measure a fractal dimension for the selected basin: a smooth (non-fractal) basin would have dimension $d_B = 1$, a fractal one $d_B > 1$.
To characterize the landscape, in practice, we repeat this measurement for the largest (in number of pixels) $n_B$ basins in a slice.
This choice is similar to the one in Ref.~\cite{Folena2025}, where the authors also show in the example of the Random Lorentz Gas that the largest basins are the most likely to be fractal.

The slices used for $\phi = 0.9$ are those of Fig.~2 of the main text, while the ones used for $\phi = 0.86$ are shown in Fig.~\ref{fig:AccuracyTime0p86}.

\subsection{Distance between kicked minima}
\label{sec:minima-distribution}

In Fig.~4$(d)$ of the main text, we plot the normalized metric distance, as defined in Ref.~\cite{Dennis2020}, as a function of kick size for a system of 1024 particles.
To obtain this data, we begin by selecting 13 random initial configurations. For each configuration, we use CVODE with $\texttt{rtol}=10^{-11}$ and $\texttt{atol}=10^{-12}$ (one order of magnitude below the values in Table~\ref{tab:cvode-params}) to locate the corresponding minimum within its basin of attraction.
We only use jammed minima by discarding any fluid states and resampling. 

For each of these 13 minima, we then apply 1000 random kicks of length $R$, choosing the direction uniformly in space, we do this for 60 $R$ values.
Of these $60$ values, 10 are sampled uniformly in log space from $0.1$ to $1$ and $50$ uniformly from $1$ to $R=L$.
In the main text, we only present the data up to $R=L/2$, confining ourselves to observe differences in intermediate kick size regimes.
After each kick, we minimize the system using CVODE, FIRE, and L-BFGS, employing the parameters listed in Tables~\ref{tab:cvode-params}, \ref{tab:fire_params}, and \ref{tab:LBFGSparameters}, respectively.
Finally, excluding any case in which the new minimum coincides with the original one, we compute the normalized metric distance between the new minimum $M'$ and the original one $M$,
\begin{align}
    d_{min}(M, M') \equiv \frac{1}{2 \langle R_i \rangle} \sqrt{\sum\limits_{i < j} \left( \bm{C}_{ij} - \bm{C}_{ij}' \right)^2}
\end{align}
with $\langle R_i\rangle$ the average particle radius and $\bm{C}_{ij}$, $\bm{C}_{ij}'$ are the ``stable contact vectors'' between particles $i$ and $j$ in minima $M$ and $M'$, respectively.
These contact vectors are defined as 
\begin{align}
    \bm{C}_{ij} = \bm{r}_{ij} \mathbb{1}(r_{ij} \leq R_i + R_j),
\end{align}
that is, the distance vector if the particles are in contact, and $0$ otherwise.
This definition ensures that the measured distance is unambiguous in spite of the presence of rattlers and of translational symmetry.

\subsection{Basin volumes}

To compute volumes of basins of attraction, we use the Markov Chain Monte Carlo method with umbrella sampling described in Refs.~\cite{Martiniani2016,Martiniani2016a,Martiniani2017a,Casiulis2023}, setting the ``oracle'' (the function that determines whether a point belongs to the basin being measured) to be either FIRE, L-BFGS, or CVODE with the parameters described in Sec.~\ref{sec:Params}.
For each method, we compute the volume of 5 basins (the same ones across methods) at $\phi = 0.9$, across sizes ranging from $N = 16$ to $N=128$.
The choice $N=128$ as the largest system size is due to the large computational cost of this method, as running the full procedure takes weeks for that system size on a full CPU node.

We here provide a list of the parameters we use for the various steps of the basin volume procedure.
The minima used for basin volumes are found by sampling the size polydispersity (here bidisperse with half the radii positive-normal with mean and standard deviation $\mu_s = 1,\sigma_s = 0.05$ and the other half positive-normal with $\mu_\ell = 1.4$ and $\sigma_\ell = 0.07$), then sampling the locations of particle centers uniformly and independently in the simulation box, and identifying the corresponding basin it belongs to via a careful CVODE minimization ($10^7$ optimization steps).

From that point, the algorithm is used with a choice of oracle that is kept constant throughout (either FIRE, L-BFGS, or CVODE).
The recorded minimum is re-quenched with that method to avoid possible small shifts of the minimum's location due to, \textit{e.g.} precision differences.
When calling the oracle, we use as many as $10^5$ optimization steps, and the method and tolerance value used for identification of the minimum are the ones described in Sec.~\ref{sec:minima-match-proc}, with a tolerance $\texttt{dtol} = 10^{-4}$.
First, the largest spring constant $k_{max}$ used in the umbrella sampling is found automatically as described in past work, by running a random walk biased by a spring attached to the minimum with an adjustable rigidity.
The rigidity is adjusted, starting from $k_0 = 5/\mu_s^2$, until it reaches a value $k_{max}$ such that $90\%$ of samples lie in the basin.
That proportion is evaluated using $10^4$ samples per value of $k$.
Then, we perform $10^5$ steps of a free random walk within the basin ($k = 0$) so as to adjust the time step (to accept $20\%$ of proposed MCMC steps on average) and measure an estimate of the extent of the basin, via the mean-square distance to the minimum of that walk.
Parallel tempering is then performed using $64$ replicas ($40$ positive ones, the $k = 0$ one, and the rest negative).
The positive values of the stiffness constants are chosen such that the mode of their distributions of distances to the center are roughly linearly spaced between $0$ and $k_{max}$, as described in Ref.~\cite{Casiulis2023}, which ensures equivalent overlaps between neighboring histograms and good scaling of the required number of replicas with $N$.
The negative replicas have linearly spaced $k$-values between $k = 0$ and $k = -0.5$, an arbitrary choice that empirically yields good behavior (meaning that the distance distribution of the most negative replica is not reduced to a point-like distribution and that histograms overlap sufficiently well).
Replica exchanges are propose every $10^2$ steps.
The Monte Carlo algorithm is run for a number of steps ranging from $5\times 10^5$ to $2\times 10^6$ steps, with a termination criterion based on an equilibration test~\cite{Martiniani2017}.
To have a reference volume, we perform direct sampling of $10^5$ independent points from a ``radial Gaussian"~\cite{Casiulis2023} around the minimum, \textit{i.e.} a distribution whose pdf of radial distances to the minimum $p(r)$ is a positive-Gaussian with a maximum at distance $0$, corresponding to a high-dimensional pdf $p(\bm{r}) \propto r^{1-d} N(0,R_{in})$ for distance vectors to the minimum, where $N(\mu, \sigma)$ is the pdf of a normal distribution with mean $\mu$ and standard deviation $\sigma$.
We choose the standard deviation $R_{in}$ to match the mean root-square distance to the minimum in the $k_{max}$ run.
Each point thus sampled is kept only if it does fall into the basin, and we record the acceptance rate of the points to correct the integral of the reference volume.
Finally, we reconstruct the volume of the basin using the pymbar implementation of MBAR~\cite{Shirts2008}.
As described in detail in Ref.~\cite{Casiulis2023}, the method actually outputs the relative dimensionless ``free energies'' (negative log-volumes) $F(k_i)$ of each replica run, and the value of interest is obtained as the difference $\Delta F$ between the reconstructed free energy of a free walker $F_0 \equiv F(k=0)$ and the known volume of the radial Gaussian, corrected by the acceptance rate of the direct sampling, in units such that the simulation box volume is $1$ (in other words, the reported volume is a fraction of the volume of configuration space).
Results are reported as intensive free energies $F_0 / N = - \log (v/\mathcal{V})^{1/N}$, with $v$ the volume of the basin and $\mathcal{V}$ the total volume of configuration space, that let us compare values across system sizes without the trivial scaling with $N$.

\subsection{Monte Carlo Simulations}
\label{sec:MonteCarlo}

In the main text, we present results obtained using samples equilibrated at a temperature $T$ as initial conditions for minimization.
Here, we provide additional information on the underlying Monte Carlo simulations used for equilibration.
All simulations are performed with the mcpele package~\cite{mcpele2024}, and rely on the Metropolis acceptance rule~\cite{Frenkel2001}: a trial move with energy change $\Delta E$ is always accepted if $\Delta E \leq 0$, and accepted with probability $e^{-\Delta E / T}$ otherwise.
Throughout this section, temperatures are expressed in units of $\varepsilon$, setting $k_B = 1$.

For the Hertzian system at $\phi = 0.9$, single-particle moves alone cannot equilibrate the system at low temperatures, as it is a dense glass-former.
We therefore complement them with swap moves, following the swap Monte Carlo strategy designed for polydisperse glass-formers~\cite{Berthier2016a, Ninarello2017}.
At each step, with probability $0.8$, we attempt a single-particle move, displacing one randomly chosen disk by a vector with components drawn uniformly in $[-\delta; \delta]$, where $\delta$ is adapted during equilibration to reach an acceptance rate in $[0.3; 0.5]$.
With probability $0.2$, we instead attempt to exchange the positions of two randomly chosen particles whose radii differ by less than a threshold, itself adapted to reach an acceptance rate in $[0.2; 0.5]$.
Keeping the move probabilities fixed ensures detailed balance, and the trial energy is fully recomputed at each step.
Each temperature is sampled by an independent run of $N = 512$ disks, started from an inherent structure obtained by an L-BFGS quench of a random initial configuration.
We simulate $16$ temperatures, roughly logarithmically spaced over $T \in [10^{-5}; 1]$.
For each run, we discard $5 \times 10^3$ sweeps ($1$ sweep corresponding to $N$ attempted moves) for adaptation and equilibration, then record one configuration every $10$ sweeps until $10^4$ configurations are collected.
We check the recorded energy time series for stationarity: at all but the lowest temperatures, the production run fluctuates around a steady thermal value, while for $T \lesssim 10^{-4}$ (of the order of the typical inherent-structure energy per particle) the system remains confined near its starting inherent structure, as expected deep in the glass phase.

For the Lennard-Jones system, we simulate $N = 128$ monodisperse particles with the full, untruncated potential, enclosed in a spherical container of radius $R = 20 \sigma$ (trial moves placing a particle outside the container are rejected), which prevents evaporation at the highest temperatures.
Since a single chain at low temperature remains trapped in one basin, we use replica-exchange (parallel tempering) Monte Carlo~\cite{Hukushima1996} with $30$ replicas at temperatures geometrically spaced between $T = 0.3$ and $T = 1$, all started from a low-lying reference minimum selected among $20$ independent L-BFGS quenches of random configurations.
Each replica performs single-particle displacement moves, with a step size adapted during the first $5 \times 10^4$ steps, and configuration exchanges between adjacent temperatures are attempted every $10^2$ steps, accepted with probability $\min \left( 1, e^{\Delta E \Delta \beta} \right)$, where $\Delta E$ and $\Delta \beta$ are the energy and inverse-temperature differences within the pair.
Production runs last $10^6$ exchange periods, after a warm-up of $5 \times 10^2$ exchange periods.

The accuracy-against-$N$ curve of Fig.~3$(a)$ of the main text is obtained from a similar protocol: for each $N \in [8; 256]$, we sample $10^4$ configurations equilibrated at $T = 1$ using single-particle Metropolis moves, this time enclosing the particles in a spherical container of radius $R = 2 N^{1/3} \sigma$, which keeps the overall density fixed across sizes.
The $10^4$ configurations per $N$ are collected from $10$ independent chains, each started from the lowest of $8$ L-BFGS quenches of random configurations, equilibrated for $2 \times 10^4$ sweeps, then recorded every $10^2$ sweeps.
Equilibrated samples are used instead of fully random points (the infinite-temperature limit used for Hertzian disks) because uniformly drawn configurations generically contain particle pairs at vanishing separations: the energy scales generated by the diverging Lennard-Jones core then cause floating-point precision issues that prevent the identification of the true steepest-descent trajectory.

The configurations sampled at each temperature are then used as the initial conditions for the accuracy measurements of the main text: each configuration is minimized with CVODE, L-BFGS, and FIRE, and we report the fraction of configurations for which each optimizer reaches the same minimum as CVODE.
The CVODE reference is run at integrator tolerances $\rtol = \texttt{atol} = 10^{-10}$ with up to $10^6$ steps, and all three methods use a gradient convergence tolerance of $10^{-10}$.
L-BFGS and FIRE use the parameters of Sec.~\ref{sec:Params}.
Minima of the Hertzian system are matched using the procedure of Sec.~\ref{sec:minima-match-proc}, restricting the statistics to jammed configurations, with a threshold $\texttt{dtol} = 10^{-3}$.
We note that the identification tolerance must be chosen jointly with the gradient convergence tolerance: the residual distance between a converged configuration and its minimum must remain small compared to $\texttt{dtol}$ for the matching to be meaningful.
For the Lennard-Jones system, two Lennard-Jones minima are declared identical if and only if the two configurations can be brought into coincidence by a global translation, rotation, reflection, and particle permutation, with a maximal residual particle mismatch smaller than $10^{-3}$.

\section{Additional data}

In this Section, we provide additional data that complements the results of the main text, but is not strictly necessary to arrive at the conclusions we draw.

\subsection{Survival probability as a function of packing fraction}

\begin{figure}[htbp]
  \centering
  \includegraphics[width=0.32\linewidth]{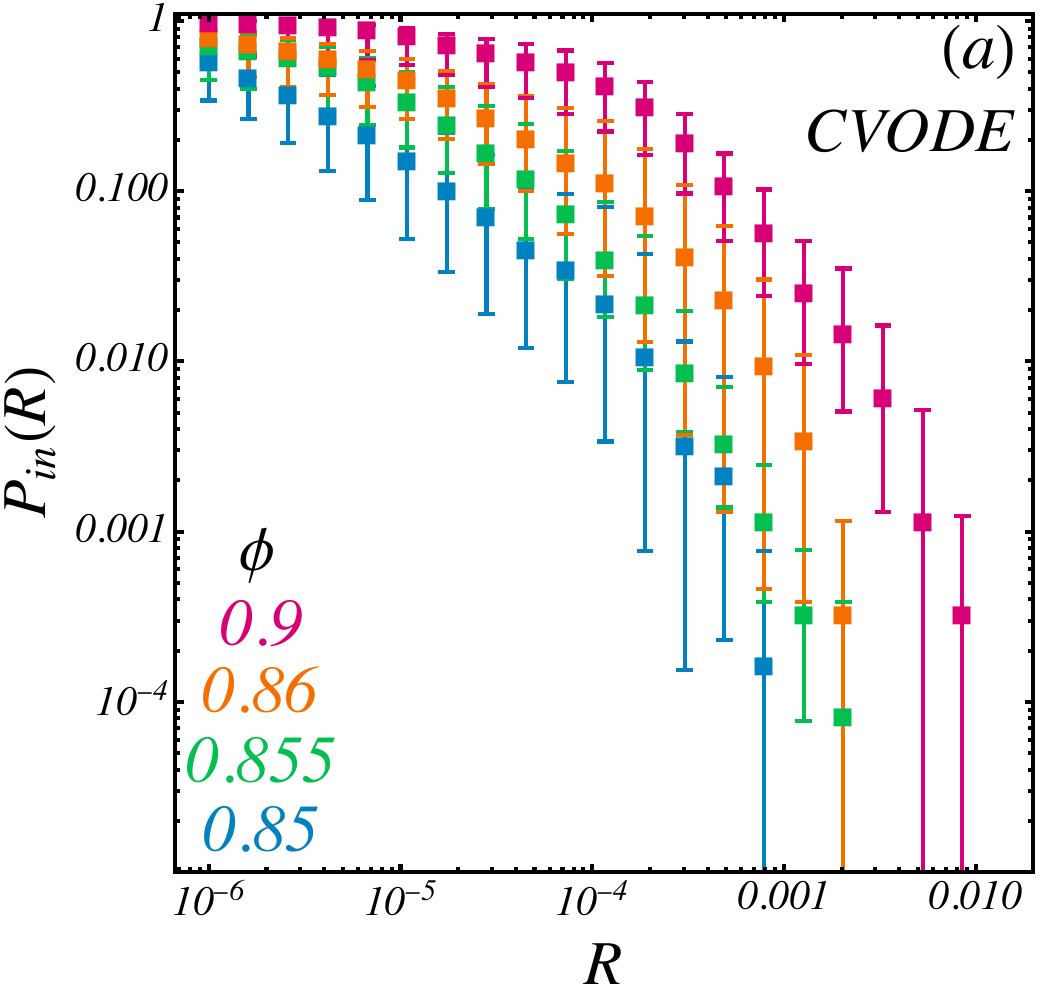}
  \includegraphics[width=0.32\linewidth]{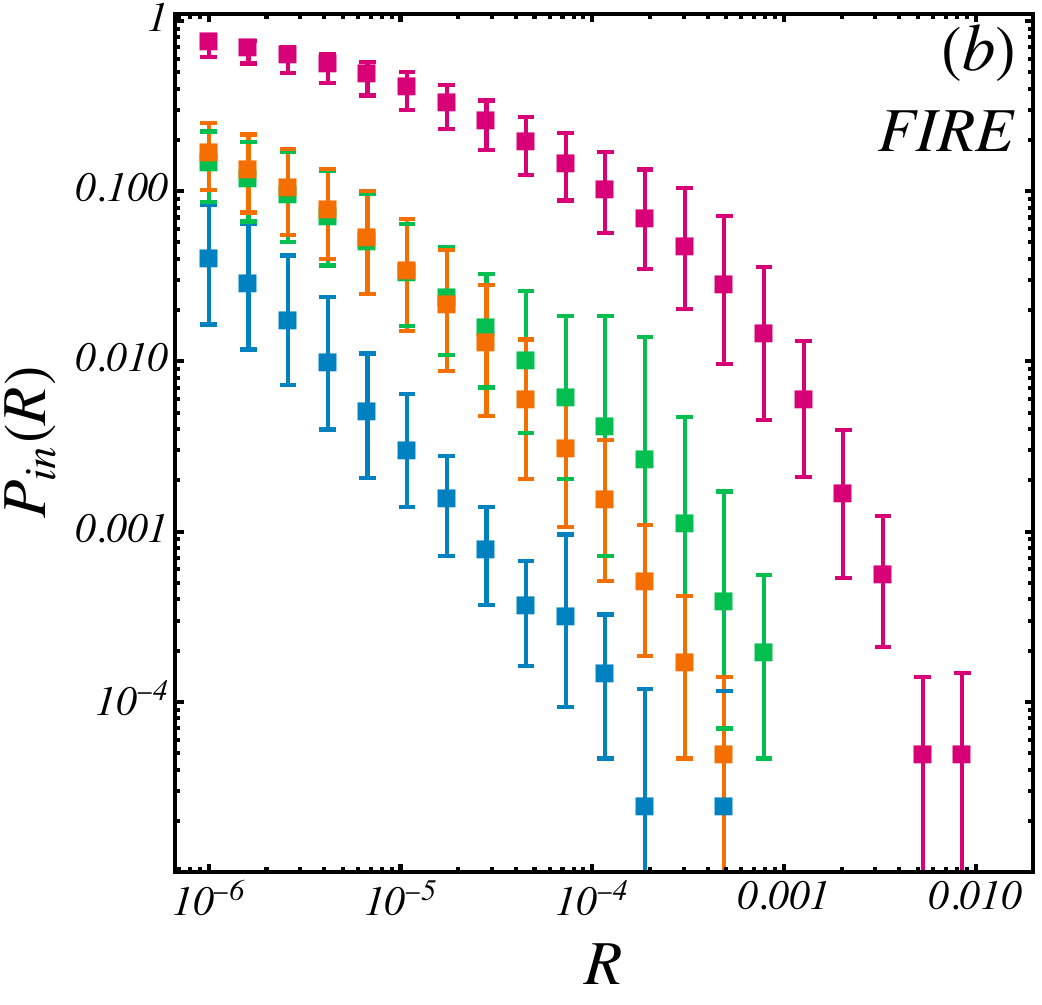}
  \includegraphics[width=0.32\linewidth]{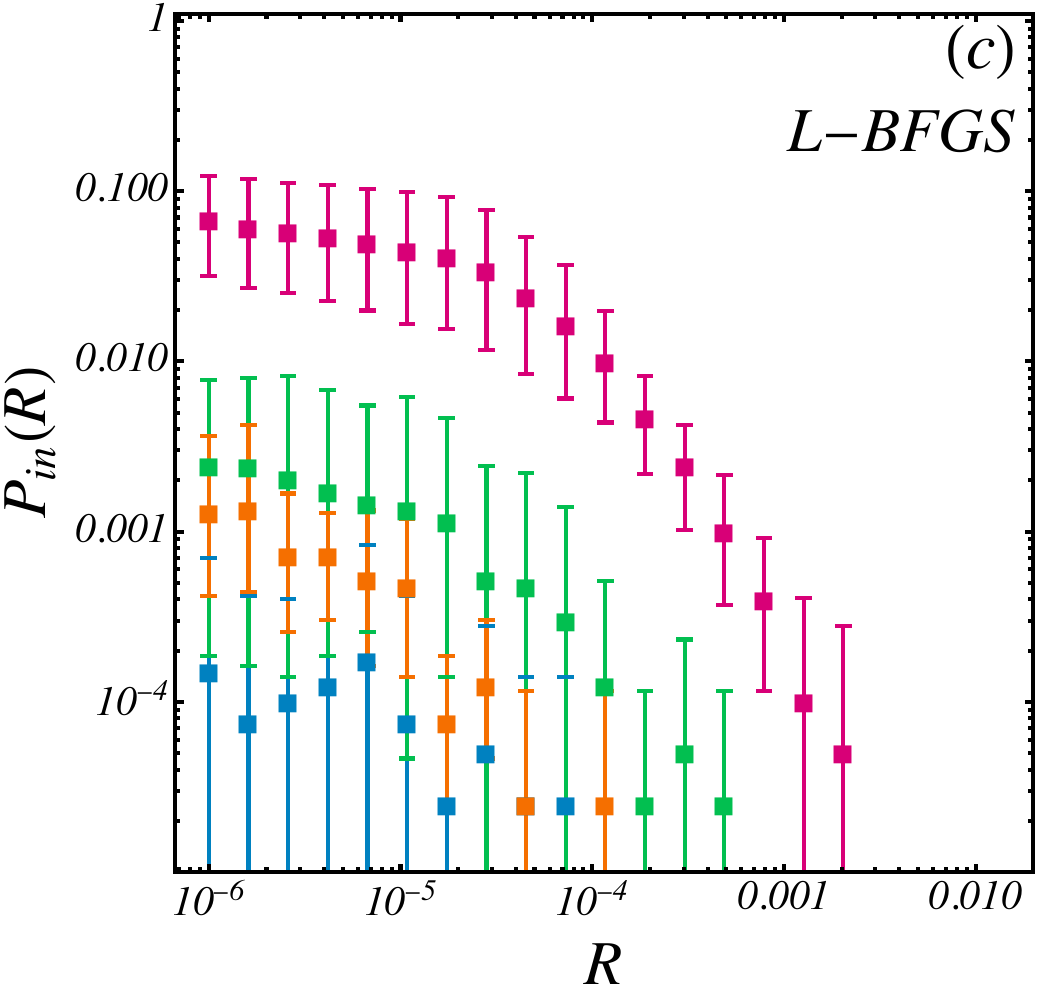}
  \caption{\textbf{Survival: towards jamming.}
  Survival probability against kicksize for a few values of density, in log-log scales, getting closer to jamming, with $(a)$ CVODE, $(b)$ FIRE, and $(c)$ L-BFGS.
  Error bars are Clopper-Pearson $95\%$ confidence intervals.}
  \label{fig:survival_vs_phi}
\end{figure}

In Fig.~4 of the main text, we report survival curves obtained across methods and at a couple of packing fractions for $N = 1024$ Hertzian disks.
In this section, we report similar measurements across a wider range of densities.
To obtain them, we use 13 random initial configurations for CVODE and 43 for FIRE and L‑BFGS.
We require more configurations for FIRE and L‑BFGS because their survival probabilities are significantly lower and their variance is higher at the same error level. For each configuration, we provide 1000 uniform random kicks of length $R$ for 30 different $R$, and measure the survival probability $P_{in}(R)$ as the fraction of points falling back to the same minimum.
The results are shown in Fig.~\ref{fig:survival_vs_phi} for $(a)$ CVODE, $(b)$ FIRE, and $(c)$ L-BFGS.
We show that, as jamming is approached, the difference between methods becomes starker, as optimizers have a lower accuracy when $\phi$ approaches $\phi_J \approx 0.842$.
In particular, optimizers approach power laws (straight lines in log-log) at densities as high as $0.85$, for which CVODE retains a stretched-exponential-looking behavior with a clear characteristic lengthscale.

\begin{figure}[htbp]
  \centering
  \includegraphics[width=1.0\linewidth]{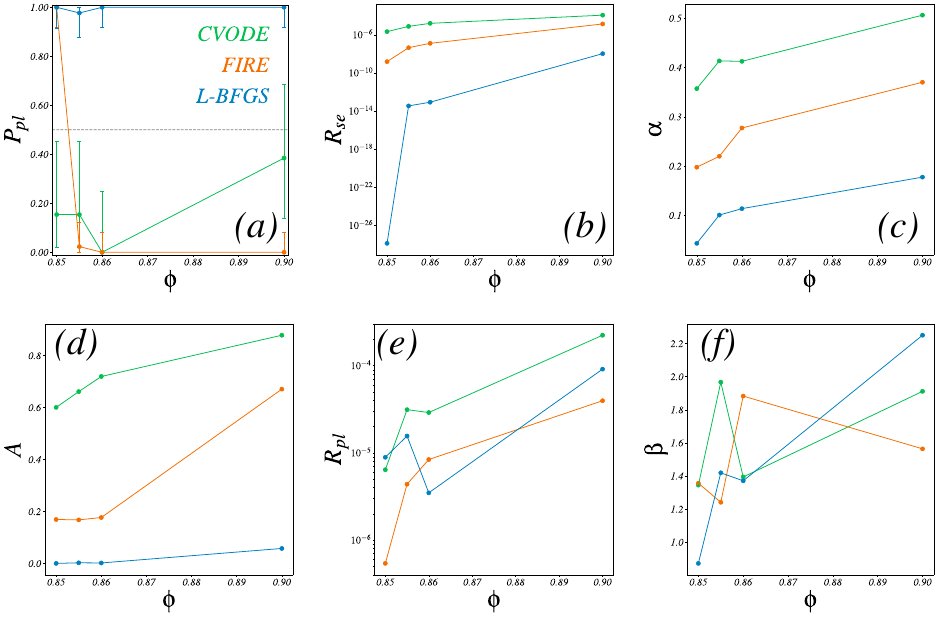}
  \caption{\textbf{Survival: Fit Comparison}
    $(a)$ Proportion $P_{pl}$ of the leave-one-out ensembles for which the power law fit is better than the stretched exponential.
    Other panels show the values of the fitting parameters we found for each fitting function and method against $\phi$, namely $(b)$$R_{se}$, $(c)$ $\alpha$, $(d)$ $A$, $(e)$ $R_{pl}$, and $(f)$ $\beta$.
    \label{fig:Survival_Fits}
  }
\end{figure}

To quantify this behavior, we fit to two models.
The first one is a stretched exponential,
\begin{align}
    P_{in}(R) = \exp\left( - \left( \frac{R}{R_{\text{se}}} \right)^\alpha \right)
\end{align}
with $R_{se}$ defining a characteristic lengthscale, and $\alpha$ a stretching exponent.
The second one is a  saturating power law
\begin{align}
    P_{in}(R) = \frac{A}{\left( 1 + \frac{R}{R_{\text{pl}}} \right)^{\beta}}
\end{align}
where $A$ is a saturating factor,$R_{pl}$ is a characteristic lengthscale, and $\beta$ a power law exponent in the tail.
To ensure that the fits accurately capture the tail, we perform weighted non-linear least-square fits in log space, with weights for each point $r_{i}$ as $1/\delta_{i}^{2}$ where $\delta_i \equiv \Delta_i / P_{\mathrm{in}}(r_i)$. Here $\Delta_{i}$ is the width of the $95\%$ confidence interval on the observed point $P_{i}$. When the Central Limit Theorem applies, and the error can be assumed Gaussian, $\Delta_{i} \propto \sigma(r_{i})$ where $\sigma(r_{i})$ would be the variance at each point.
To compare the fit quality, we compute the difference between the weighted residual sum of squares, $\Delta \mathrm{WRSS}$, defined as
\begin{equation}
  \label{eq:wrss}
    \Delta \mathrm{WRSS}_{\log} \equiv \mathrm{WRSS}_{\text{power law}} - \mathrm{WRSS}_{\text{stretched exp}}
\end{equation}
as a function of $\phi$.
If the power law fit is better, $\Delta \mathrm{WRSS}_{\log}  < 0$ and if the stretched exponential fit is better, $\Delta \mathrm{WRSS} > 0$.
We then construct leave-one-out (or ``jackknife''~\cite{Efron1979}) ensembles of the basins and find the proportion of the ensemble $P_{pl}$ where $\Delta \mathrm{WRSS}_{\log} < 0$, i.e the proportion
where the data is better described by a power law, with 95\% Clopper-Pearson confidence intervals.
The resulting $P_{pl}$ is plotted in Fig.~\ref{fig:Survival_Fits} as a function of $\phi$.
We show that for L-BFGS, $P_{pl} \approx 1$ at all $\phi$, suggesting power law behaviour at all packing fractions.
For FIRE, $P_{pl}\approx 0$ at large $\phi$, but becomes $1$ close to jamming, suggesting an emergent power-law behaviour close to jamming.
However, when using CVODE, we find that $P_{pl} < 0.5$ at all values of $\phi$, and does not seem to grow nor decrease systematically with $\phi$.
In other words, CVODE reveals that the seemingly power-law behaviour of survival curves was yet another mirage linked to the use of inappropriate optimization methods.

For completeness, in Fig.~\ref{fig:Survival_Fits}$(b)$-$(f)$, we report the values of the fitting parameters we found for each $\phi$ for both models (stretched exponential and power law), and across methods.
We note in particular that the scale $R_{se}$ of the stretched exponential fits with CVODE (Fig.~\ref{fig:Survival_Fits}$(b)$, green symbols) remains nearly constant as jamming is approached, while the L-BFGS one for instance (blue symbols) becomes vanishingly small.
This is an additional sign that basins are well-behaved shapes with finite length scales.

\subsection{Jamming point}\label{sec:jamming}

\begin{figure}[htbp]
  \centering
    \includegraphics[width=0.3\textwidth]{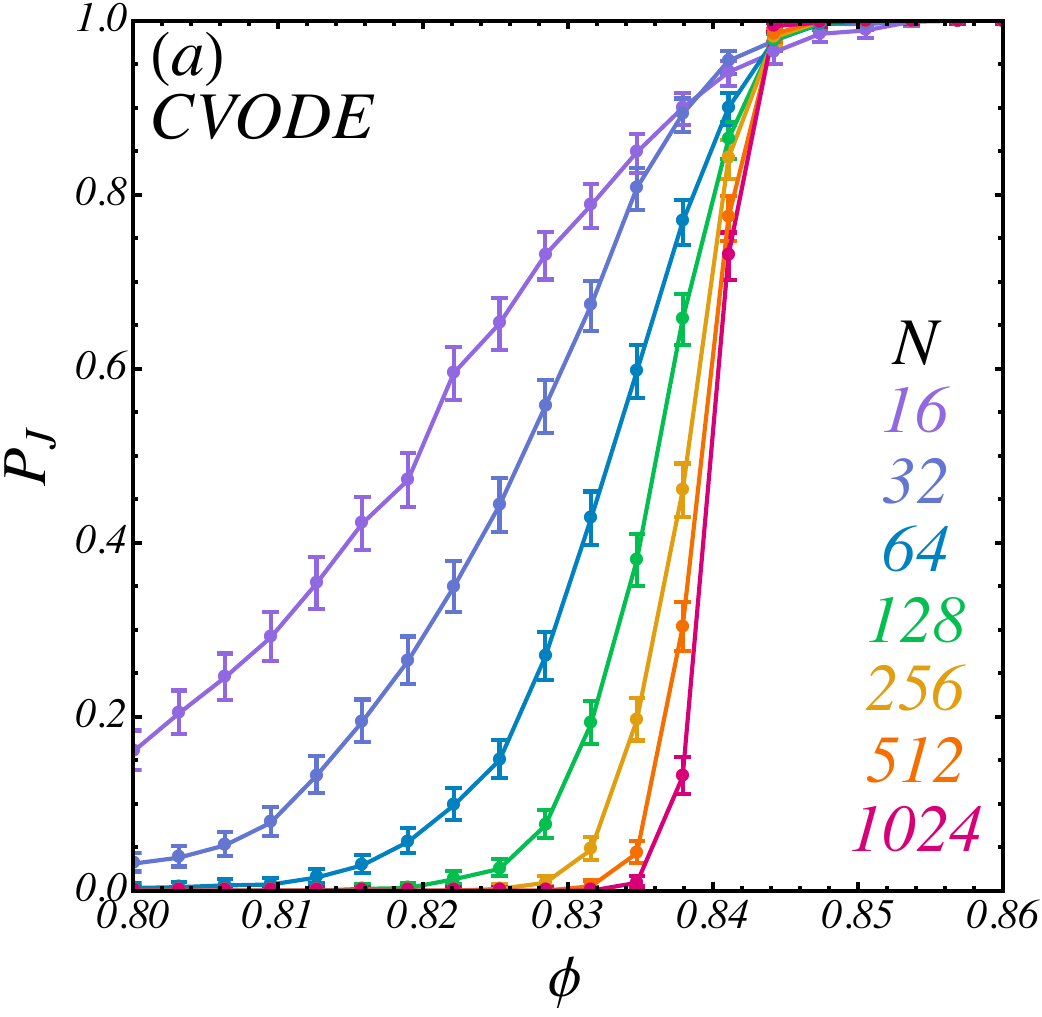}
    \includegraphics[width=0.3\textwidth]{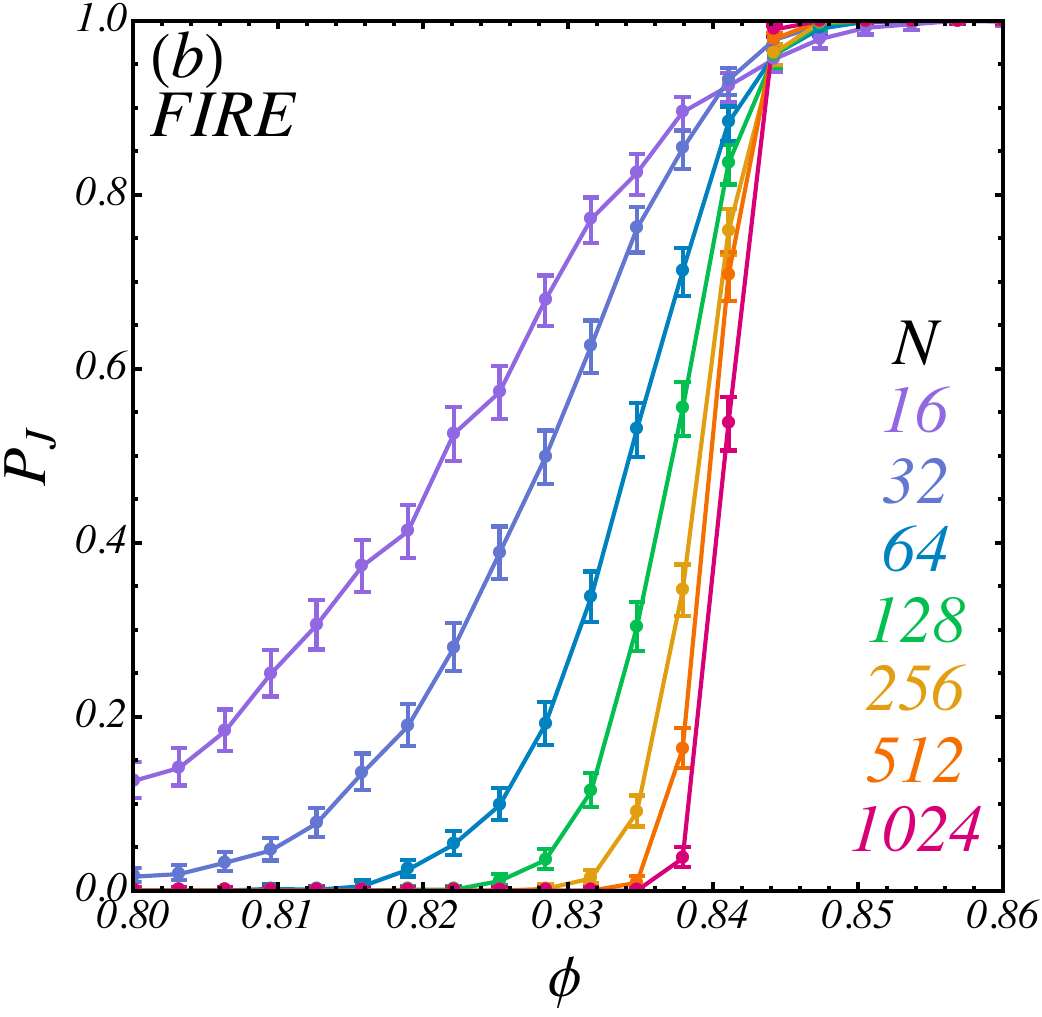}
    \includegraphics[width=0.3\textwidth]{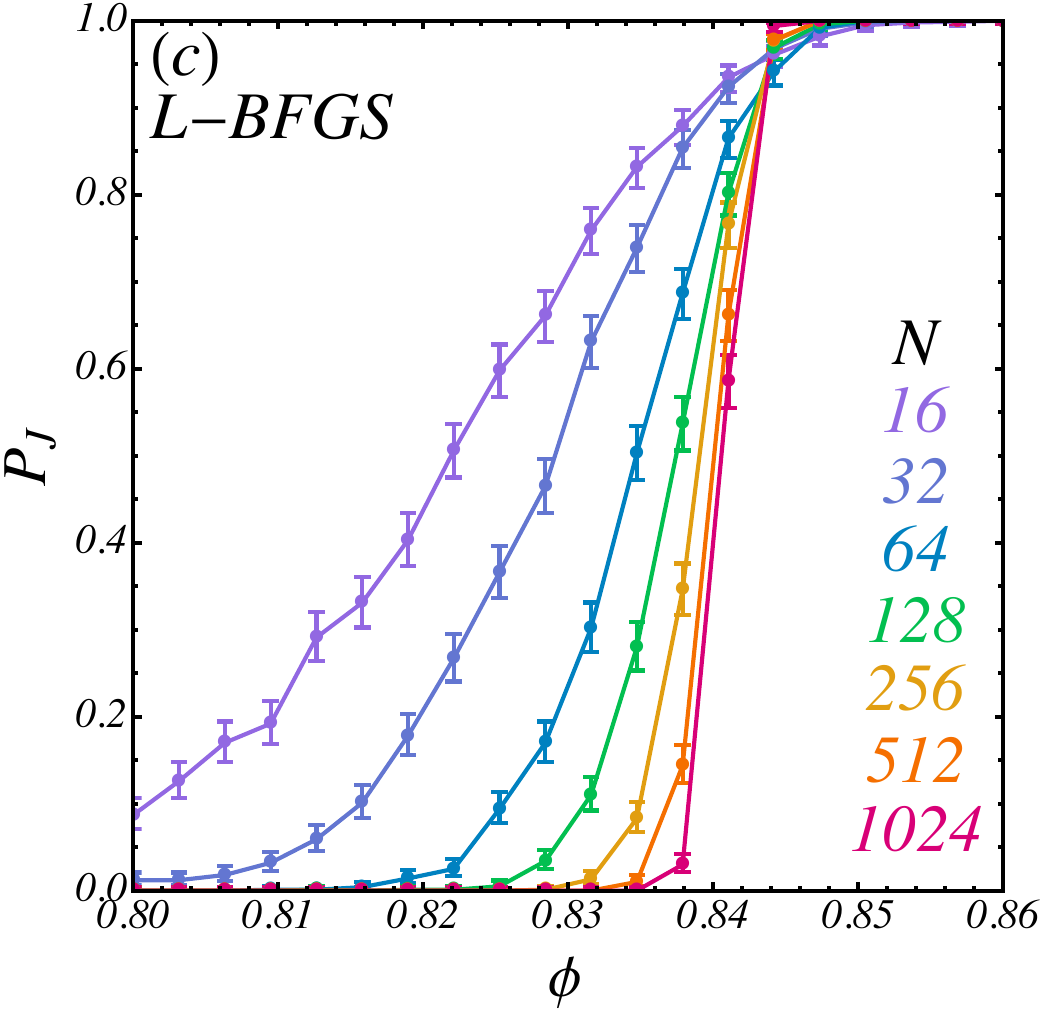}
    \caption{\textbf{Jamming transition across sizes.}
    Fraction of jammed states as a function of packing fraction, across system sizes (color within panels), using $(a)$ CVODE, $(b)$ FIRE, and $(c)$ L-BFGS.
    Error bars are Clopper-Pearson $95\%$ confidence intervals.
    }
    \label{fig:SI_JammingFigures}
\end{figure}

In this section, we present additional data on the location of the jamming transition as a function of system size and optimizer choice.
We consider a few system sizes, $N=8$, $16$, $32$, $64$, $128$, $256$, $512$, $1024$ and, for each $N$, and a range of packing fractions $\phi \in [0.80;0.86]$.
For each $(N, \phi)$ combination, we draw a set of $10^3$ random initial conditions uniformly in $[0;L]^{N d}$, and perform relaxations from them using FIRE, L-BFGS, and CVODE to generate a list of minima (note that the figure in the appendix of the main text uses $10^4$ points per density instead).
We then measure the fraction $P_J$ of these minima that is jammed, as per the criterion given in Sec.~\ref{sec:minima-match-proc}.
The results are reported in Fig.~\ref{fig:SI_JammingFigures}.
Across $(a)$ CVODE, $(b)$ FIRE, and $(c)$ L-BFGS, $P_J$ displays a finite-size transition from $0$ at low $\phi$ to $1$ at high $\phi$, with a transition that gets sharper as $N$ increases, as expected at jamming~\cite{OHern2003}.
We show that using FIRE or L-BFGS leads to a systematic bias of the curve $P_J(\phi)$ towards lower values (fewer jammed states, more liquid states), meaning that $\phi_J$ is overestimated.
However, this bias is most pronounced at small $N$, so that previous reports of $\phi_J$ values relying on large $N$ or on $N$ scalings were likely not noticeably affected by their choice of optimizer.

To evaluate the jamming density presented in the main text, we perform a fit to a function of the form
$P(\phi) = \sigma \left( a x + b \right)^{p}$ where $\sigma(x)$ is the sigmoid function, and we estimate a jamming packing fraction $\phi_J$ as the packing fraction at which $P_J = 1/2$.
As mentioned in main text, we perform a finite-size scaling analysis of $\phi_J$ with $N$ to assess the effect of the choice of method on critical properties.
To do so, we collect $\phi_J(N)$ values and fit them, for each method, to a power law of the form $|\phi_J(N) - \phi_J^\infty| = C N^{- \theta}$ with fitting parameters $\phi_J^\infty$ the asymptotic $\phi_J$, $\theta$ a critical exponent, and $C$ a proportionality constant.
We find that CVODE yields $\phi_{J \mathrm{(CVODE)}}^\infty \approx 0.8415$, whereas FIRE and CVODE yield $\phi_{J \mathrm{(FIRE)}}^\infty \approx 0.8423$, $\phi_{J \mathrm{(LBFGS)}}^\infty \approx 0.8422$   but slightly different exponents, $\theta_{\mathrm{CVODE}} \approx 0.673$, $\theta_{\mathrm{FIRE}} \approx 0.668$ and $\theta_{\mathrm{LBFGS}} \approx 0.674$.
These values are compatible with results from similar scaling analysis~\cite{OHern2003,Vagberg2011}.
We note that the probability of observing a jammed state is always higher with CVODE at all densities, as compared to using optimizer methods.

\begin{figure*}[htbp]
    \centering
    \includegraphics[width=0.24\textwidth]{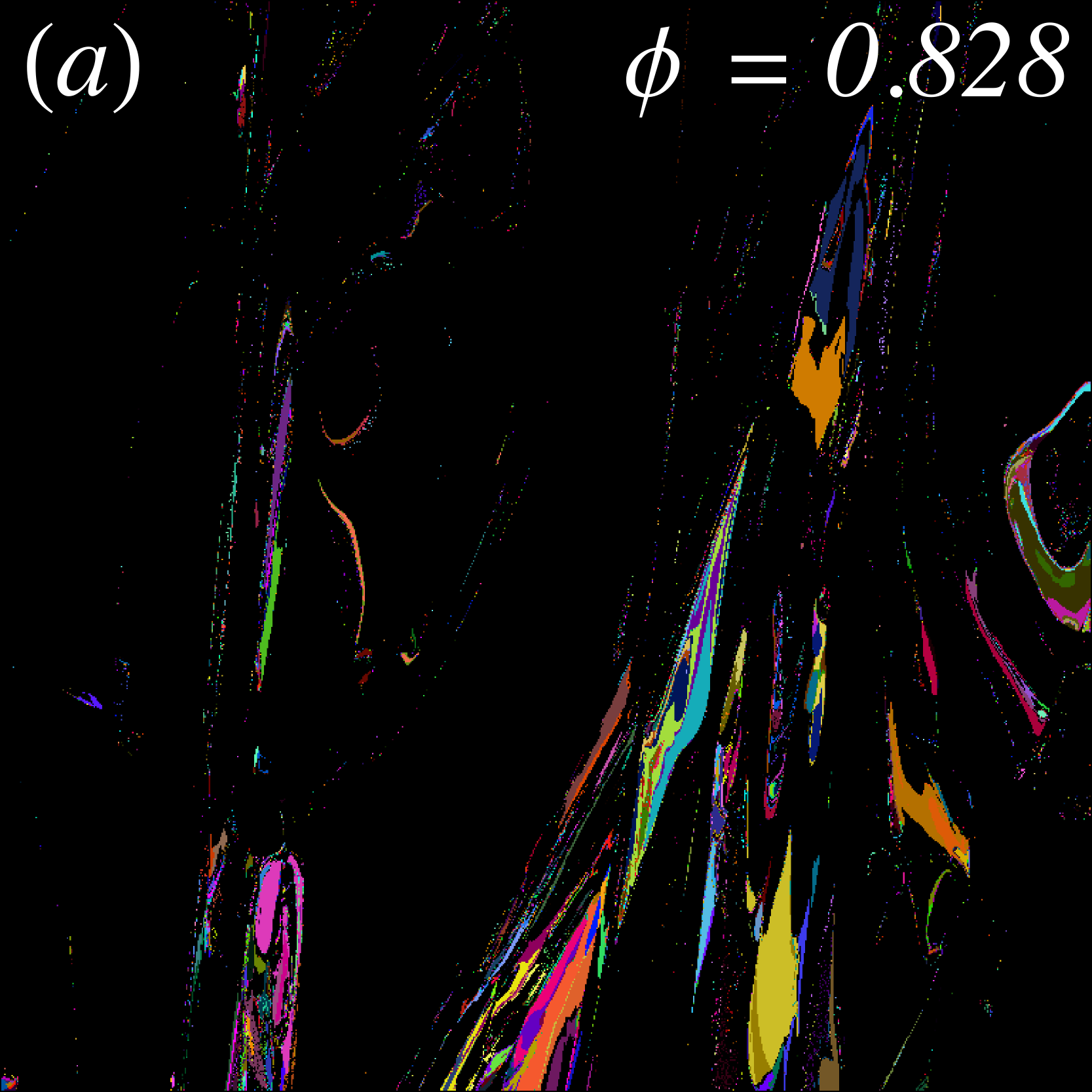}
    \includegraphics[width=0.24\textwidth]{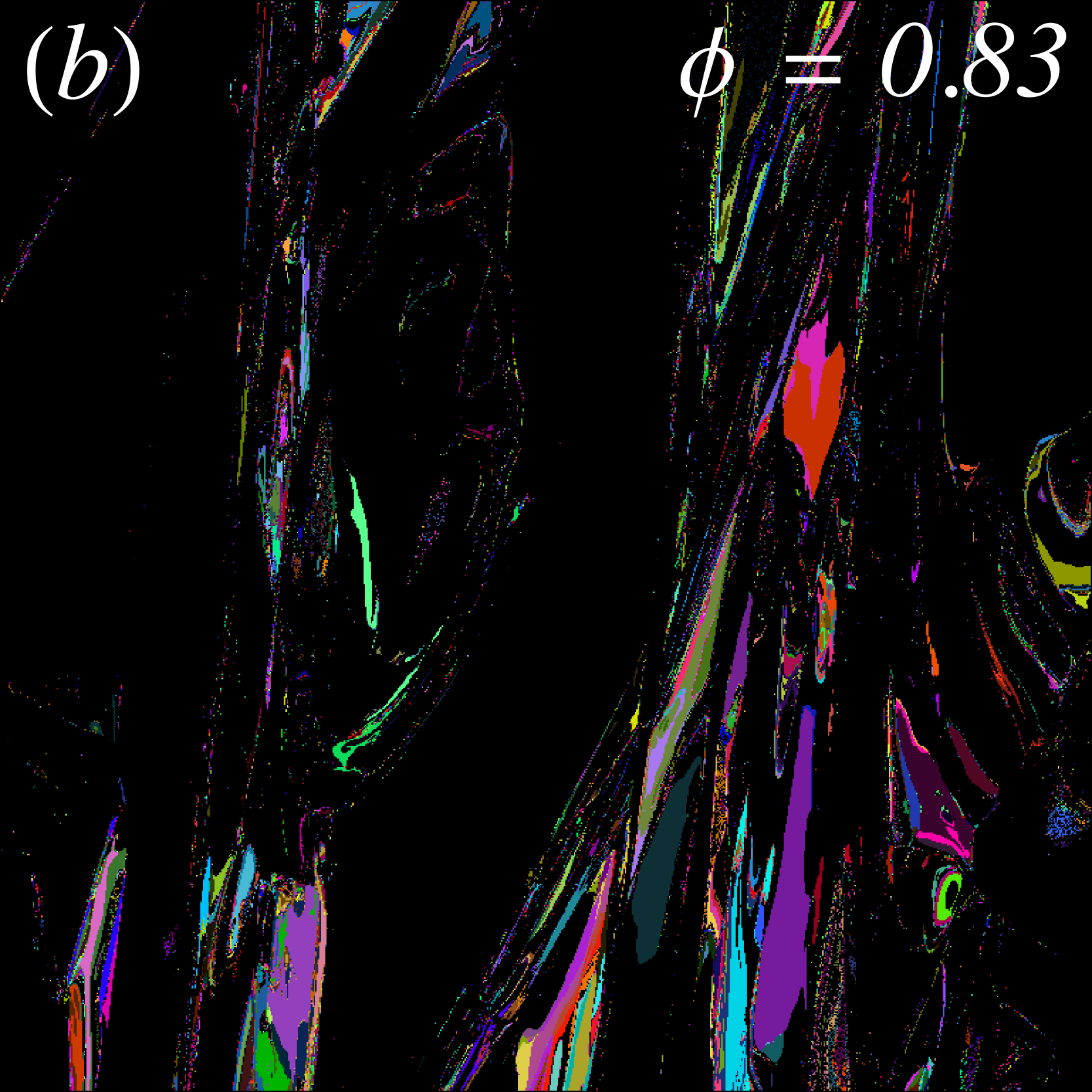} 
    \includegraphics[width=0.24\textwidth]{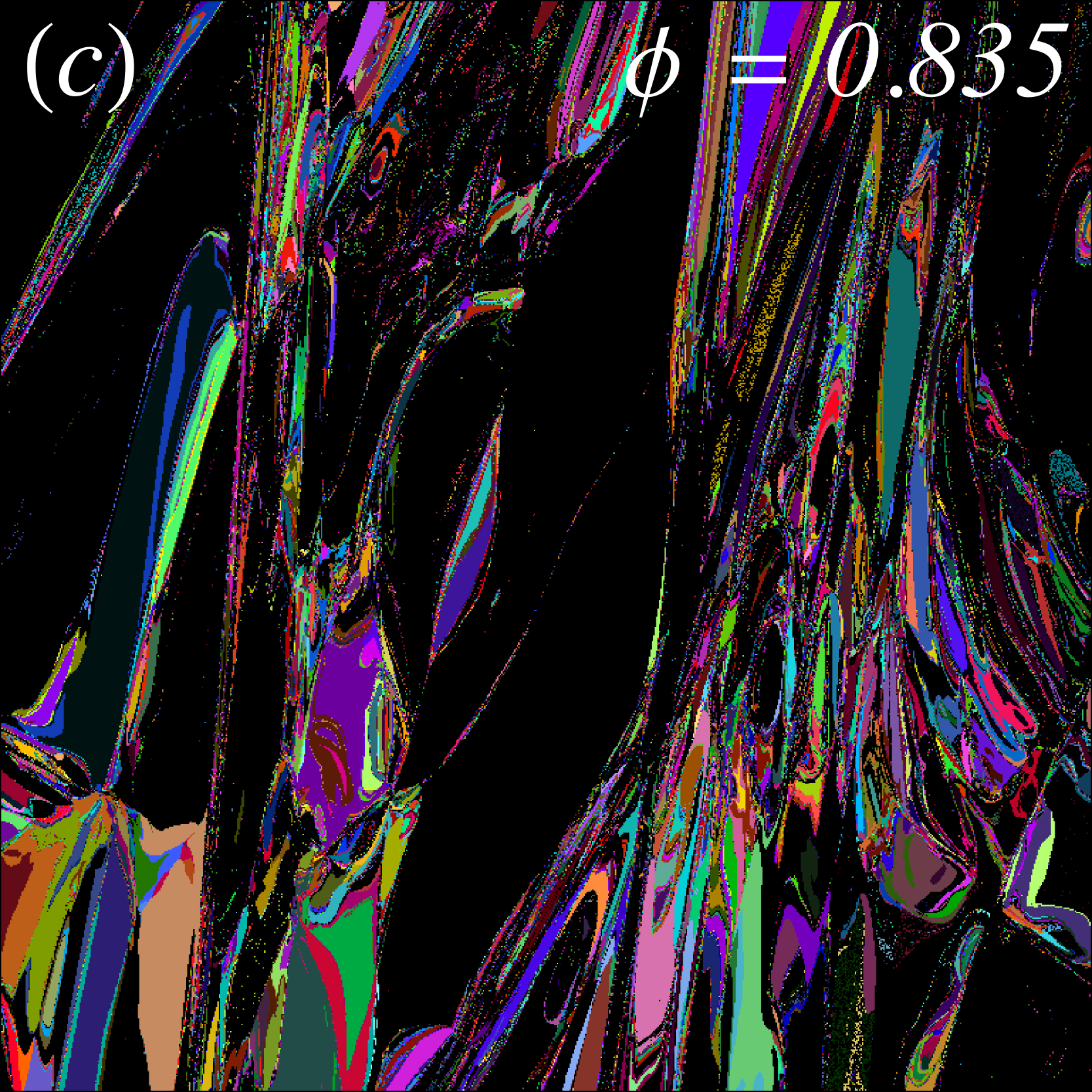}
    \includegraphics[width=0.24\textwidth]{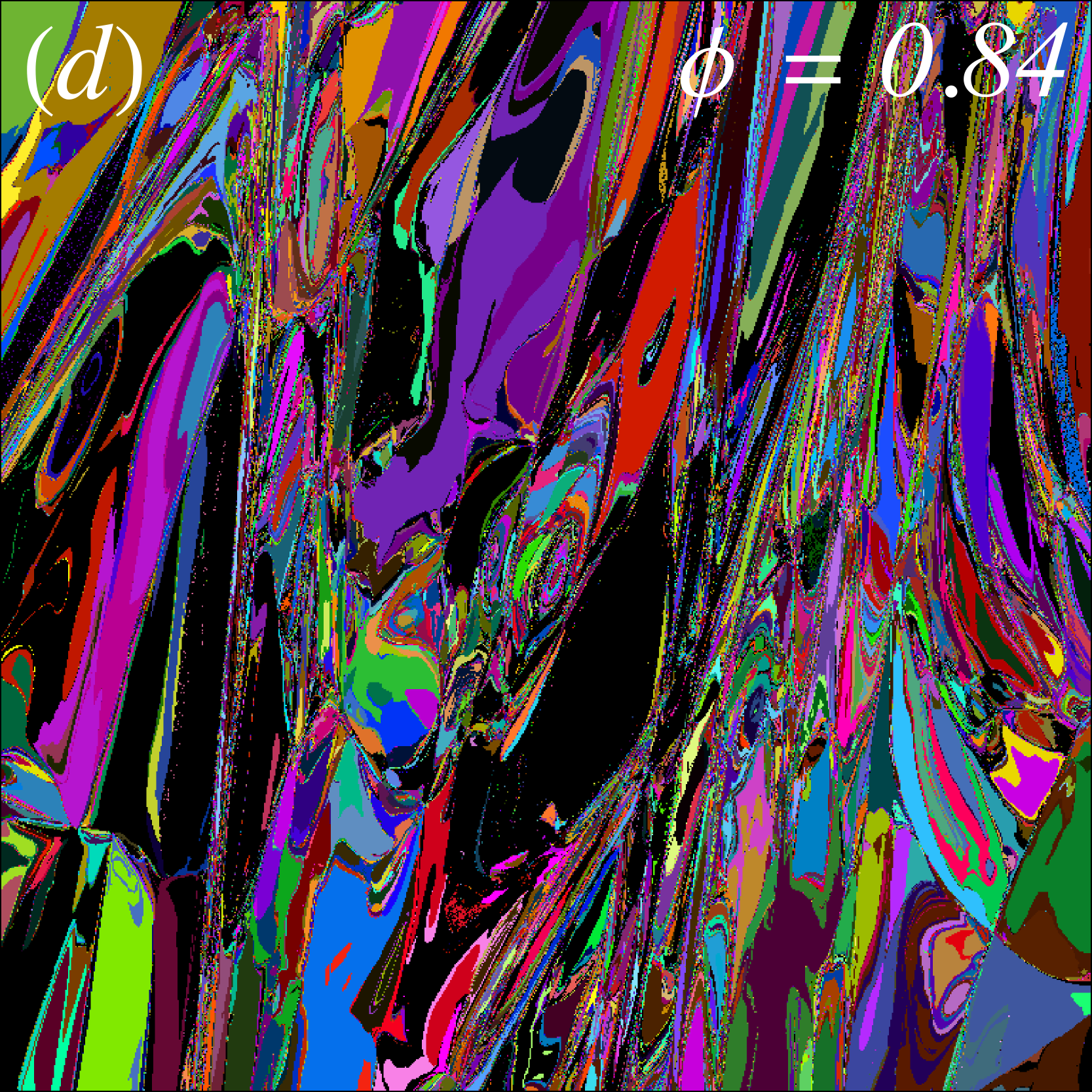}\\
    \includegraphics[width=0.24\textwidth]{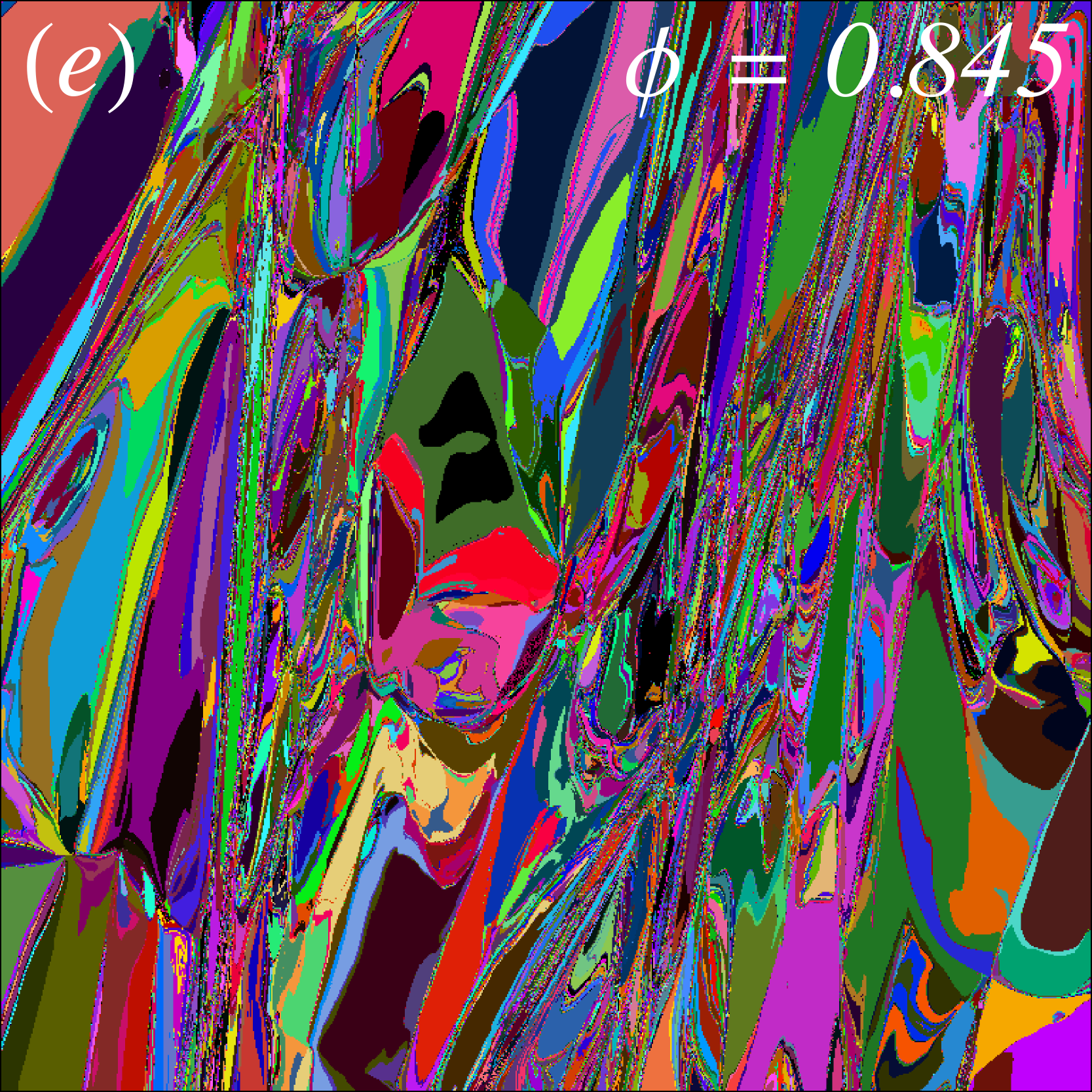}
    \includegraphics[width=0.24\textwidth]{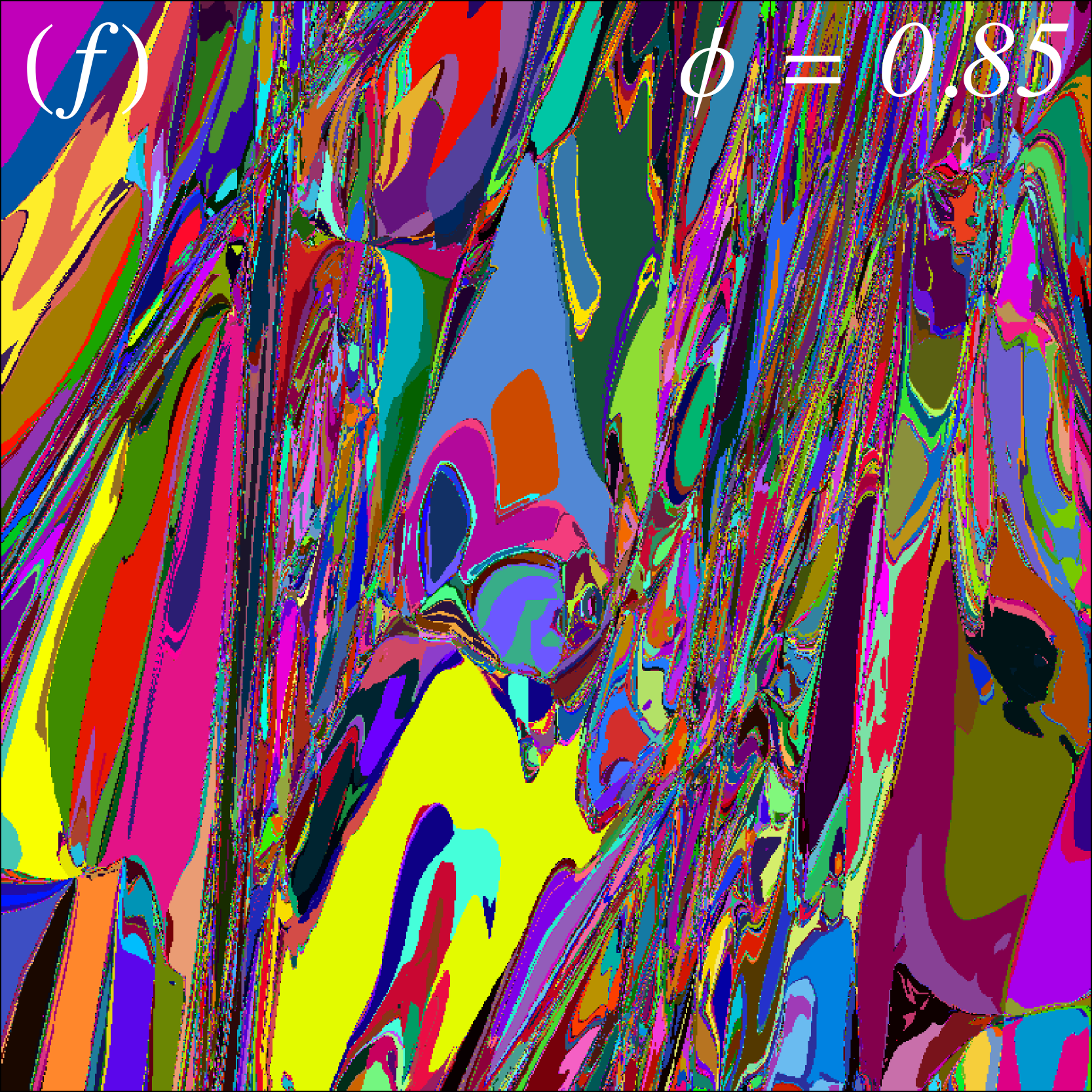}
    \includegraphics[width=0.24\textwidth]{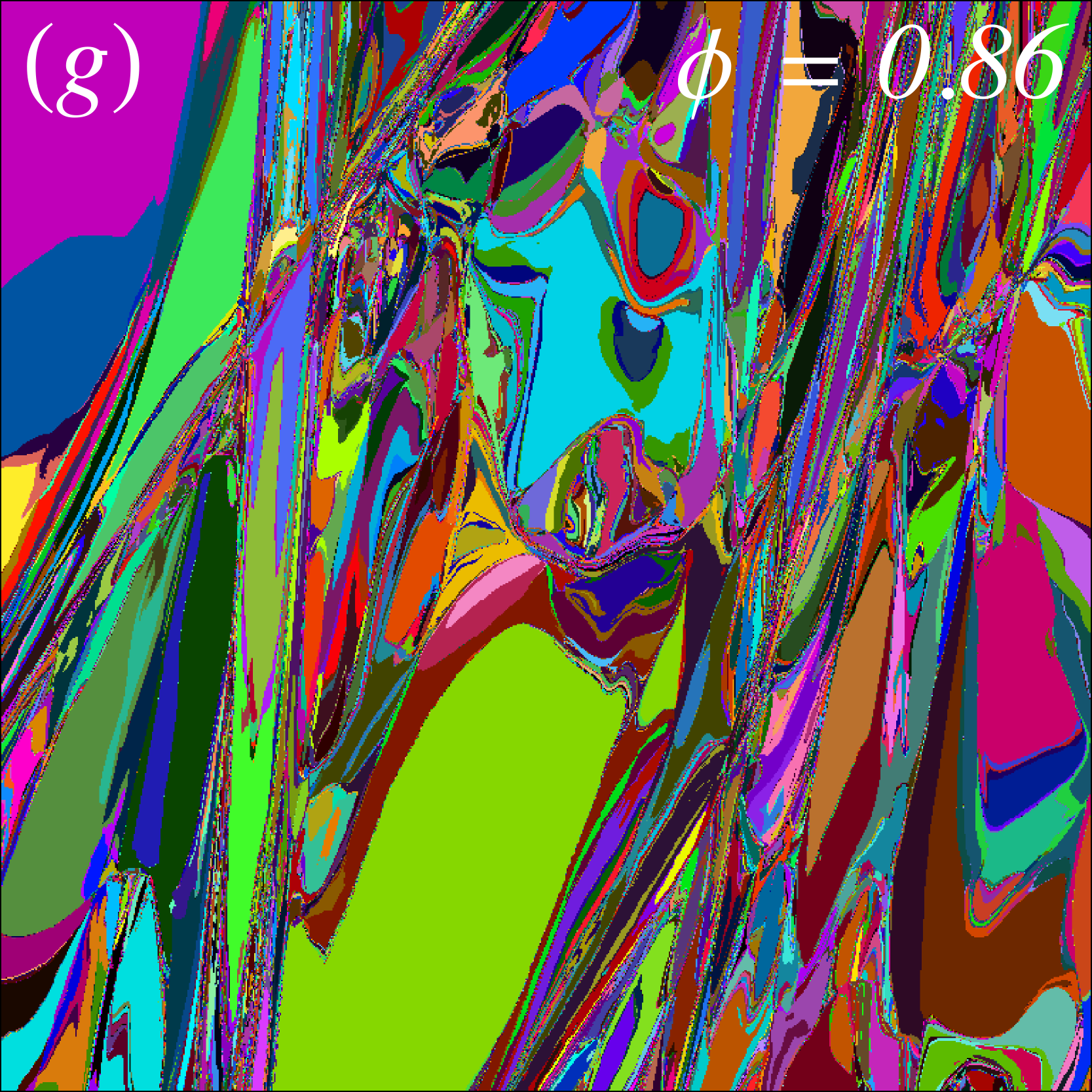}
    \includegraphics[width=0.24\textwidth]{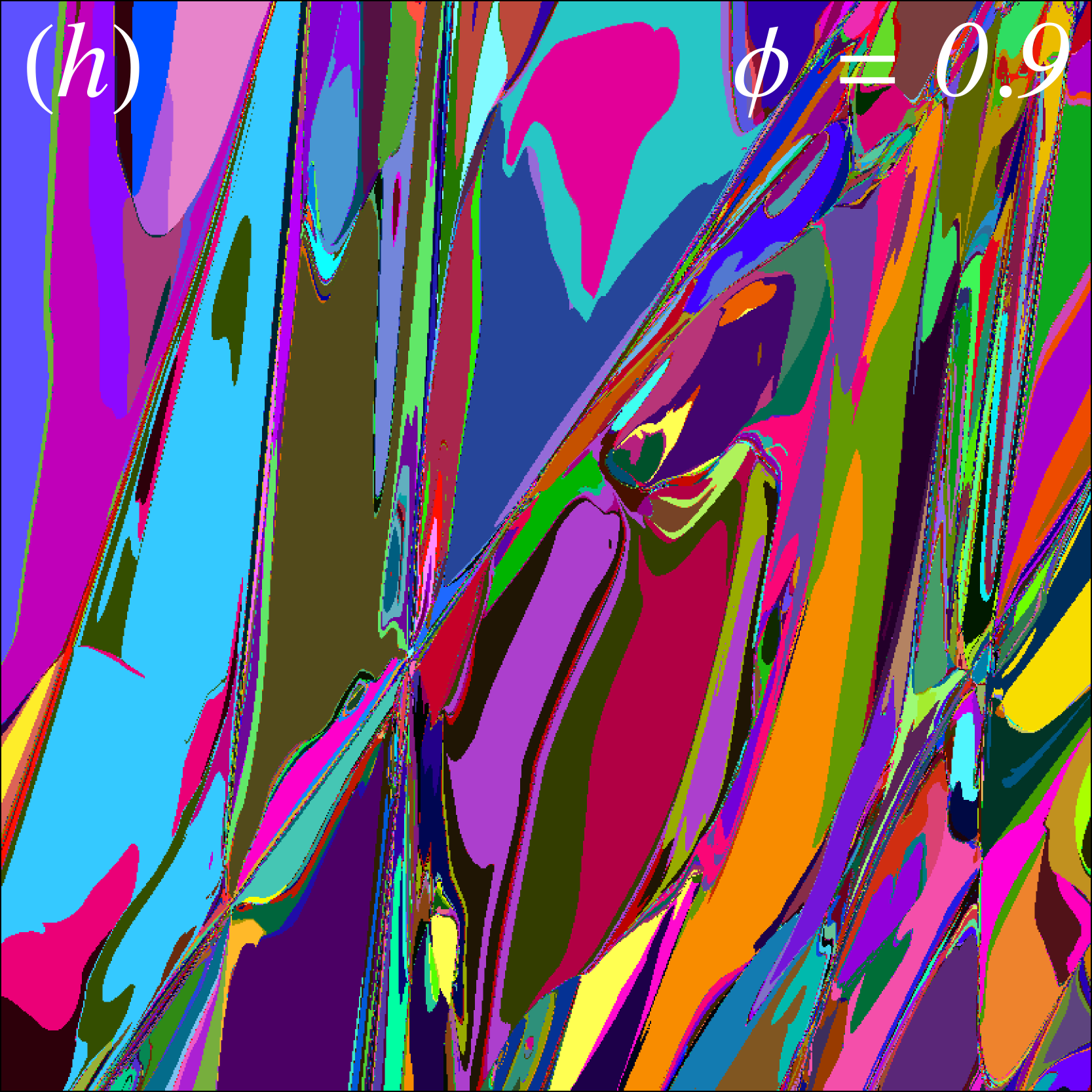}

    \caption{\textbf{Density and landscape slices.}
    Slices of the energy landscape for $N=128$ and at $\phi$ equal to $(a)$ $0.828$, $(b)$ $0.830$, $(c)$ $0.835$, $(d)$ $0.84$, $(e)$ $0.845$, $(f)$ $0.85$, $(g)$ $0.86$, $(h)$ $0.9$.
    Individual basins are encoded by colors, while liquid states are shown as black pixels.
    }
    \label{fig:density_slices}
  \end{figure*}

We further illustrate the effect of density on basins by reproducing the cuts from Fig.~2 of the main text at several $\phi$, but using the same relative coordinates $\left\{\bm{r}_i/L\right\}_{i=1..N}$ and the same size polydispersity up to a global factor on all diameters, so that the slices correspond to the same initial point positions but different global dilations of the system.
The results are shown in Fig.~\ref{fig:density_slices}.
In this figure, colors are \textit{not} matched across slices as minima move across densities, and black indicate non-jammed (fluid) states.
Going from $\phi = 0.828$ to $\phi = 0.86$, we show that the configuration space goes from being mostly made of fluid states and sparsely populated by basins to being tiled entirely by basins, at a packing fraction $\phi \approx 0.84$ comparable to the final plateau from Fig.~\ref{fig:SI_JammingFigures} for $N=128$.

Furthermore, the pictures show that the basins are all very slender at small densities, but become bulkier as density grows, highlighting the difficulties associated with both integrating steepest descent and measuring volumes at low densities.
Interestingly, such slices taken at the same relative positions of configuration space display a sequential fragmentation of configuration space into basins as $\phi \to \phi_J^-$, reminiscent of thin sheet crumpling, see in particular illustrations in Ref.~\cite{Andrejevic2021}.
This suggests that the fragmentation of configuration space into basins as a function of density could be modeled by similar dynamical processes.
Also note that the number of basins is maximal near jamming: as $\phi$ grows, for $\phi < \phi_J$, the liquid states get replaced by basins of jammed states, while for $\phi > \phi_J$ basins merge.

\subsection{Slices at $N = 8$}

\begin{figure*}[htbp]
    \centering
    \includegraphics[width=0.32\textwidth]{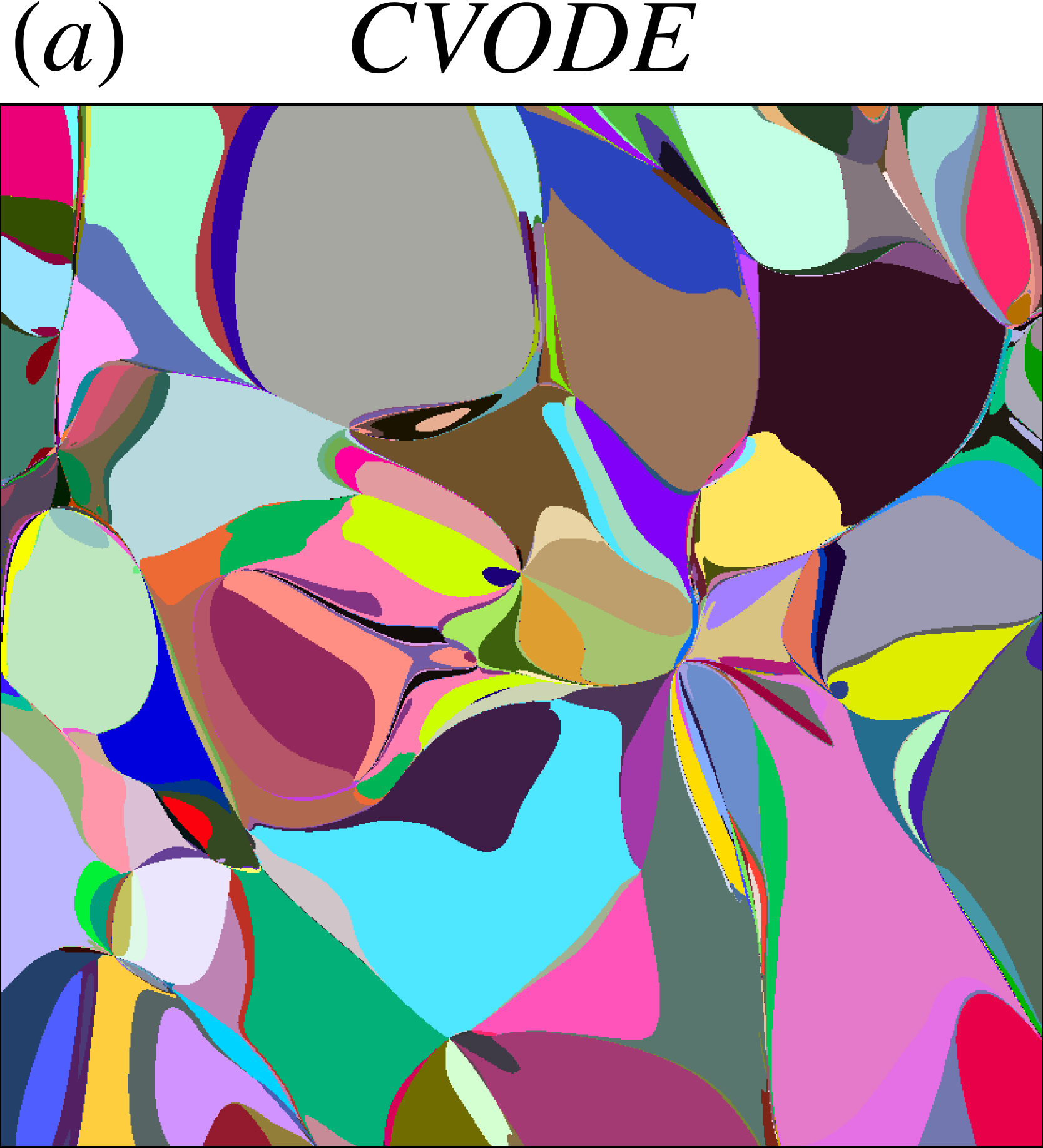}
    \includegraphics[width=0.32\textwidth]{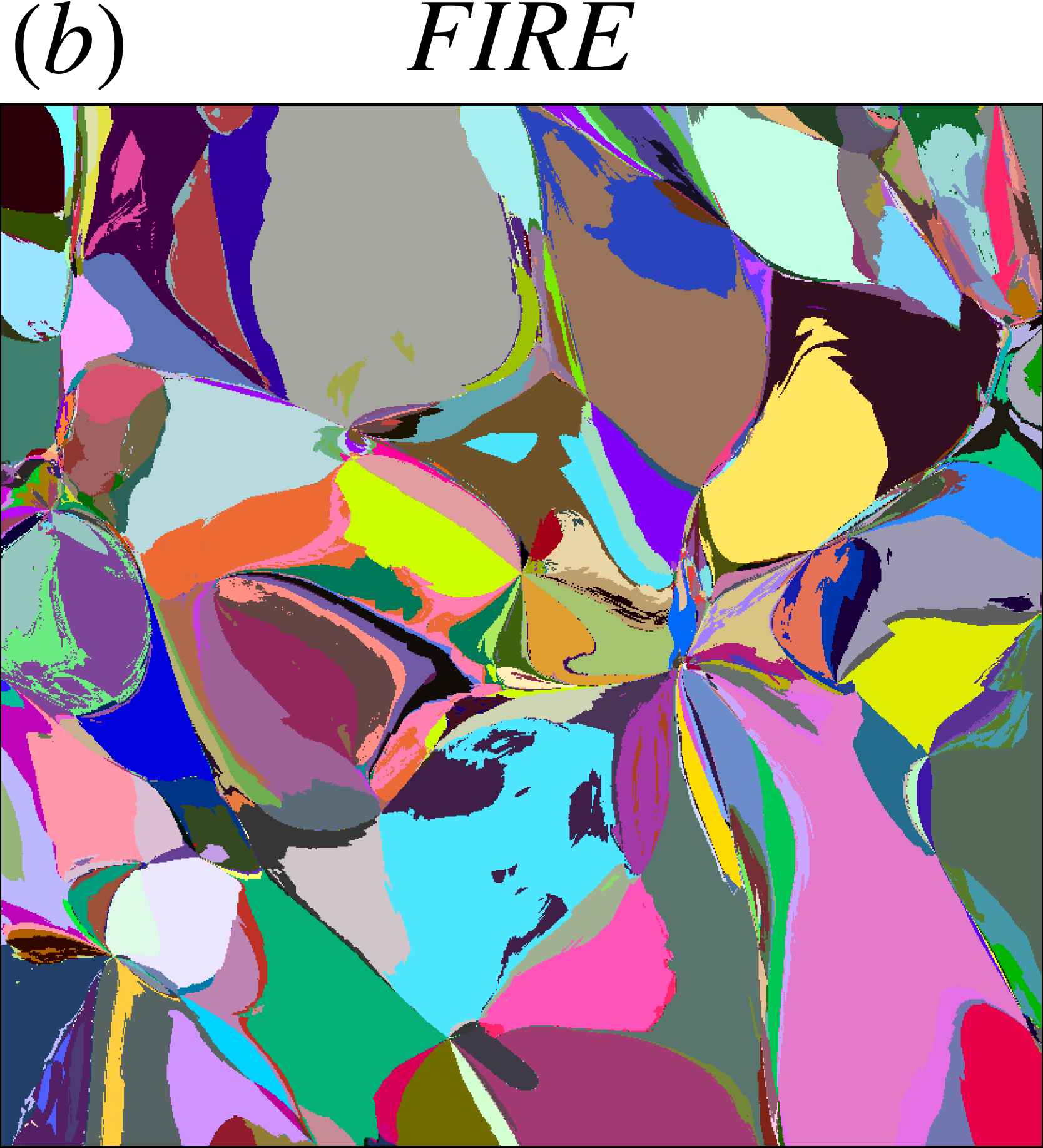}
    \includegraphics[width=0.32\textwidth]{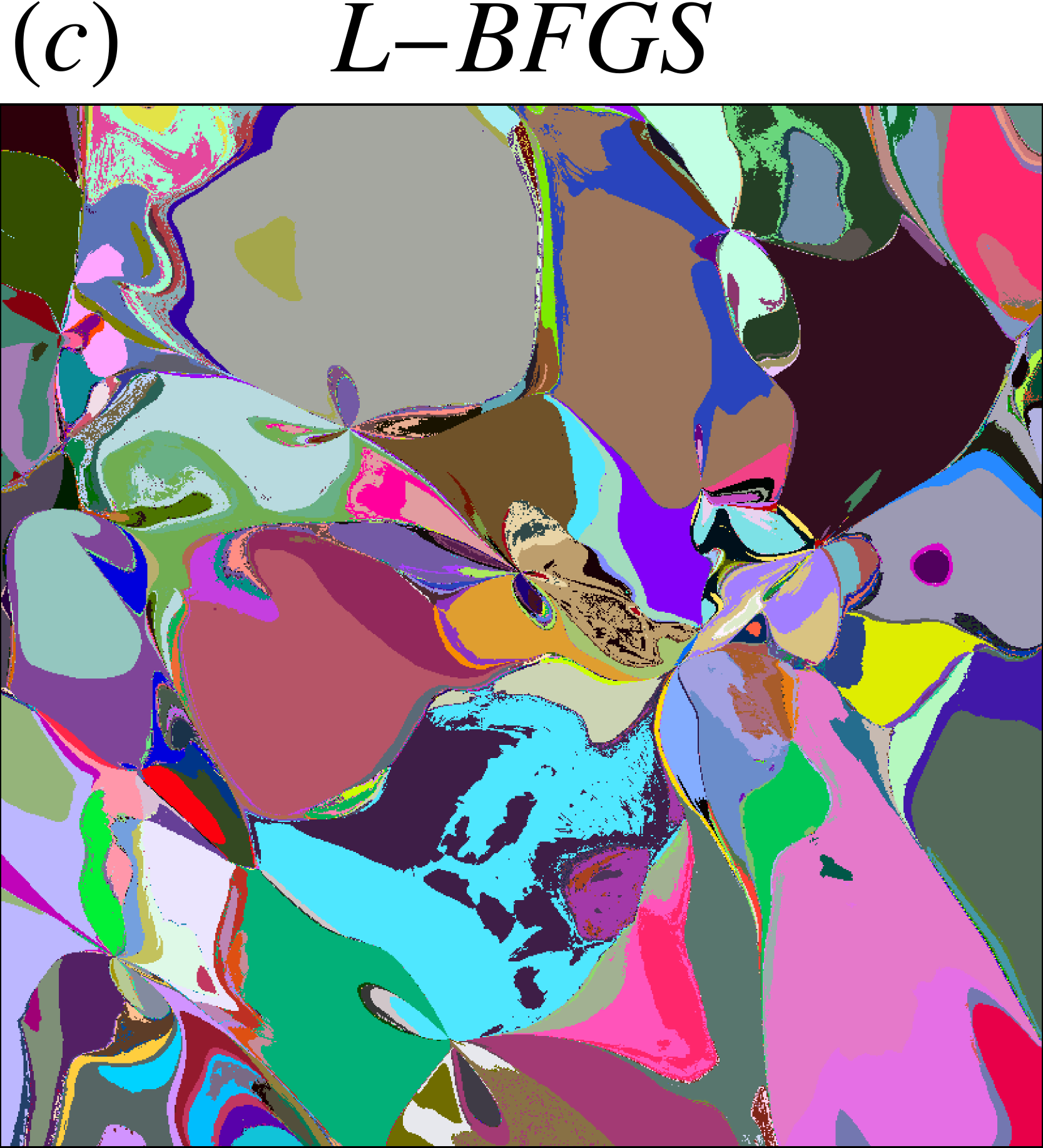}
    \caption{\textbf{Small-system Slices.}
    Random $2d$ slice of the energy landscape of $N = 8$ particles at $\phi = 0.9$ using $(a)$ CVODE, $(b)$ FIRE, and $(c)$ L-BFGS.
    Minima are matched across slices then color-coded.
    }
    \label{fig:8_slice}
\end{figure*}

To complement the data shown in main text, in Fig.~\ref{fig:8_slice}, we show a slice of the energy landscape of an $N=8$ particle system at $\phi = 0.9$, measured using $(a)$ CVODE, $(b)$ FIRE, and $(c)$ L-BFGS.
Across these slices, identical minima are color-matched to highlight the error in basin tagging introduced by optimizers.
Even at this very small $N$, at which the accuracies of FIRE and L-BFGS are fairly high (about $70\%$ and $40\%$, respectively, as per Fig.~2 of the main text), there are noticeable basin deformations.
Furthermore, as noted in several other parts of this paper, optimizers make basins look less regular, more ``fractal-like'', even at modest $N$, in a way reminiscent of Refs.~\cite{Wales1992,Asenjo2013}.

\subsection{Accuracy of GMIN's L-BFGS}

\begin{figure}
    \centering
    \includegraphics[width=0.5\linewidth]{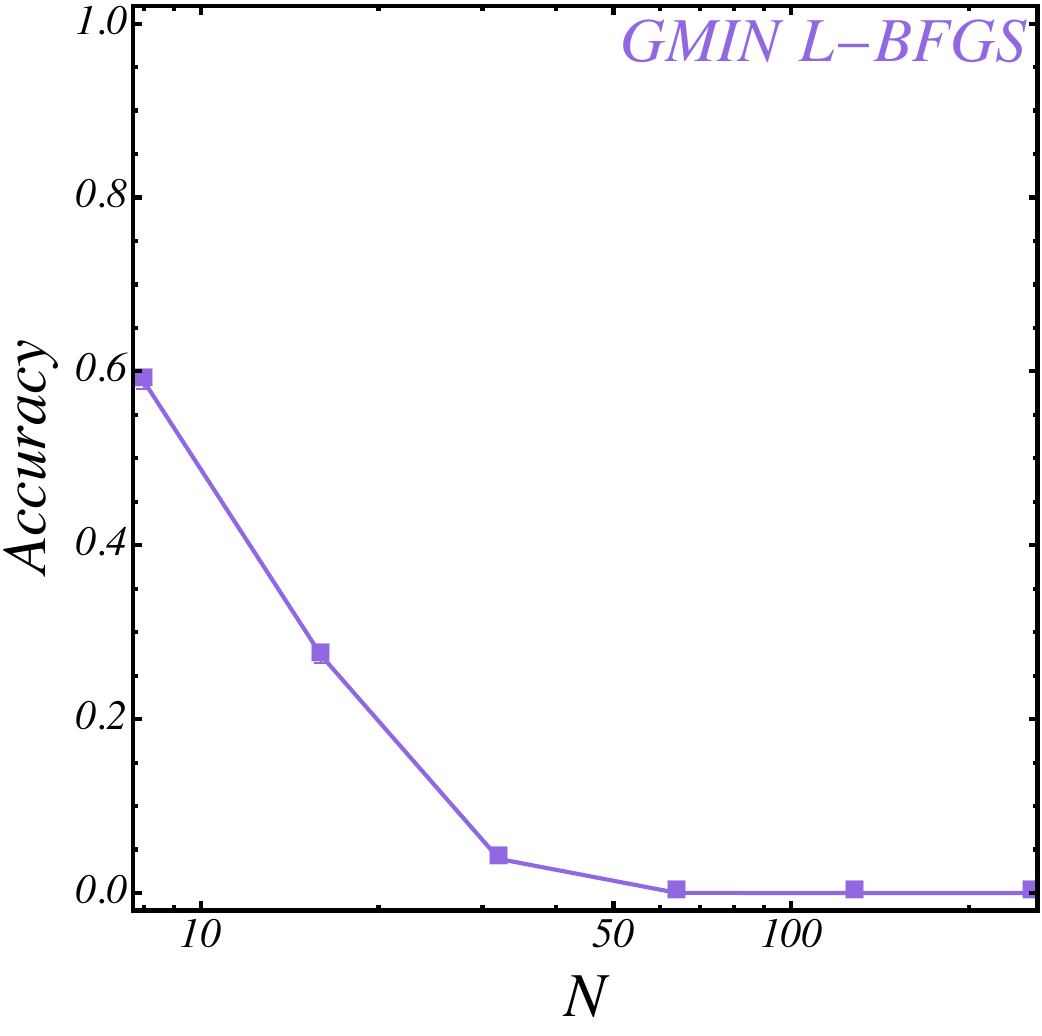}
    \caption{\textbf{GMIN Accuracy.}
    Accuracy measurement, analogous to Fig.2 of the main text, for the GMIN implementation of L-BFGS for Hertzian disks at $\phi = 0.9$.}
    \label{fig:GMIN_Acc}
\end{figure}

In this section, we provide data on the accuracy of a broadly used optimizer in the context of energy landscapes, the L-BFGS implementation found in the GMIN~\cite{GMIN} library.
Analogously to the results of Fig.2 of the main text, these are obtained by sampling $10^4$ points uniformly in configuration space, then minimizing the energy for each of these points using both CVODE at a low tolerance ($\rtol = \texttt{atol} = 10^{-10}$, stricter than the tolerances used in the main text, see Tab.~\ref{tab:cvode-params}) and GMIN's L-BFGS.
To run GMIN's native Fortran L-BFGS routine, \texttt{MYLBFGS}, on the same potential implementation as all other methods in this work, we interfaced it with the pele library~\cite{pele2024}; this interface is available in the \texttt{gmin-interface} branch of pele.
We use GMIN's default optimizer settings: history size $\texttt{M} = 4$, initial inverse-Hessian scaling $\texttt{DGUESS} = 0.1$, maximum step size $\texttt{MAXBFGS} = 0.4$, and maximum allowed energy rise per step $\texttt{MAXERISE} = 10^{-4}$, together with a convergence tolerance on the gradient of $10^{-10}$, as in Sec.~\ref{sec:Params}, and a maximum of $2 \times 10^5$ iterations.
We perform the measurement for our Hertzian disk model at $\phi = 0.9$ and across values of $N$.
We then report the fraction of points that fall into the same minimum by both methods as a function of $N$ in Fig.~\ref{fig:GMIN_Acc}.
We show that the accuracy decays similarly to those shown in the main text, although the GMIN implementation of L-BFGS here yields a larger accuracy than the one of the main text.

\begin{figure}
    \centering
    \includegraphics[width=0.5\linewidth]{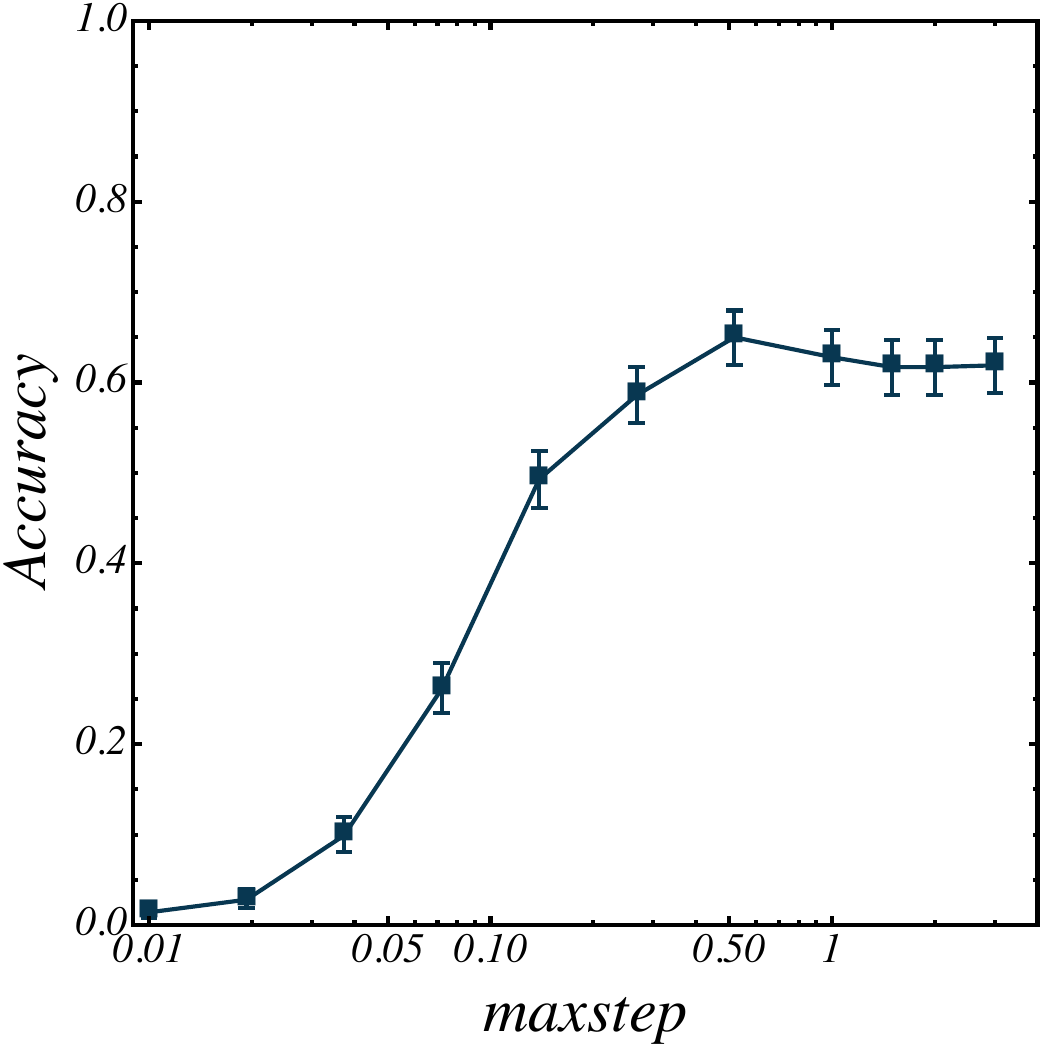}
    \caption{\textbf{Effect of the maximum step size.}
    Accuracy of the pele implementation of L-BFGS for Hertzian disks at $N = 8$ and $\phi = 0.9$, as a function of the step cap \texttt{maxstep}, measured over $10^3$ points sampled uniformly in configuration space.
    Error bars are Clopper-Pearson $95\%$ confidence intervals.
    At GMIN's default cap, $\texttt{maxstep} = 0.4$, the accuracies of the two implementations nearly match ($62\%$ vs $57\%$).}
    \label{fig:Maxstep}
\end{figure}

This difference is mostly attributable to the maximum step size: GMIN caps L-BFGS steps at $\texttt{MAXBFGS} = 0.4$, four times the value $\texttt{maxstep} = 0.1$ used in the main text (Tab.~\ref{tab:LBFGSparameters}).
The latter value follows the recommendation of Ref.~\cite{Asenjo2013}, namely a step cap of one tenth of the equilibrium pair distance, motivated by the observation that, in the systems studied there, smaller steps led to more accurate basins of attraction.
To quantify the effect of this parameter, we sweep \texttt{maxstep} in the pele implementation of L-BFGS at $N = 8$, using $10^3$ random points, as shown in Fig.~\ref{fig:Maxstep}: the accuracy grows monotonically with the step cap, from about $1\%$ at $\texttt{maxstep} = 10^{-2}$ to a plateau of approximately $62\%$ for $\texttt{maxstep} \gtrsim 0.5$.
At a matched cap, $\texttt{maxstep} = 0.4$, the pele implementation reaches an accuracy of $62\%$, close to the $57\%$ obtained with GMIN, so that the remaining implementation differences (history size $\texttt{M} = 4$ instead of $1$, and a tolerance for small energy rises, $\texttt{MAXERISE} = 10^{-4}$ instead of $10^{-10}$) only have a comparatively small effect on accuracy.
Notably, the trend we observe is opposite to that reported in Ref.~\cite{Asenjo2013}: for Hertzian disks, smaller steps make basin identification less, not more, accurate.
The results suggest that optimizer parameter prescriptions tuned on one landscape do not transfer to another, and that no choice of step cap turns an optimizer into a faithful basin-mapping tool: even at the best cap, the accuracy remains far from $1$ at $N = 8$, and still decays rapidly with $N$.

\subsection{Accuracies and times at $\phi = 0.86$ \label{sec:accuracytimephi86}}

\begin{figure}
    \centering
    \includegraphics[width=0.48\linewidth]{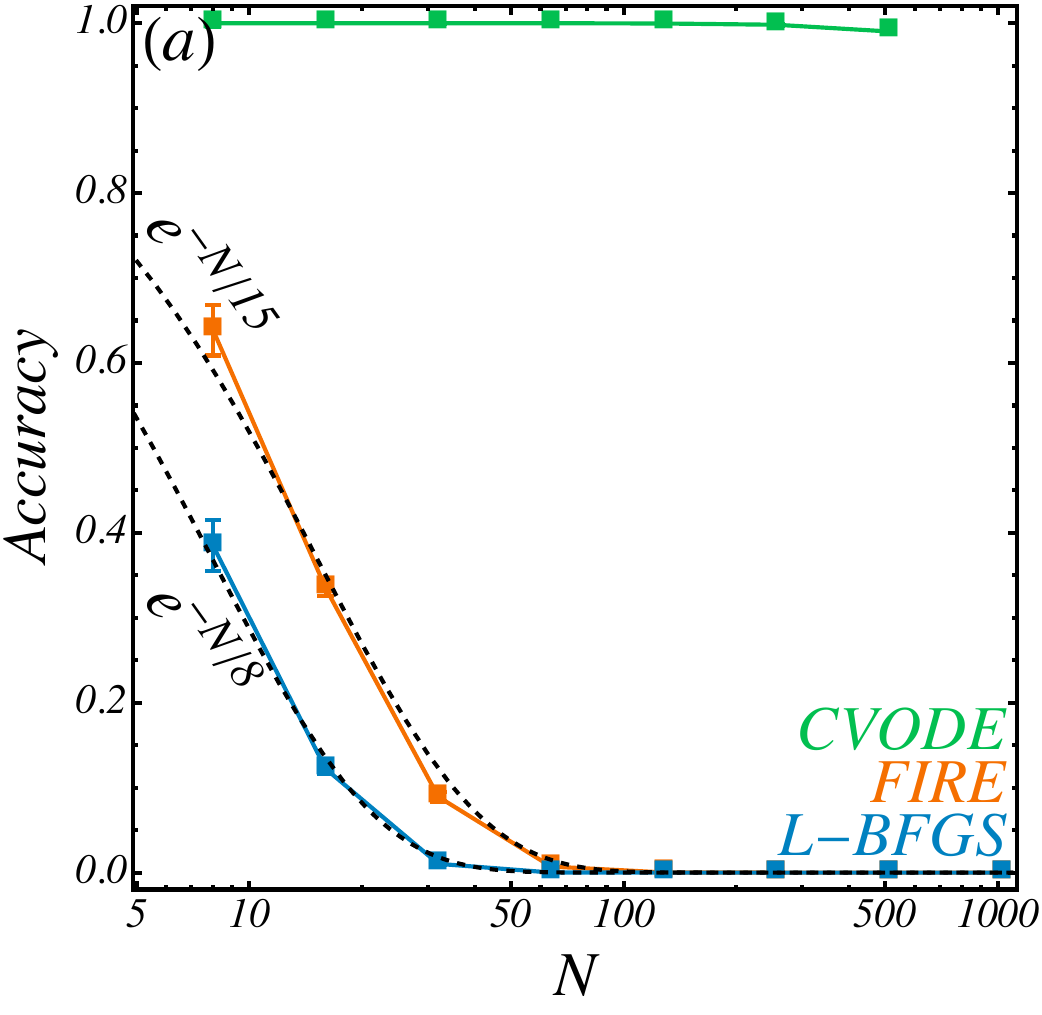}
    \includegraphics[width=0.48\linewidth]{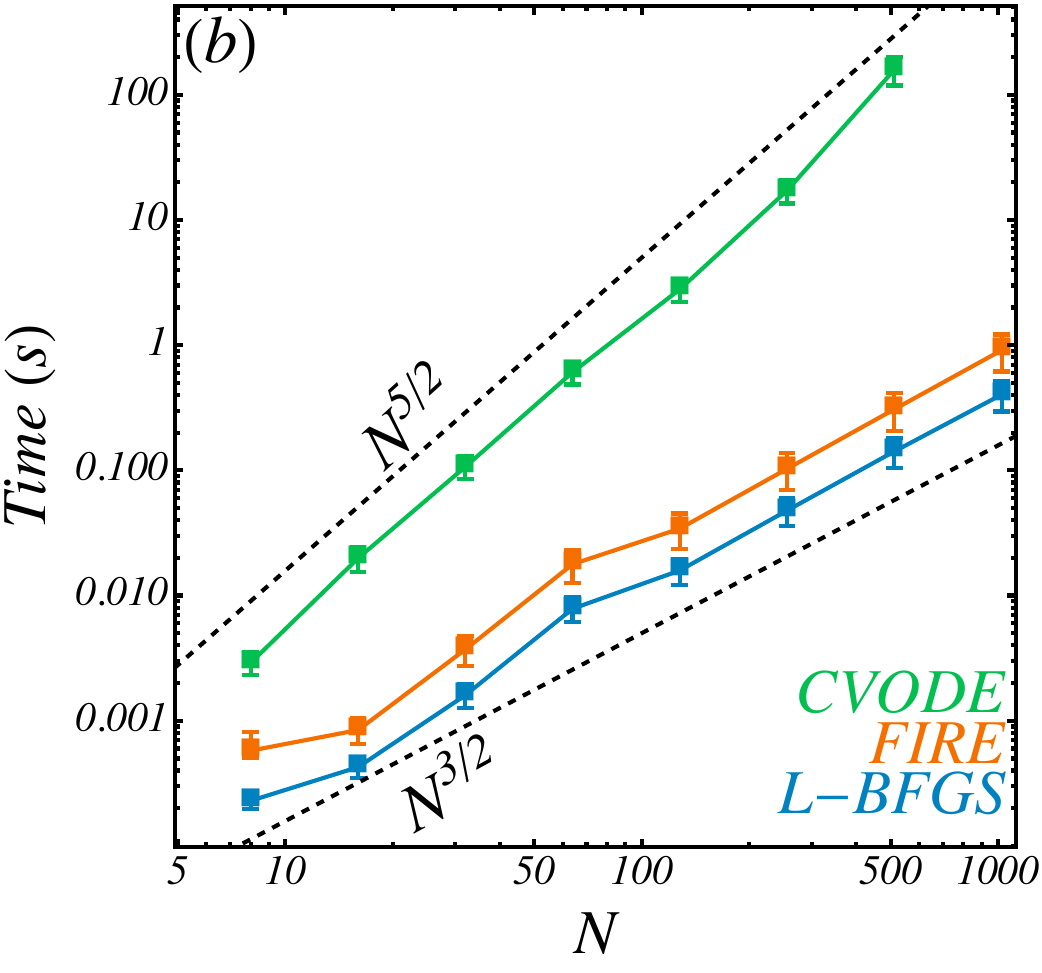} \\
    \includegraphics[width=0.32\linewidth]{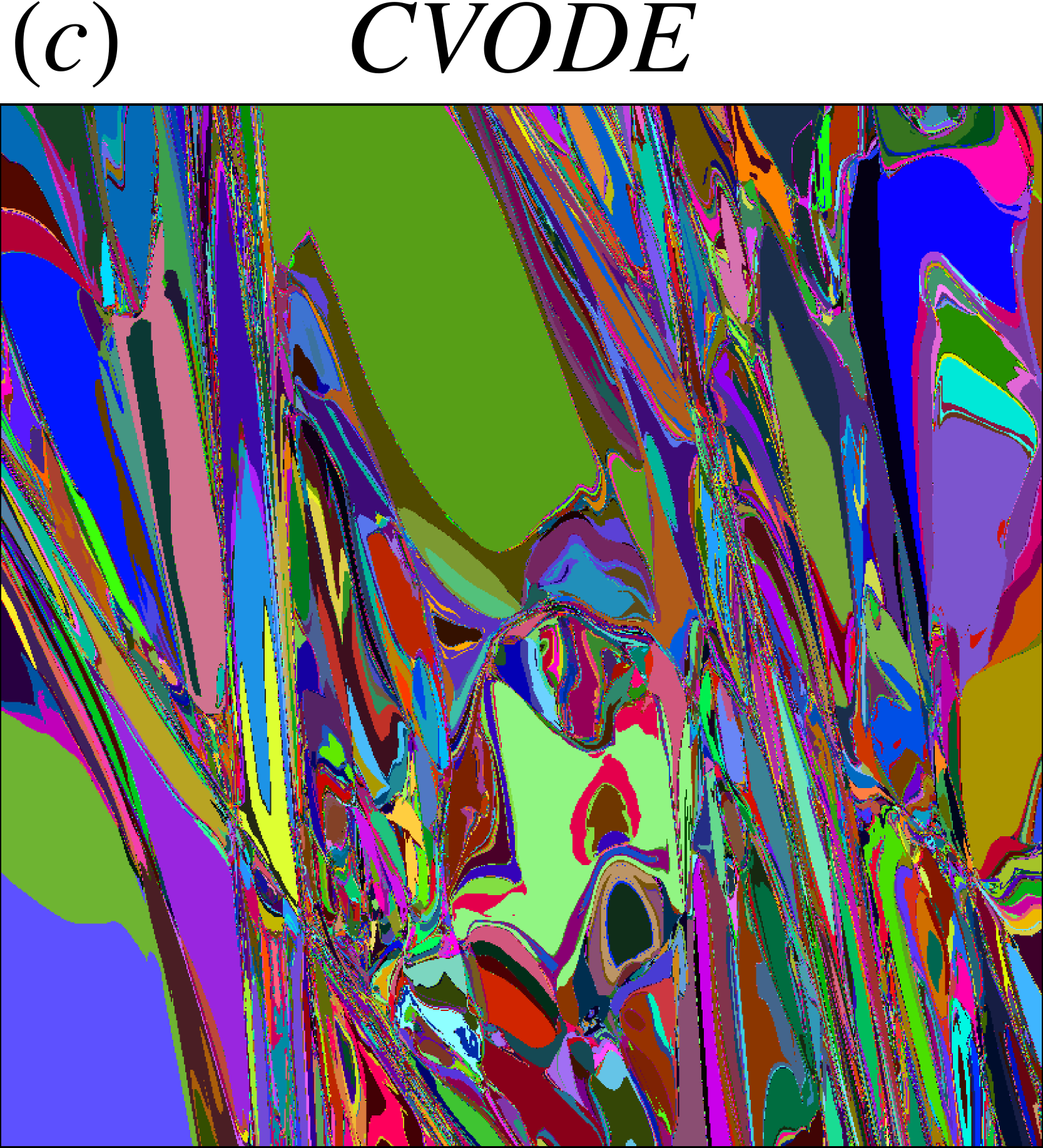}
    \includegraphics[width=0.32\linewidth]{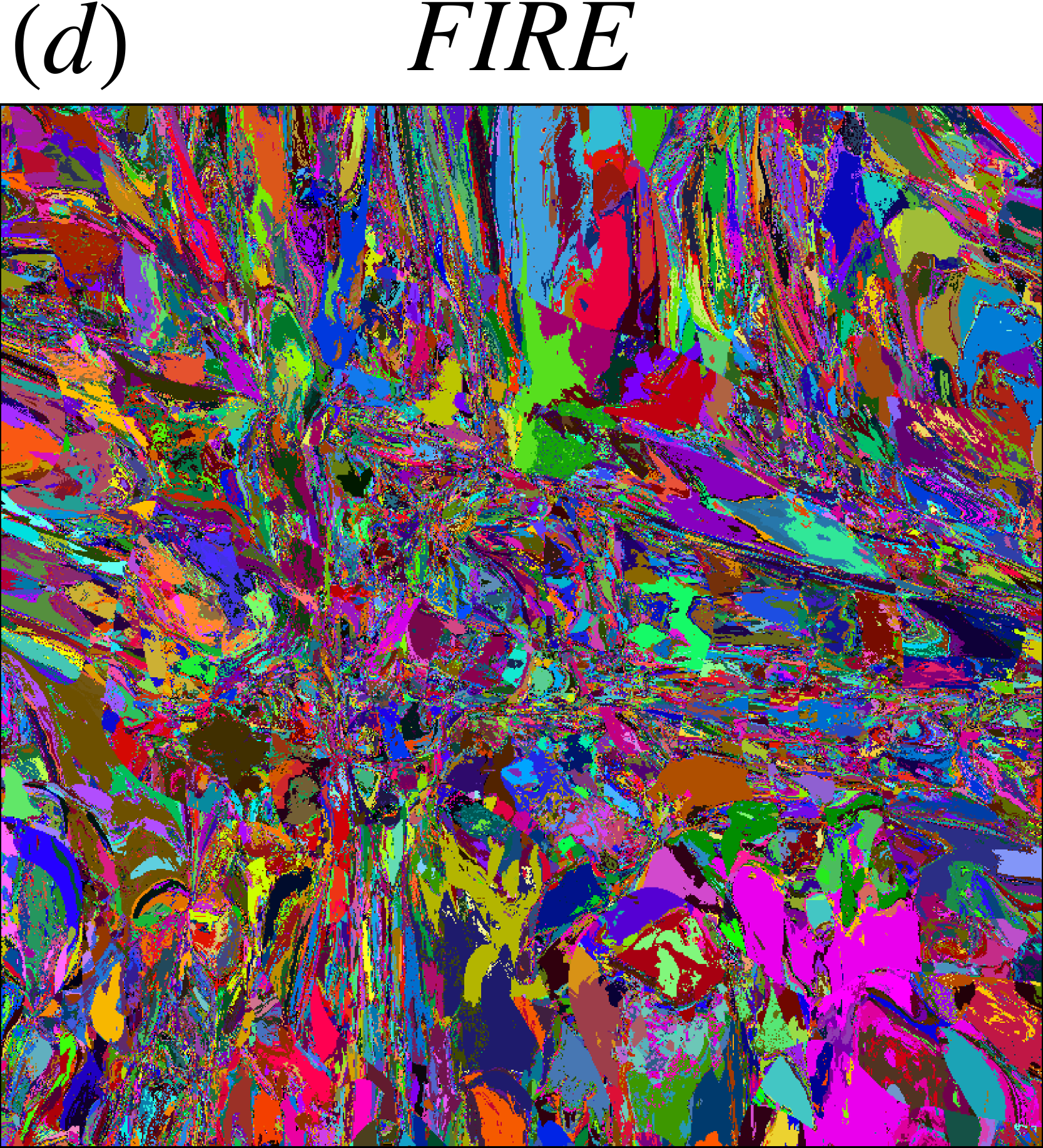}
    \includegraphics[width=0.32\linewidth]{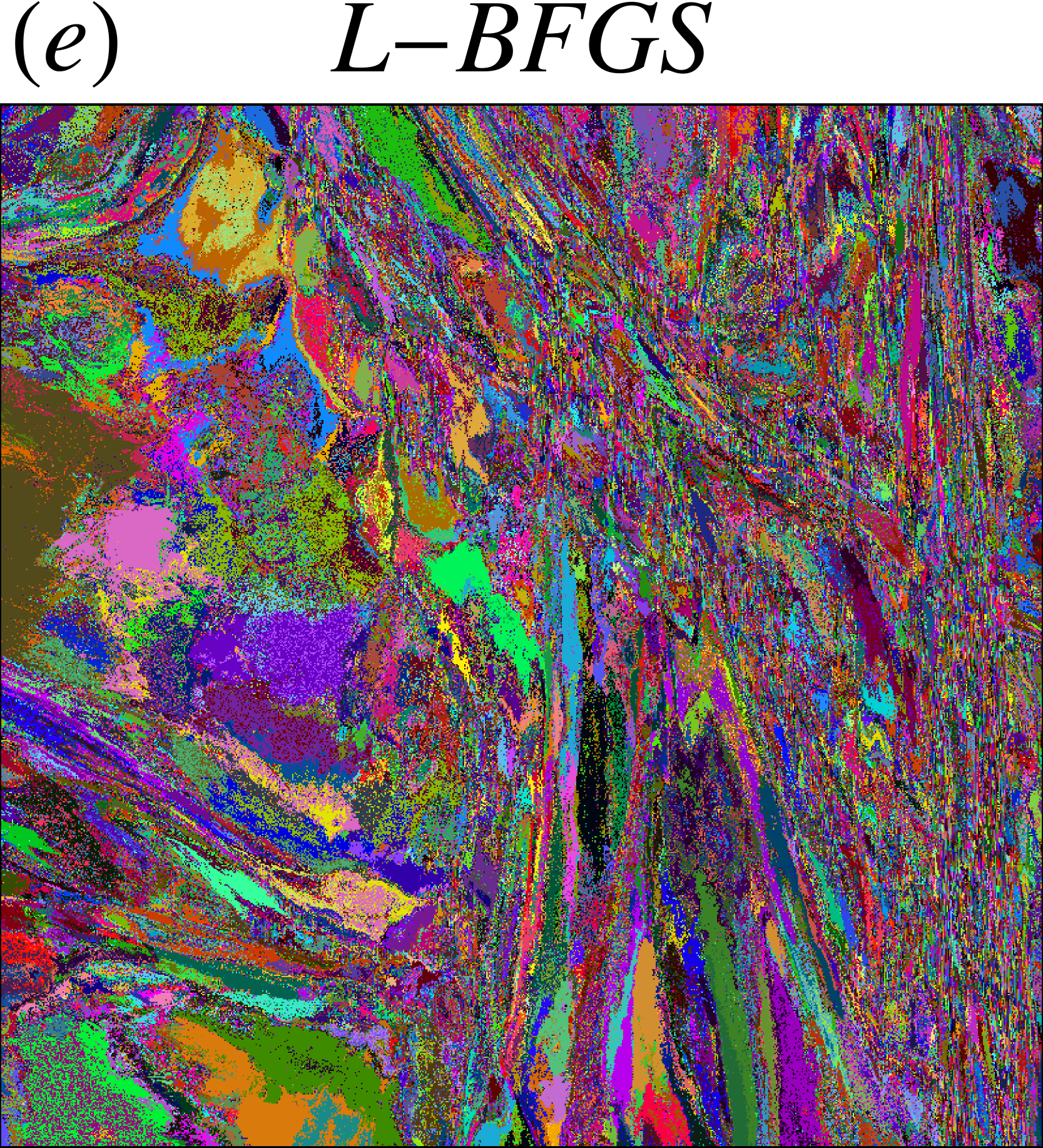}
    \caption{
    \textbf{Optimizers: fast but inaccurate at every $\phi$.}
    $(a)$ Average accuracy of algorithms compared to low-tolerance CVODE, computed over $10^4$ random uniform initial conditions.
    Error bars are $95\%$ confidence intervals, obtained using a Clopper-Pearson estimator.
    Dashed lines are exponential fits, and the long-dashed line is a stretched exponential fit.
    $(b)$ Corresponding scalings of the average computation times, with error bars obtained from a Student-T $95\%$ confidence interval.
    $(c)-(e)$ Like in Fig.~2 of the main text, we show $800\times 800$ pixels slice for $N=128$ particles for $(c)$ CVODE, $(d)$ FIRE, and $(e)$ L-BFGS.    \label{fig:AccuracyTime0p86}
    }
\end{figure}

In Fig.~\ref{fig:AccuracyTime0p86}, we present results analogous to those of Fig.~2 of the main text, but at $\phi = 0.86$ and focus on optimizers compared to CVODE.
We do not consider GD due to the much higher cost associated with it at large $N$.
We show that $(a)$ accuracies and $(b)$ times follow the same trends as those presented at $\phi = 0.9$, namely exponentially decaying accuracies for optimizers, and time scalings $N^{3/2}$ for optimizers but $N^{5/2}$ for CVODE.
Furthermore, landscape slices in Figs.~\ref{fig:AccuracyTime0p86}$(c)-(e)$ are qualitatively similar to those of the main text.
In other words, the results of the main text are not specific to $\phi = 0.9$ but generic to the overcompressed regime of soft spheres.

\subsection{Accuracy dependence on volume fraction}
\label{sec:accuracy-phi-dependence}

To complementing the fixed-$\phi$ analyses presented in the main text and the previous subsection, we here examine how the accuracy of L-BFGS and FIRE optimizers varies as a function of volume fraction $\phi$.

\begin{figure}[h]
    \centering
    \includegraphics[width=0.8\linewidth]{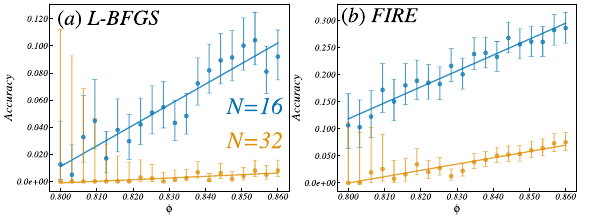}
    \caption{
    \textbf{Optimizer accuracy dependence on volume fraction.}
    Accuracy of $(a)$ L-BFGS and $(b)$ FIRE optimizers compared to high-precision CVODE solutions as a function of volume fraction $\phi$.
    Data points show the fraction of random initial conditions that yield identical minima to the CVODE reference, ignoring fluid states, for system sizes $N = 16$ (blue circles) and $N = 32$ (orange circles).
    Solid lines represent linear regression fit guidelines.
    Error bars are Clopper-Pearson $95\%$ confidence intervals.
    }
    \label{fig:AccuracyPhi}
  \end{figure}

Then, like in other parts of the paper, we measure an accuracy by comparing the minima obtained by optimizers to those found by CVODE using the same data to find the location of the jamming transition in Sec.~\ref{sec:jamming}.
Note that in this measurement, we only consider jammed states obtained with the CVODE reference and ignore the fluid states.
The results are shown in Fig.~\ref{fig:AccuracyPhi}.
We show a degradation of accuracy as $\phi$ decreases.
This confirms the intuition that since the landscape becomes flatter near jamming, the associated numerical ODE solving problem becomes more sensitive to initial conditions, and optimizers have a stronger tendency to be inaccurate.
For $N$'s larger than those shown in Fig.~\ref{fig:AccuracyPhi}, the accuracy is so low near jamming that the measurements are dominated by noise.
Note that our statement made in the main text at $\phi = 0.9$, that for $N\gtrsim64$ optimizers have a vanishing accuracy, is thus shifted to significantly lower $N$ as $\phi$ decreases.
For L-BFGS, for instance, we show that for $\phi < 0.86$ there is essentially zero accuracy already at $N = 32$.

% L-BFGS, N=16: slope=1.517602, R²=0.8664, p=0.0000
% L-BFGS, N=32: slope=0.122688, R²=0.6495, p=0.0000
% 
% FIRE, N=16: slope=2.945237, R²=0.9393, p=0.0000
% FIRE, N=32: slope=1.164045, R²=0.8707, p=0.0000

\subsection{Slices at $N = 128$}
\begin{figure}[h!]
    \centering
    \includegraphics[width = 0.32\columnwidth]{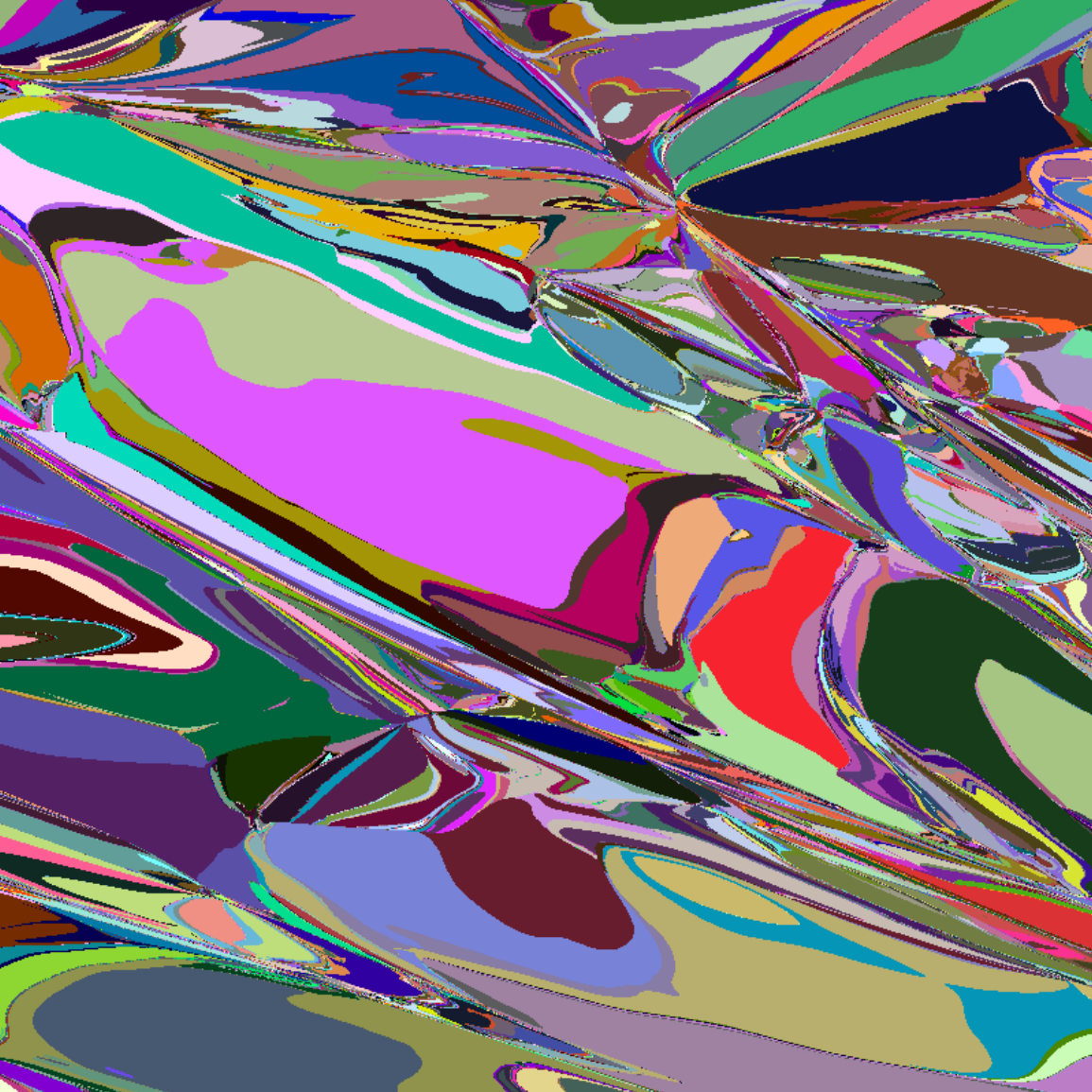}
    \includegraphics[width = 0.32\columnwidth]{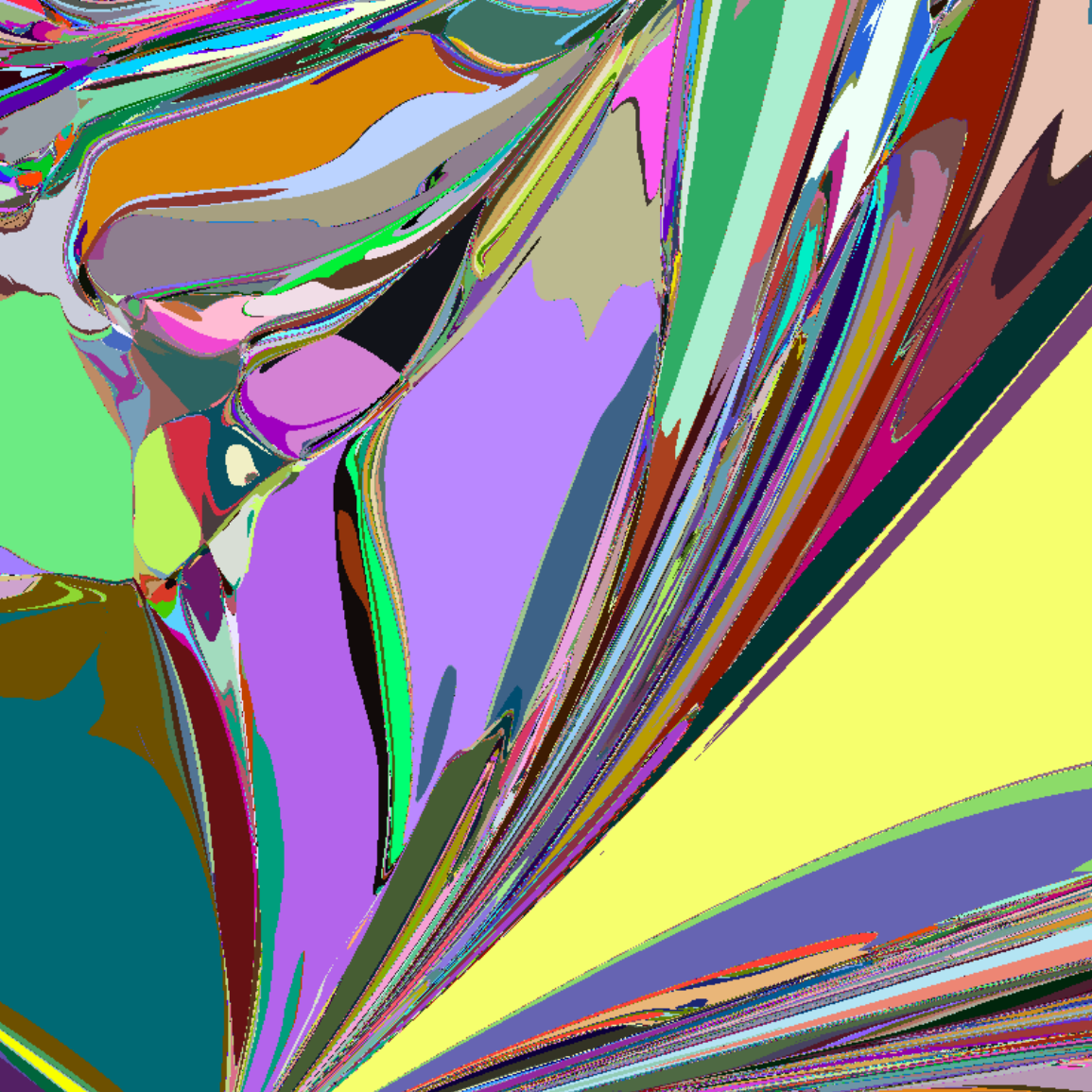}
    \includegraphics[width = 0.32\columnwidth]{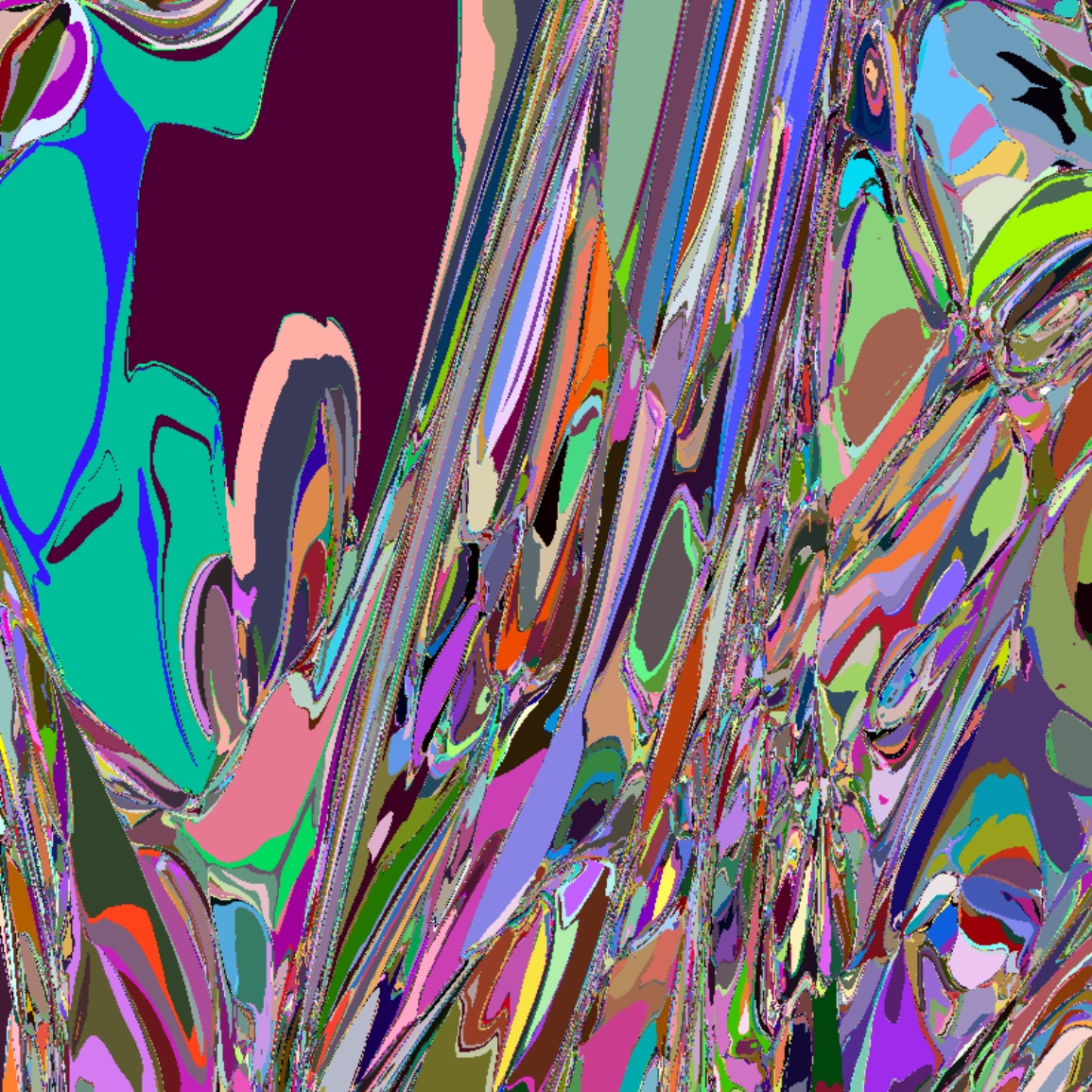} \\
    \includegraphics[width = 0.32\columnwidth]{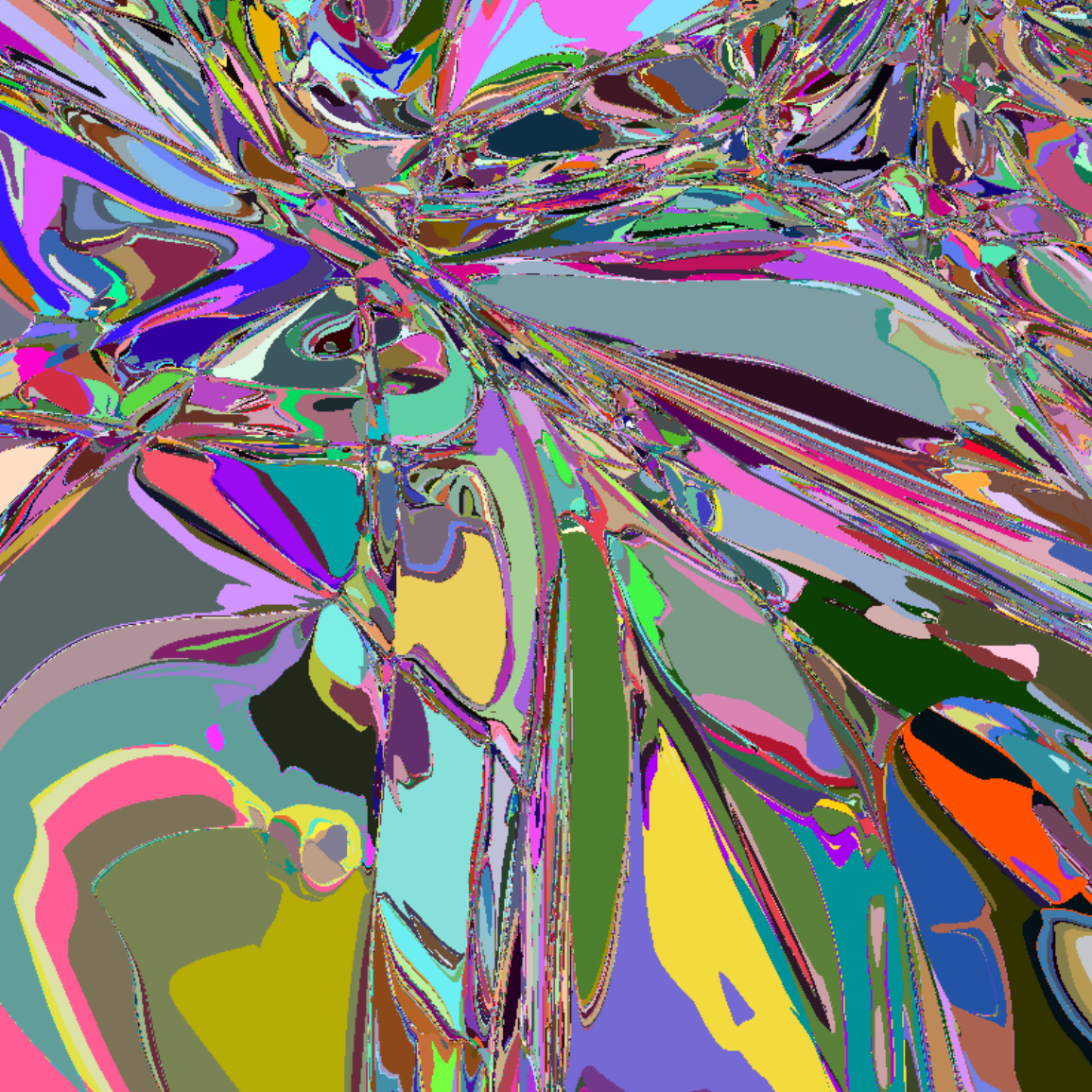}
    \includegraphics[width = 0.32\columnwidth]{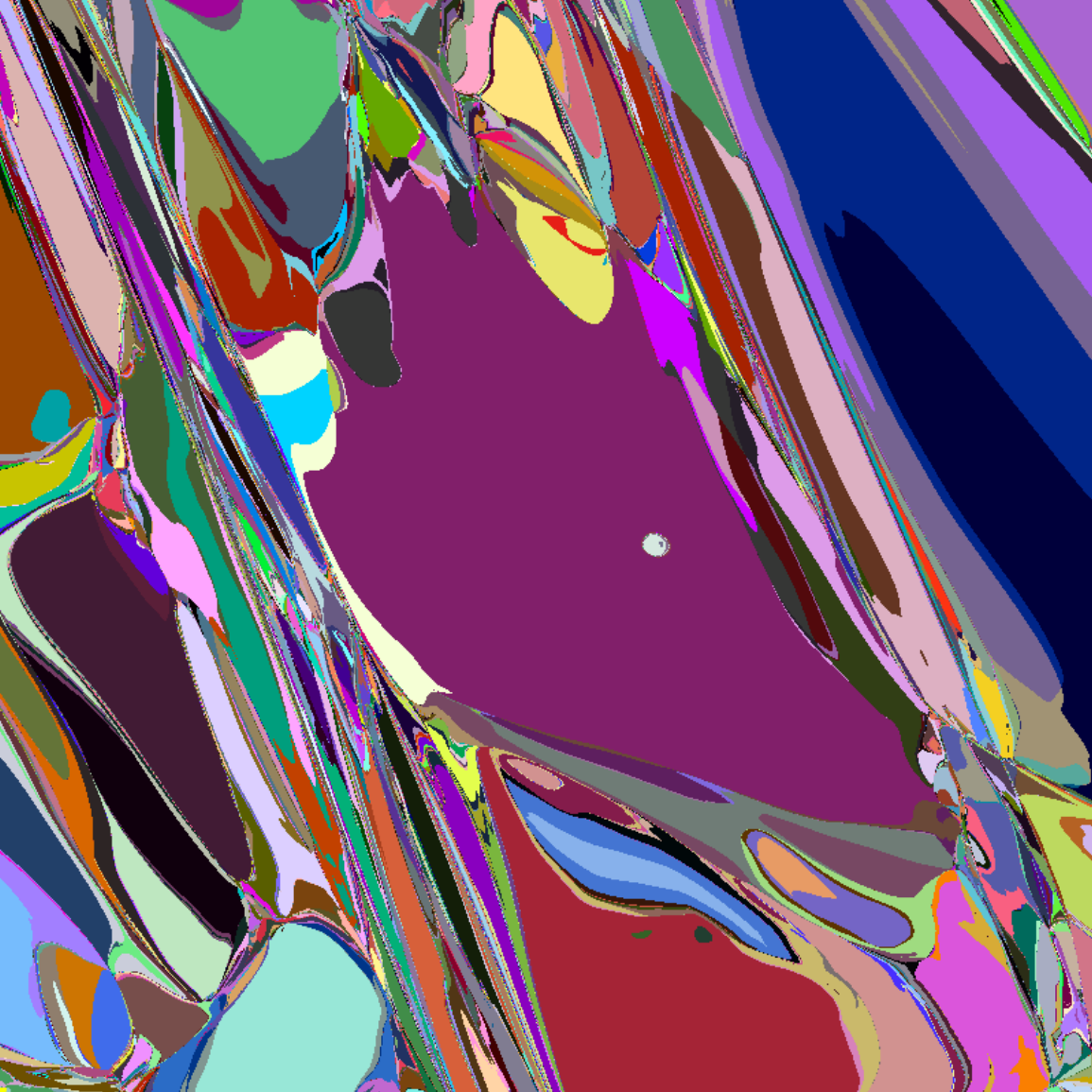}
    \includegraphics[width = 0.32\columnwidth]{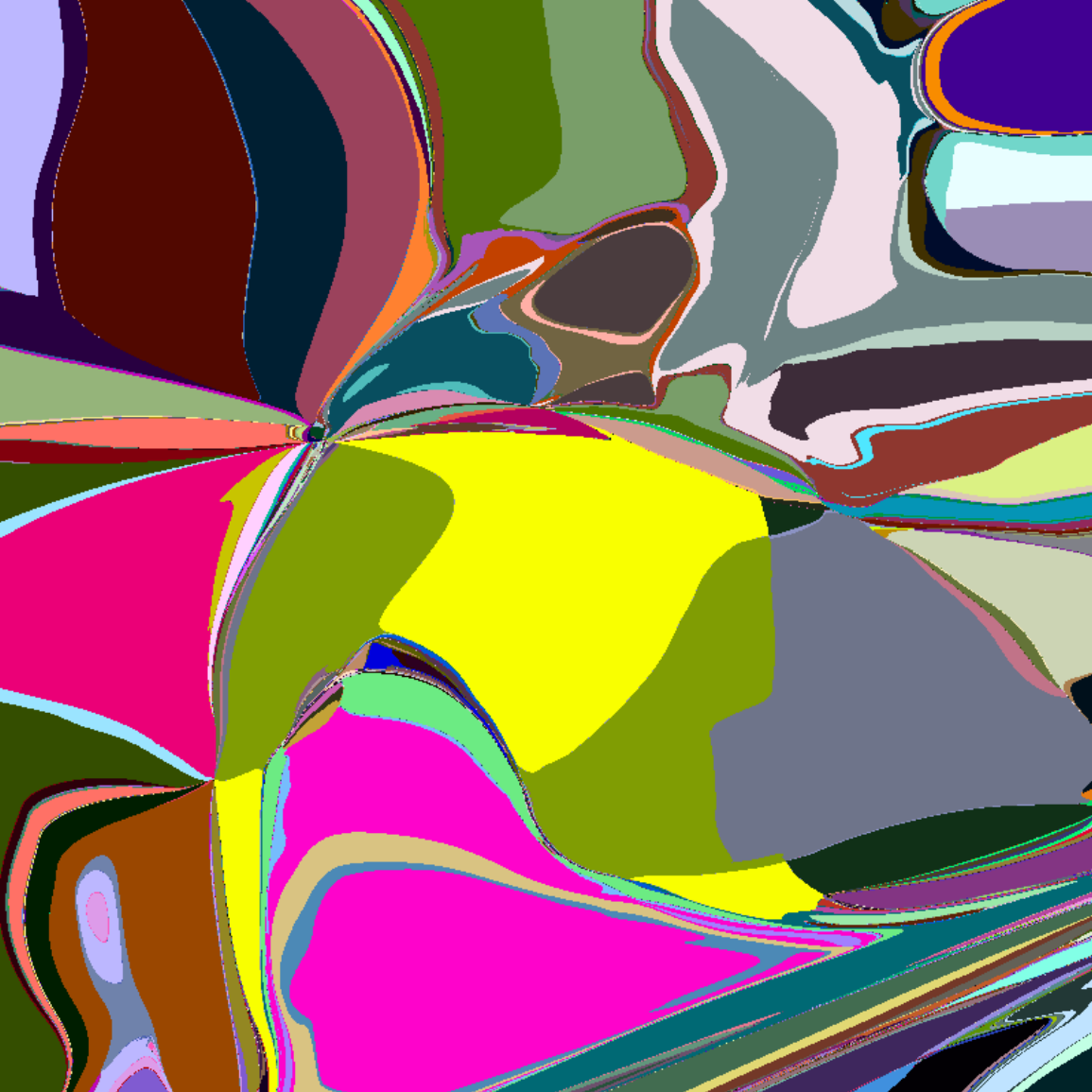} \\
    \includegraphics[width = 0.32\columnwidth]{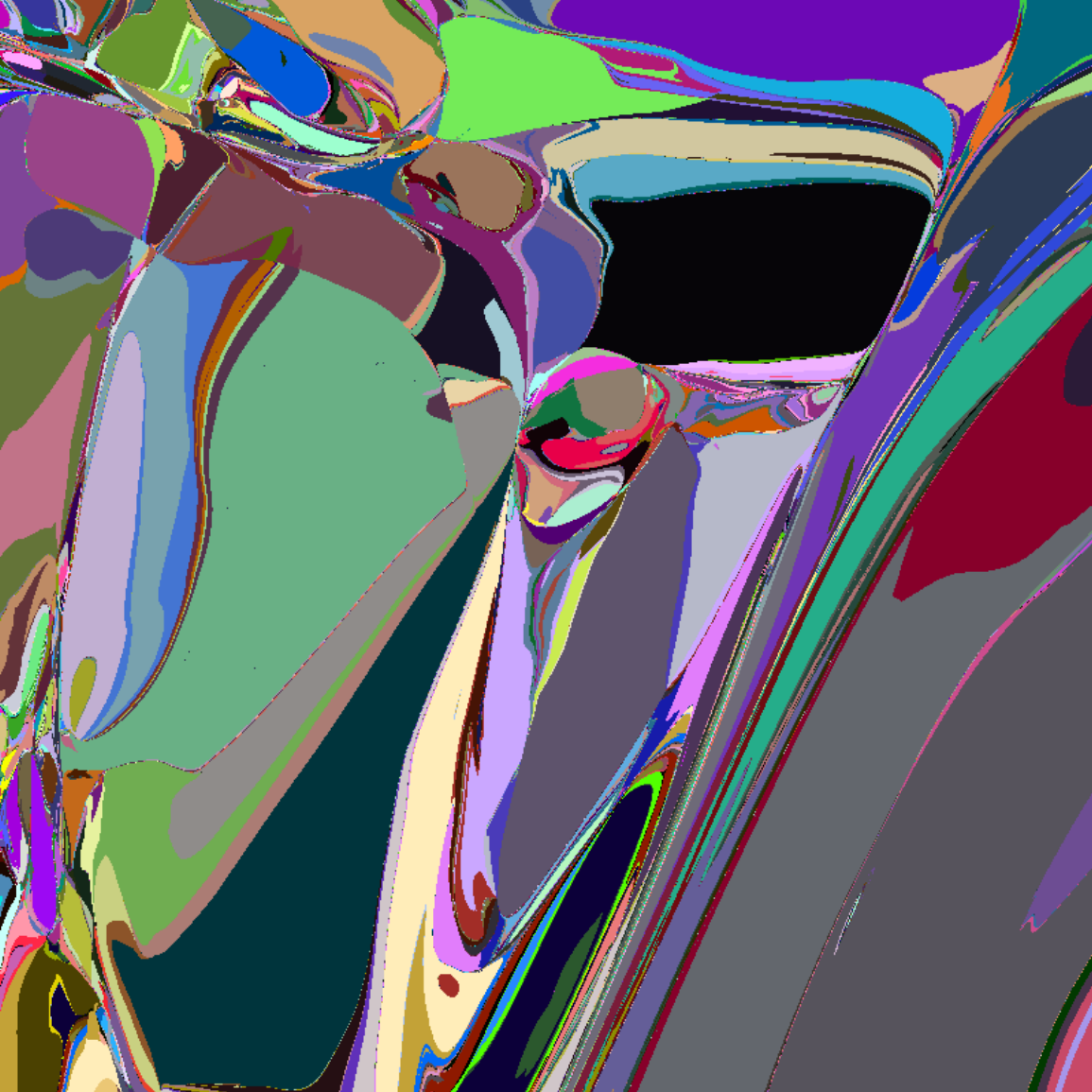}
    \includegraphics[width = 0.32\columnwidth]{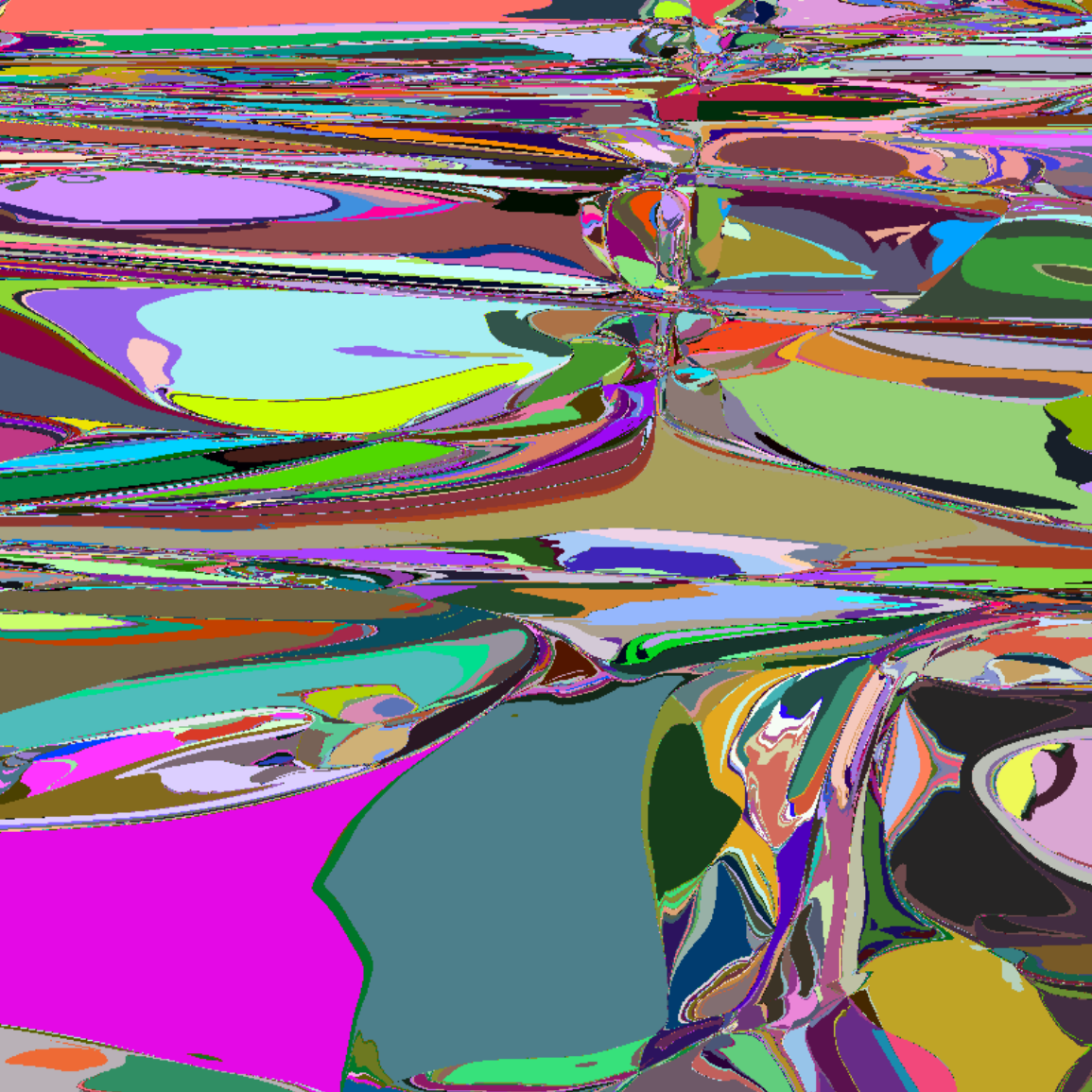}
    \includegraphics[width = 0.32\columnwidth]{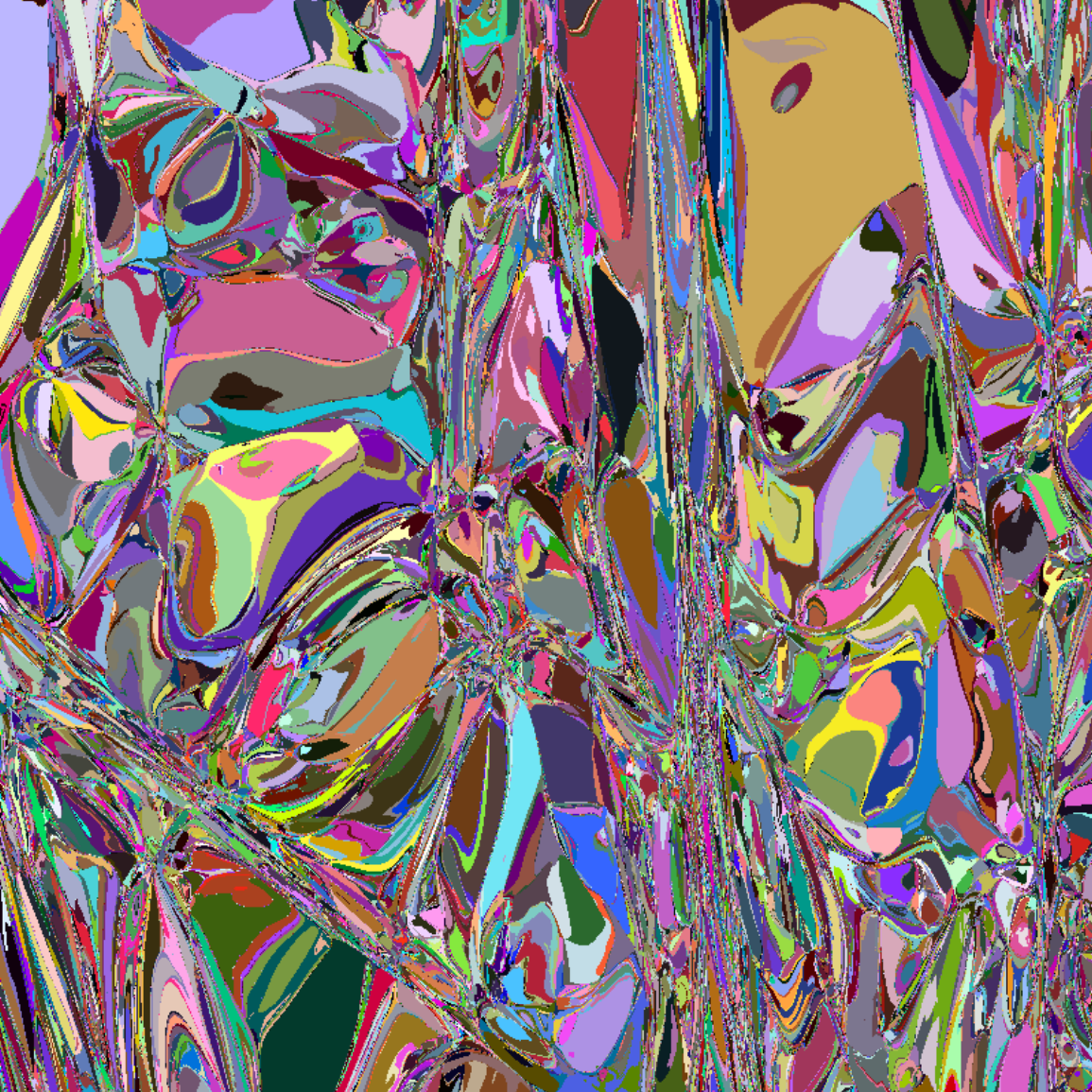}
    \caption{\textbf{Gallery of energy landscape slices.}
    Slices of the landscape of $N = 128$ particles at $\phi = 0.9$, obtained for the same parameters as in Fig.~2 of the main text but with different random seeds.}
    \label{fig:LandscapeSliceGallery}
\end{figure}

In Fig.~\ref{fig:LandscapeSliceGallery}, we present additional slices of the energy landscape for $N = 128$ particles at $\phi = 0.9$, obtained for the same parameters as in Fig.~2 of the main text but with different random seeds.

\bibliographystyle{apsrev4-2}
\bibliography{PostDoc-StefanoMartiniani, supp, supp_extra}